\documentclass[11pt,graphicx,amsmath]{article}
\usepackage{amsmath}
\usepackage{graphicx}
\usepackage{bm}
\usepackage[dvips]{color}
\usepackage{amssymb}
\usepackage{amsfonts}
\usepackage{comment}
\usepackage{cite}
\usepackage{todonotes}
\usepackage{caption}
\usepackage{subcaption}

\def\be{\begin{equation}}
\def\ee{\end{equation}}
\def\ba{\begin{eqnarray}}
\def\ea{\end{eqnarray}}
\def\nn{\nonumber}

\def\bl#1\el{\begin{align}#1\end{align}}
\def\l{\left}
\def\r{\right}

\title{ \bf Adiabatic regularization and Green's function
             of a scalar field in de Sitter space:
             Positive energy spectrum   and no trace anomaly}

\author{\small    \,  Yang  Zhang\thanks{yzh@ustc.edu.cn},  \,
            Xuan Ye  ,
            and Bo Wang   \\
 \small  Department of  Astronomy,  Key Laboratory
               for Researches in Galaxies and Cosmology, \\
 \small    School of Astronomy and Space Sciences, \\
 \small  University of Science and Technology of China,  Hefei, Anhui, 230026,  China \\
 }

 \date{}

\def\be{\begin{equation}}
\def\ee{\end{equation}}
\def\ba{\begin{eqnarray}}
\def\ea{\end{eqnarray}}
\def\nn{\nonumber}
\def\bl#1\el{\begin{align}#1\end{align}}

\begin{document}

\maketitle

\begin{abstract}

In the conventional adiabatic regularization
  the vacuum ultraviolet   divergences
of a quantum field in curved spacetime
are removed by subtracting  the $k$-mode of the stress tensor to the 4th-order.
For a  scalar field in de Sitter space,
we find that the 4th-order regularized spectral energy density is negative.
Moreover,   the 2nd-order regularization for   minimal coupling ($\xi=0$)
and  the 0th-order  regularization for  conformal coupling ($\xi=\frac16$)
yield a positive and UV-convergent spectral
energy density and power spectrum.
The regularized stress tensor in the vacuum is maximally symmetric
and can   drive  inflation,
while its $k$-modes representing the primordial fluctuations
are nonuniformly distributed.
Conventional regularization of a Green's function in position space
is generally plagued by a log IR divergence.
Only in the massless case with $\xi=0$ or $\frac16$,
we can directly regularize the Green's functions and obtain  vanishing results
that agree with the  adiabatic regularization results.
In this case, the regularized power spectrum and stress tensor
are both zero, and no trace anomaly exists.
To overcome the log IR divergence problem in
the massive cases with $\xi=0$ and $\frac16$,
we perform Fourier transformation of the regularized power spectra
and obtain the regularized analytical Green's functions
which are IR-  and UV-convergent.

\end{abstract}

\

{\bf inflationary universe},
{\bf mathematical and relativistic  aspects of cosmology},
{\bf quantum fields in curved spacetimes}


\large

\section{Introduction}\label{section1}

To combine quantum field theory with general relativity,
a   quantum field must first be place in classical curved spacetime.
Quantum fluctuations in the vacuum state contain ultraviolet (UV) divergences
arising  from  sums of the high $k$-modes
\cite{UtiyamaDeWitt1962,FeynmanHibbs1965}.
These divergences manifest in
physical quantities such as the stress tensor and power spectrum,
and  must be removed.
In curved spacetime,
 the UV divergences cannot simply be omitted as in   Minkowski  spacetime,
because the finite part of the fluctuations has important gravitational effects.
For example,
  vacuum fluctuations in the perturbed inflaton field and perturbed metric
were  the origins of large-scale structures \cite{Peebles1993,WangZhang2017}.
They also induced  CMB anisotropies and polarization
in the cosmic  microwave background \cite{MaBertschinger1995,ZhaoZhang2006}.

Proper removal of   UV divergences
is the most important priority when introducing
quantum fields in classical curved spacetimes.
Among the several existing regularization approaches
are     dimensional regularization
\cite{CandelasRaine1975,DowkerCritchley1976,Brown1977,Bunch1979},
  point-splitting \cite{Christensen1976,AdlerLiebermanNg1977,
Christensen1978,BunchDavies1978,BunchChristensenFulling1978,Wald1978},
and the zeta function \cite{DowkerCritchley1976,Hawking1977}
that can be also written as a  path-integral  \cite{Fujikawa1980}.
These methods are applied to the Green's function,
and the stress tensor can be constructed from derivatives of the Green's function.
In  these methods the Green's function of a massive field
is often expressed in terms of the DeWitt-Schwinger proper-time integration
      \cite{Schwinger1951,DeWitt1975},
and they are virtually  equivalent \cite{DowkerCritchley1976}.
In particular,
dimensional regularization is known to agree with
the zeta function regularization \cite{DowkerCritchley1976,Hawking1977}.
 UV divergences are isolated by expanding
the Green's function in position space over a small distance $\sigma$.
However, the $\ln \sigma $ term in this process is both UV- and infrared (IR)-divergent.
Subtracting the $\ln \sigma $ term achieves  a UV-convergent Green's function,
but  also introduces an IR divergence
in the  originally IR-convergent Green's function
(as  demonstrated below).
This difficulty plagues all regularization methods based on
the Green's function in position space.
Previous researchers have selected
a particular term (of 4th adiabatic order)
from the series expansion of the DeWitt-Schwinger proper-time integration,
and taken its massless limit $m=0$ while neglecting other terms.
The result is
the so-called trace anomaly \cite{DowkerCritchley1976,Christensen1976,Hawking1977,
Christensen1978,Brown1977,BunchChristensenFulling1978,Wald1978}
of a conformally-coupling scalar field.
However, the series expansion of
the DeWitt-Schwinger proper-time integration is defined only at $m\ne 0$
and is actually  undefined at $m=0$
\cite{DeWitt1975,Christensen1976,Christensen1978}.

The UV-divergence problem can also be solved by
adiabatic regularization \cite{ParkerFulling1974,FullingParkerHu1974,
HuParker1978,BLHu1978,Birrell1978,
Bunch1978,Bunch1980,AndersonParker1987,BunchParker1979,
BirrellDavies1982,ParkerToms,Parker2007,MarkkanenTranberg2013}
which   directly  handles  the $k$-modes of a quantum field.
The spirit of this method is analogous to Einstein's equivalence principle,
that is, the leading high $k$ modes regard  curved spacetime as a flat spacetime.
These high $k$ modes are UV divergent and should be subtracted
like the UV-divergent term in quantum field theories of flat spacetime.
The UV divergences of the $k$-modes  next to the leading mode
are subtracted by the adiabatic subtraction terms
 formed from the Wentzel--Kramers-Brillouin (WKB) approximations.
For this purpose,  a minimal subtraction rule
that minimizes the  adiabatic subtraction
and retains the convergent terms is required.
This  method  removes UV divergences
more effectively than the Green's function-based methods;
in particular, it avoids the log-term difficulty
of the Green's function in position space,
without causing  IR divergence of  a massive scalar field.
In   conventional adiabatic regularization,
the order of the adiabatic expansion is
the order of the time derivatives of the scale factor $a(\tau)$
in Robertson-Walker (RW) spacetime.
Meanwhile,  the minimal subtraction rule \cite{ParkerFulling1974}
commonly  prescribes
the 4th-order and the 2nd-order subtractions for the stress tensor
and the power spectrum of a scalar field, respectively.
In principle these  prescriptions will  sufficiently remove
 the UV divergences.
Parker \cite{Parker2007} explicitly obtained
a 2nd-order regularized power spectrum
for a minimally-coupling massive scalar field during  de Sitter inflation.
However,  an IR- and UV-convergent regularized spectral stress tensor
of a massive scalar field during de Sitter inflation
remains elusive.

For a massless field,
the adiabatic order matches the power of $k$,
such  as $(\frac{1}{k})^n \rightarrow (a'/a)^n$.
In our previous work on
  relic gravitational wave  (RGW) \cite{WangZhangChen2016},
we implemented adiabatic regularization
in which the 4th-order regularization
is both sufficient and necessary for  removing   the $k^4, k^2, k^0$  divergences
from the  stress tensor,
whereas  the 2nd-order regularization is sufficient and necessary
for  removing   the  $k^2, k^0$  divergences from  the power spectrum.
The slight IR distortions at $k=0$ caused by  the subtraction terms
 can be avoided by the inside-horizon scheme \cite{ZhangWangJCAP2018}.

The adiabatic subtraction terms of a massive  field
contain powers of  $\omega=(k^2+m^2 a^2)^{1/2}$, rather than powers of $k$.
The 4th-order subtraction term for the stress tensor
is constructed from   time-derivatives of $a(\tau)$ up to fourth-order.
Therefore, it  contains   powers of
$\omega^{1}$, $\omega^{-1}$, $\omega^{-3}$, ...  $\omega^{-13}$
\cite{Bunch1980,AndersonParker1987,BirrellDavies1982,ParkerToms}.
Similarly,   the 2nd-order  subtraction term for the power spectrum
is constructed from  time-derivatives of $a(\tau)$  up to second order
and thus contains  powers of $\omega^{-1}$, ...  $\omega^{-7}$ \cite{Parker2007}.
When the (positive or negative)  powers  of $\omega$
are further expanded to high-order terms of $k$,
the subtraction terms involve   infinitely many terms in powers of $k^{-1}$.
Two consequences arise from this property.
First,  the 4th-order subtraction term   will cancel not only the
$k^4, k^2, k^0$  divergences,
but also some of the convergent $k^{-2}, k^{-4} $ terms
in   $\langle T^\mu\, _\nu \rangle_{k}$ for a massive field.
Therefore,   the regularized spectral energy density becomes
  negative (non-positive-definite).
Second,  the    lower-order (0th- and 2nd-order) subtraction terms
also contain $k^4, k^2, k^0$ terms,
whose  coefficients
might be exactly poised to  cancel the respective divergences
and yield a positive, UV-convergent spectral energy density.

Both consequences are observed in actuality.
Specifically,
for a minimal coupling ($\xi=0$) massive scalar field in the vacuum state
in de Sitter space,
the 2nd-order adiabatic regularization yields
simultaneously yields  a positive UV-convergent  power spectrum
and  a  positive UV-convergent spectral energy density
which respects  the covariant conservation.
In  a conformally-coupling ($\xi=\frac16$) massive scalar field,
the 0th-order regularization achieves a similar result.
These two  prescriptions achieve the   proper
  adiabatic regularization for their  respective cases.
In  both  cases,
the 4th-order regularization is an incorrect prescription,
because  it gives a negative spectral energy density
and a negative power spectrum.
For a massless scalar field  with $\xi=0$ and   $\xi=\frac16$,
the regularized stress tensor and power spectrum are zero,
and,  no conformal trace anomaly appears in  the $\xi=\frac16$ case.
In the literature \cite{Bunch1980,AndersonParker1987}
a trace anomaly appeared  in the 4th-order adiabatic regularization
of a massive scalar field with $\xi=\frac16$ in the limit $m=0$.
The present paper will demonstrate the inappropriateness
     of  the  4th-order regularization
in a scalar field  with $\xi=\frac16$.
  The  scalar field is assumed  in  de Sitter space,
in which the exact modes
and the adiabatic subtraction terms of various orders
 are explicitly given.
Taking the difference yields
the regularized spectral stress tensor and power spectrum, respectively,
in each order.

For a general coupling  $\xi \in (0, \frac16)$,
we show by trial and error   that
 no  regularization of a fixed adiabatic order
can  simultaneously provide a positive
 spectral energy density and a positive power spectrum
 with UV convergence.

To  complement the adiabatic regularization,
we also  regularize
the Green's function in position space.
 We first demonstrate
the $\ln \sigma $ IR divergence in a coupling massive scalar field.
Only when  $m=0$ and  $\xi=\frac16$ or $\xi=0$,
the Green's function involves  only one (or two) UV-divergent term
and is regularized to zero,
confirming that the adiabatic regularization yields  zero regularized spectra.
We also examine the literature-reported  procedures for obtaining the trace anomaly
by the Green's function method,
and show that the result is an artifact by invalid treatments
of the massless limit of the  4th-order term.
Furthermore,
we   perform Fourier transformations
on the adiabatically regularized
power spectra of the massive scalar field with $\xi=0$ and $\frac16$,
and obtain the respective regularized Green's functions
with  IR- and UV- convergence.
This     regularization
overcomes the log IR-divergence difficulty
in the  Green's function method.

The remainder of this  paper is organized as follows.

Section \ref{section2}  briefly introduces the exact solution of
a massive scalar field during  de Sitter inflation, namely,
the spectral stress tensor and the power spectrum
in the vacuum state.

Section  \ref{section3} explores
 the 4th-, 2nd-, and 0th-order  regularization prescriptions for $\xi=0$.
 Only the 2nd-order regularization is proven  successful,
yields a zero stress tensor and a zero power spectrum
in the massless case.

Section  \ref{section4} explores
 the 4th-, 2nd-, and 0th-order regularization prescriptions for  $\xi=\frac16$.
In this case, only the 0th-order regularization  works successfully,
and,  in the massless case,  yields a zero stress tensor and a zero power spectrum
with  no trace anomaly.

In Sect.\ref{section5},  we explore the possible regularizations
     for a general   $\xi \in (0, \frac16)$,
    and  show that negative spectra always occur.

Section \ref{section6}  regularizes  the Green's function,
and pinpoints   the log IR difficulty and
the invalid treatments  leading  to the trace anomaly.
In this section, we also  perform Fourier transformations of
the regularized power spectra for $\xi=0$  and $\frac16$,
thereby  obtaining  the IR- and UV-convergent Green's functions.

Section  \ref{section7}  discusses the results and   concludes the paper.

Appendix \ref{sectionA}   lists the  high $k$ expansions of the exact modes.
Appendix \ref{sectionB} lists  the 0th-, 2nd-, and 4th-order adiabatic modes
                 used in the context of this paper.
Finally,  \ref{sectionC} lists  the 0th-, 2nd- and 4th-order subtraction terms
        for the stress tensor,
        and demonstrates  the  zero four-divergence  of  each order.
In this paper, we set $c=\hbar= 1$.

\section{ The power spectrum and stress tensor
        of scalar field during de Sitter inflation}\label{section2}

\begin{figure}[htb]
\centering

\subcaptionbox{}
    {%
        \includegraphics[width = .48\linewidth]{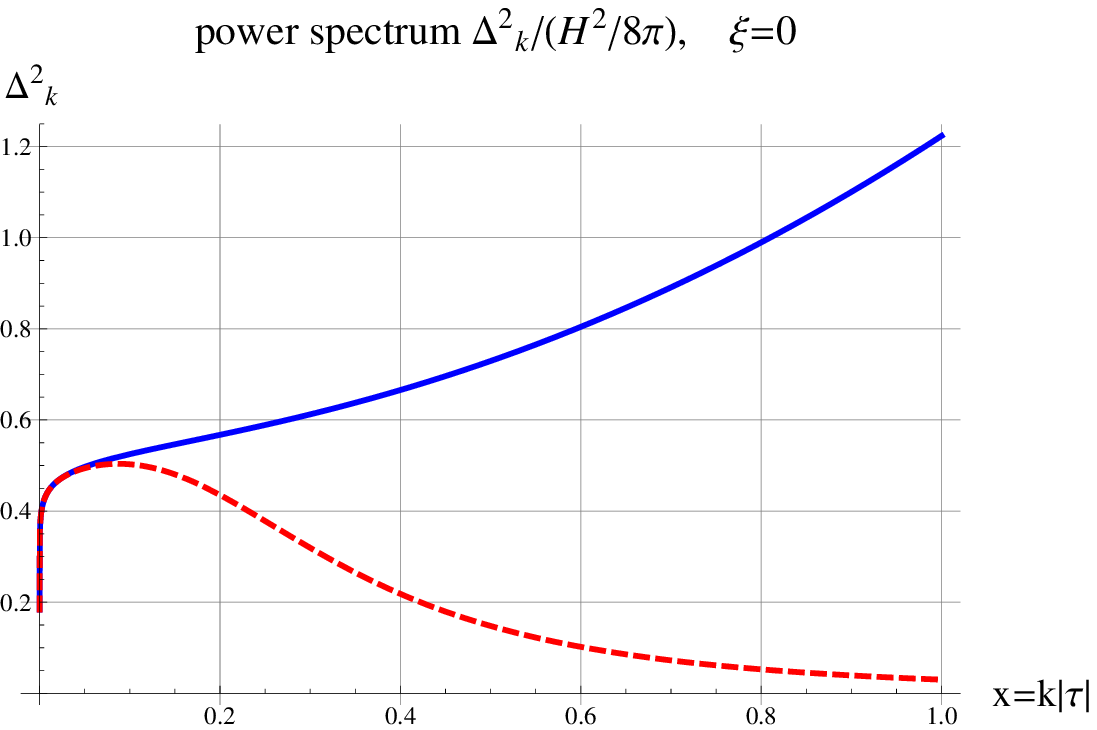}
    }
\subcaptionbox{}
    {%
        \includegraphics[width = .48\linewidth]{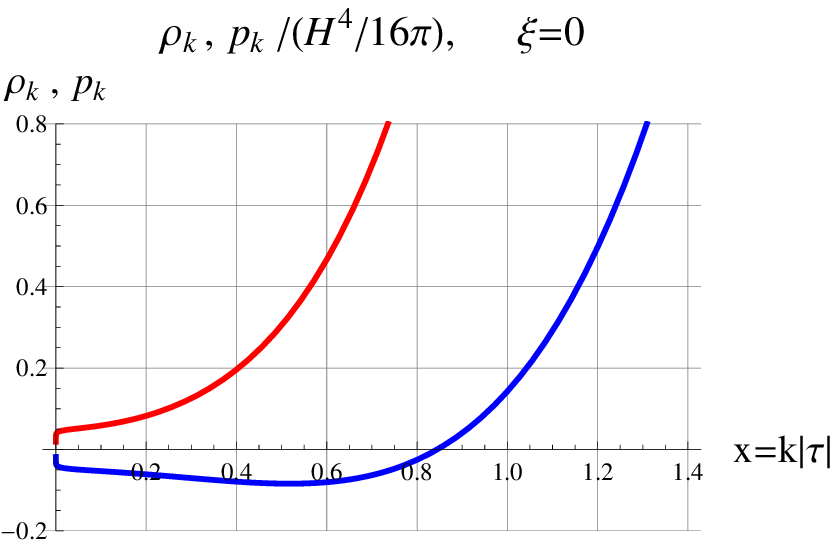}
        }
\subcaptionbox{}
    {%
        \includegraphics[width = .48\linewidth]{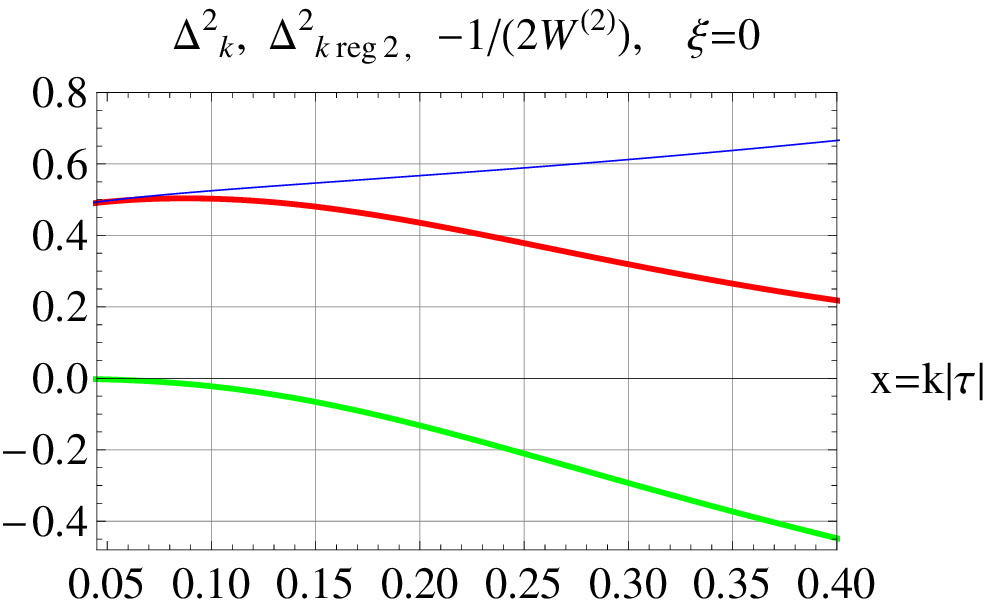}
        }
\caption{  The plot is at a time $|\tau|=1$ in de  Sitter space
     and for the model $\frac{m^2}{H^2}=0.1$ and $\xi=0$.
     (a) Blue Solid: the unregularized power spectrum $\Delta^2_k $
    is IR-convergent at low $k$, and is $k^2$ UV-divergent  at high $k$.
    Red Dash:  the regularized $\Delta^2_{k\, reg} $
    is  $k^{-2}$ UV-convergent  at high $k$,
    agreeing with that of Ref\cite{Parker2007}.
    (b)    Red: the  unregularized spectral energy density.
           Blue: the  unregularized spectral pressure.
    (c)   Blue: the unregularized  $\Delta^2_k $.
          Red: the regularized  $\Delta^2_{k\, reg}$.
          Green: the subtraction term,
        which is much lower than the unregularized.
    }
     \label{Figure1}
\end{figure}

The metric of  a flat Robertson-Walker spacetime  is
\be \label{metric}
ds^2=a^2(\tau)[d\tau^2- \delta_{ij}   dx^idx^j],
\ee
with  the conformal time $\tau$.
 The Lagrangian density of  a scalar field $\phi$   is
\be
{\cal L} =\frac12 \sqrt{-g}(g^{\mu\nu}\phi_{,\mu}\phi_{,\nu}-m^2\phi^2-\xi R\phi^2) ,
\ee
and the field  equation is
\be    \label{fieldequxi}
(  \Box +m^2 + \xi R   )\phi =0
\ee
where
$\Box = \frac{1}{a^4} \frac{\partial}{\partial \tau}
(a^2 \frac{\partial}{\partial \tau}) -\frac{1}{a^2} \nabla^2$
is the generalized D'Alembertian operator,
$m$ is the mass, $\xi $ is a  coupling constant,
$R= 6 a''/a^{3}$ is  the scalar curvature.
Write
\be
\phi  ({\bf r},\tau) = \int\frac{d^3k}{(2\pi)^{3/2}}
        \left[ a _{\bf k}   \phi_k(\tau)  e^{i\bf{k}\cdot\bf{r}}
    +a^{\dagger}_{\bf k} \phi^{*}_k(\tau) e^{-i\bf{k}\cdot\bf{r}}\right]
\ee
where  $ a_{\bf k}, a^{\dagger}_{\bf k'} $ are
the annihilation and creation operators
and satisfy the canonical commutation relations.
Since the field equation is linear,
the $k$-modes are independent of each other.
Let $\phi_k(\tau) =  v_k(\tau)/a(\tau)$.
The   equation of $k$-mode $v_k$   is
\be\label{equvk}
v_k'' + \Big( k^2 +m^2 a^2 + ( \xi  -\frac16 ) a^2 R  \Big) v_k = 0 .
\ee
It should be mentioned
that the  conformally-coupling massless case ( $\xi=\frac16$ and $m=0$)
is special in that
Eq.(\ref{equvk}) reduces to
$v_k'' +   k^2    v_k = 0 $
which is a wave equation in Minkowski spacetime.
We consider the de Sitter inflation with a scale factor
\be \label{inflation}
a(\tau)=\frac{1}{H |\tau| },\,\,\,\,-\infty<\tau\leq \tau_1,
\ee
and   $R  = 12 H^2$,
where $H$ is a constant, and
$\tau_1$ is the ending time of inflation.
(This de Sitter inflation can be driven by several possible alternative sources,
such as  the cosmological constant,
the quantum effective Yang-Mills condensate \cite{Zhang1994},
and some type of coherent scalar inflaton field.
As we shall see later in Sects. \ref{section3} and \ref{section4},
it can be also driven by the regularized stress tensor
of the massive scalar field $\phi$  in the vacuum.)
Then
\be\label{equxi}
v_k'' + \Big[ k^2 +  \big( \frac{m^2}{H^2}
       +   12 \xi -2 \big) \tau^{-2} \Big] v_k = 0 .
\ee
The general solution of Eq.(\ref{equxi}) is
$ \propto x^{1/2} H^{(1)}_{ \nu } (x)$, $x^{1/2} H^{(2)}_{\nu} (x)$,
where  $ H^{(1)}_{\nu} (x)=  H^{(2)\, *}_{\nu} (x)$
are the Hankel functions, the variable   $x \equiv  k |\tau|$,  and
\be
\nu \equiv  \big(\frac94  - \frac{m^2}{H^2} -12 \xi \big)^{1/2} .
\ee
For concreteness   in this paper we consider $\nu \leq  3/2$
and  $\xi \geq 0$.
The normalized solution is taken to be
\be  \label{u}
v_k (\tau )  \equiv  \sqrt{\frac{\pi}{2}}\sqrt{\frac{x}{2k}}
  e^{i \frac{\pi}{2}(\nu+ \frac12) } H^{(1)}_{\nu} ( x)
\ee
which approaches  the positive-frequency mode
$v_{k }(\tau ) \rightarrow \frac{1}{\sqrt{2k} }e^{-i  k\tau }$
in the high $k$ limit.
Corresponding to this,   the Bunch-Davies (BD) vacuum  state
is defined as the state vector $|0 \rangle$ such that
\be\label{ask}
a_{\bf k}  |0 \rangle =0, ~~~
      {\rm for\   all} \  {\bf k} .
\ee
The Green's function in the BD vacuum is defined as
\ba\label{Greenunregdf}
G({\bf r}, {\bf r}')
   =  \langle0|  \phi(\textbf{r},\tau) \phi (\textbf{r}',\tau) |0\rangle
   = \frac{1}{|r-r'|} \int_0^\infty  \frac{\sin( k|r-r'|)}{k^2}
          \Delta^2_k (\tau)  \, d k
\ea
for the equal-time $\tau=\tau'$ case,
and  the auto-correlation  function  is
\ba\label{vevcorr}
\langle0|  \phi(\textbf{r},\tau) \phi (\textbf{r},\tau) |0\rangle
         =   \frac{1}{(2\pi)^3} \int d^3k \,  |\phi_k(\tau)|^2
         = \int_0^{\infty} \Delta_k^2 (\tau)\frac{dk}{k}  ,
\ea
and the   power spectrum associated with the Green's function
 is
\ba \label{BunchDaviesSpectrum}
\Delta_k^2 (\tau)
&  = & \frac{ k^{3}}{2  \pi^2 a^2 }   |v_k(\tau)|^2
   =  \frac{H^2 }{8 \pi  }     x^3
      \big|  H^{(1)}_{\nu} ( x)   \big|^2
\ea
 which is nonnegative by definition.
At low $k$ the power spectrum
is IR convergent  for the massive field.
At  high $k$ (i.e.,  $x \gg 1$),  by  (\ref{vksq}),
$\Delta_k^2   \propto
   k^{3} \big( \frac{1}{2k}  +\frac{4\nu^2 -1}{16  k^3 \tau^2}  \big)
   + O(k^{-2})$
leads to  quadratic and  logarithmic
UV divergences in the $k$-integration of the auto-correlation (\ref{vevcorr}).
The spectrum is shown by solid line in Figure \ref{Figure1}(a).
In this paper   for illustration the plots are at a time $|\tau|=1$
and with $\frac{m^2}{H^2}=0.1$ except specified otherwise.
These  UV divergences are to be removed
by the following  subtraction
\be
\Delta^2_{k\, \, reg} =\Delta^2_{k}- \Delta^2_{k\, \, sub}
\ee
where $\Delta^2_{k\, \, sub}$ is
a subtraction term  of certain adiabatic order,
formed from the WKB approximate  solution. (See Appendix \ref{sectionB}.)
The conventional prescription \cite{Parker2007}
adopts the 2nd-order adiabatic subtraction to the power spectrum with $\xi=0$.
For other values of coupling $\xi$,  nevertheless,
the 2nd-order subtraction will generally lead to a negative power spectrum,
as we shall see in later sections.
Therefore,  we shall try different order of subtraction for different $\xi$.
For this purpose,
we propose the following criteria
for an  adequate regularized  power spectrum:
(1) UV convergent, (2) IR convergent, (3) nonnegative.

The energy momentum tensor is given by
\cite{DeWitt1975,DowkerCritchley1976,Bunch1980,AndersonParker1987}
\ba
T_{\mu\nu} &  = &  (1-2\xi) \partial_ \mu \phi \partial_ \nu \phi
  +(2\xi - \frac12) g_{\mu\nu } \partial^\sigma  \phi \partial_\sigma  \phi
  -2\xi \phi_{;\mu\nu} \phi
  \nn \\
&&  + \frac12 \xi g_{\mu\nu} \phi \Box \phi
 -\xi (R_{\mu\nu}-\frac12 g_{\mu\nu} R + \frac32 \xi R g_{\mu\nu}) \phi^2
+ (\frac12 -\frac32 \xi )g_{\mu\nu }m^2 \phi^2
\ea
which satisfies the conservation law $T^{\mu\nu}_{~~ ;\nu} = 0$
using the field equation (\ref{fieldequxi}).
The trace is
\be\label{traceTmunu}
T^\mu\, _\mu = (6\xi -1) \partial^ \mu \phi \partial_ \mu \phi
          +\xi (1-6\xi) R \phi^2
          +2 (1-3\xi) m^2 \phi^2 ,
\ee
and,  for  $\xi=\frac16$ it reduces to
\be\label{traceTmunu16}
T^\mu\, _\mu = m^2 \phi^2 .
\ee
This expression   reveals that
the trace of  stress tensor of the conformally-coupling massless scalar field
is  zero  by definition.
The energy density in the BD vacuum state is given by the expectation value
\be \label{energyspectr}
\rho  = \langle T^0\, _0 \rangle =\int^{\infty}_0   \rho_k \frac{d k}{k}
\ee
where the spectral energy density  is
\ba \label{rhok}
\rho_k  & = & \frac{ k^3}{4\pi^2 a^2}
 \Big( |\phi_k'|^2 + k^2  |\phi_k|^2    +m^2 a^2 |\phi_k|^2
   +6\xi \frac{a'}{a}(\phi_k^*\phi_k'+ \phi_k\phi_k ^*\, ')
   +  6\xi \frac{a'\, ^2}{a^2}|\phi_k|^2 \Big) \nn \\
&= & \frac{ k^3}{4\pi^2 a^4}
 \Big[ |v_k'|^2 + k^2  |v_k|^2 +m^2 a^2 |v_k|^2
  + (6\xi-1) \Big(  \frac{a'}{a} (v'_k v^*_k + v_k v^*\, '_k  )
    -  \frac{a'\, ^2}{a^2}  |v_k|^2  \Big) \Big]  \nn \\
\ea
which is nonnegative.
The pressure in the  vacuum  is
\ba
p =  -\frac13    \langle T^i\, _i \rangle
             =  \int^\infty_0   p_k \frac{dk}{k}
\ea
where  the spectral pressure is
{\allowbreak
\ba \label{sprectpressure}
p_k & = &   \frac{k^3 }{4\pi^2a^{2}}
   \bigg[|\phi'_k|^2-\frac13k^2|\phi_k|^2 -a^2m^2|\phi_k|^2
     \nn \\
&& + 2\xi \Big(\frac{a'}{a}(\phi_k\phi^{*'}_k+\phi_k'\phi^{*}_k)
   + \frac{a'^2}{a^2}|\phi_k|^2-2 |\phi'_k|^2+2 (k^2+a^2m^2)|\phi_k|^2
   + 6(\xi-\frac16)\frac{a''}{a}|\phi_k|^2\Big)\bigg]
   \nn \\
& = & \frac{k^3}{4 \pi^2 a^4}
  \Bigg[   \frac13 |v_k'|^2 + \frac13 k^2  |v_k|^2 - \frac13 m^2 a^2 |v_k|^2
  + 2(\xi-\frac16)\Big( -2 |v_k'|^2 + 3 \frac{a'}{a} (v'_k v^*_k + v_k v^*\, '_k  )
         \nn \\
&&  - 3(\frac{a'}{a})^2 |v_k|^2
    +  2(k^2 + m^2 a^2) |v_k|^2
    + 12 \xi  \frac{a''}{a} |v_k|^2  \Big)\Bigg]  .
\ea
The  spectral pressure can take both positive and negative values,
unlike the spectral energy density.
}
The trace of stress tensor      is
\ba
\langle T^\mu\, _\mu \rangle
& = &    \int   \langle T^\mu\, _\mu \rangle_{k} \,  \frac{dk}{k}  ,
\ea
the spectral trace is
\ba\label{tracekgen}
\langle T^\mu\, _\mu \rangle_{k}
& = & \frac{k^3}{2\pi^2a^{2}}
\Big[  (6\xi-1) (|\phi'_k|^2-k^2|\phi_k|^2)
  + 6\frac{a''}{a}\xi (1- 6\xi)|\phi_k|^2
  +2 (1-3 \xi )a^2m^2|\phi_k|^2\nn \Big] \nn \\
& = & \frac{k^3}{2\pi^2 a^4}
 \Bigg[ m^2 a^2 |v_k|^2
  + (6\xi-1)\Big( |v_k'|^2 - \frac{a'}{a} (v'_k v^*_k + v_k v^*\, '_k  )
       +  \frac{a'\, ^2}{a^2} |v_k|^2\nn \\
&& -(k^2 + m^2 a^2) |v_k|^2  - 6\xi \frac{a''}{a}|v_k|^2 \Big) \Bigg].
\ea
In the above the sums of the $k$-modes
contain no product terms between different $k$ since  the scalar field is linear.
The stress tensor is sensitive to the coupling   $\xi$.
For the minimal  coupling  $\xi=0$,
\ba\label{energykxi0}
\rho_k  & = &  \frac{ k^3}{4\pi^2 a^2}
 \Big(  |(\frac{v_k}{a})' |^2  + k^2  |\frac{v_k}{a} |^2
 +m^2 a^2 |\frac{v_k}{a} |^2  \Big) \nn \\
& =& \frac{ k^3}{4\pi^2 a^4}
 \Big[ |v_k'|^2  - \frac{a'}{a} (v'_k v^*_k + v_k v^*\, '_k)
   +(\frac{a'}{a})^2 |v_k|^2 + k^2 |v_k|^2 +m^2 a^2 |v_k|^2  \Big]  ,
         \\
p_k & = &   \frac{k^3}{4 \pi^2 a^2}
  \Big[   |(\frac{v_k}{a})' |^2  - \frac13 k^2 |\frac{v_k}{a}|^2
  - m^2 a^2 |\frac{v_k}{a}|^2  \Big] \nn \\
&= & \frac{ k^3}{4\pi^2 a^4}
 \Big[ |v_k'|^2  - \frac{a'}{a} (v'_k v^*_k + v_k v^*\, '_k)+(\frac{a'}{a})^2 |v_k|^2
    -\frac13 k^2  |v_k|^2 - m^2 a^2 |v_k|^2  \Big]. \label{pk}
\ea
For the conformal  coupling $\xi= \frac16$,
\ba \label{rhok16}
\rho_k
&= & \frac{ k^3}{4\pi^2 a^4}
 \Big[ |v_k'|^2 + k^2  |v_k|^2 +m^2 a^2 |v_k|^2     \Big] , \\
p_k
&= & \frac{ k^3}{12 \pi^2 a^4}
 \Big[ |v_k'|^2 + k^2  |v_k|^2 - m^2 a^2 |v_k|^2     \Big] . \label{pk16}
\ea
The spectral energy density and pressure
are plotted in Figure \ref{Figure1}(b) for $\xi=0$ during  de Sitter inflation,
$\rho_k $ is positive in the whole range,
$p_k $ is positive at high $k$ and negative at low $k$
for $k|\tau| <0.82$ (outside the horizon).

At low $k$,  $ \rho_k $ and $ p_{k}$ are IR convergent,  dominant by the mass term.
At high $k$, $ \rho_k $ and $ p_{k}$
contain  quartic, quadratic, and logarithmic UV divergences
which are to be removed by the following subtraction
\be
\langle T^\mu\, _{\nu} \rangle_{k\, \, reg}
   =  \langle T^\mu\, _{\nu} \rangle_k
    -\langle T^\mu\, _{\nu } \rangle_{k\, \, A}
\ee
where $\langle T^\mu\, _\nu \rangle_{k\,\, A}$ is  a  subtraction term
constructed from the WKB  solution  (see Appendix \ref{sectionC}).
The conventional prescription \cite{ParkerFulling1974,Bunch1980,AndersonParker1987}
adopts the 4th-order adiabatic subtraction for the stress tensor.
However,  as we shall demonstrate in later sections, the 4th-order subtraction
will generally lead to a negative spectral energy density.
Therefore, we seek alternative prescriptions
with different adiabatic orders which largely depend on the coupling $\xi$.
For this,  we propose the following criteria for
an adequate regularized stress tensor of the scalar field:
(1)   UV convergent,
(2)   IR  convergent,
(3)  respect the covariant conservation,
(4) the spectral energy density be nonnegative.
In Ref.\cite{Wald1978} the five axioms for
the regularized stress tensor
did not include the positiveness of the spectral energy density.
However,  in frame work of general relativity,
the energy of the cosmic matter must be nonnegative as a source of gravitational field.
For instance, the Friedmann equation $(\frac{\dot a}{a})^2=\frac{8\pi G}{3} \rho$
requires that $\rho$ be nonnegative, otherwise it will lead to inconsistency.
Thus we  include  the  criterion (4).
We shall show that these four criteria  on the stress tensor
and  the three criteria  on the power spectrum
can be satisfied by proper regularization
for $\xi=0$ and $ \frac16$ respectively.

\section{The regularization of  scalar field with   $\xi=0$}\label{section3}

We consider the massive scalar field with $\xi=0$,
i.e.,   $\nu=(\frac94-\frac{m^2}{H^2})^{1/2}$,  in  de Sitter space.
This  important case
includes  a class of scalar inflaton fields in inflation models
and has direct applications to cosmology of the early Universe.
The power spectrum (\ref{BunchDaviesSpectrum})
is to be regularized to the adiabatic 2nd-order \cite{Parker2007}
as the following
\ba \label{ParkerPS}
\Delta^2_{k\, \, reg} & = &  \frac{ k^{3}}{2  \pi^2 a^2 }
      \Big( |v_k(\tau)|^2 -|v_k^{(2)}(\tau)|^2 \Big)
      =   \frac{ k^{3}}{2  \pi^2 a^2 }
      \Big( |v_k(\tau)|^2 -  (2 W_k^{(2)})^{-1} \Big)
\ea
where  $( W_k^{(2)})^{-1}$ is  given by (\ref{W2xi}).
The resulting regularized power spectrum at $|\tau|=1$
is plotted in dashed line in  Figure \ref{Figure1}(a)
for de Sitter inflation.
 The result  agrees with that of Ref\cite{Parker2007}.
It is revealing to examine the  behavior of spectrum at high $k$.
By (\ref{vksq}), in terms of $k$,
\be \label{vksqtps}
\Delta^2_k  =  \frac{ k^{2}}{4 \pi^2 a^2 }
     \Big(1 +\frac{4\nu^2 -1}{ 8  x^2}
     +\frac{3 (4 \nu ^2-1 ) (4 \nu ^2-9  )}{128 x^4}  \\
  +\frac{5 (4 \nu ^2-1 )  (4 \nu ^2-9  ) (4 \nu ^2-25 )}{1024 x^6}
         + ... \Big) ,
\ee
and,   by (\ref{W2xi}) and using
\be
\omega =\frac{1}{|\tau|}\Big( x +\frac{9-4 \nu ^2}{8\text{  }x}
-\frac{\left(4 \nu ^2-9\right)^2}{128\text{  }x^3}
-\frac{\left(4 \nu ^2-9\right)^3}{1024\text{  }x^5}
-\frac{5 \left(4 \nu ^2-9\right)^4}{32768\text{  }x^7}+... \Big),
\ee
\be
 (2 \omega)^{-1} \simeq
 |\tau| \Big(\frac{1}{2 x} +\frac{4 \nu ^2-9}{16 x^3}
+\frac{3 \left(4 \nu ^2-9\right)^2}{256 x^5}
+\frac{5\left(4 \nu ^2-9\right)^3}{2048 x^7}
 +...\Big),
\ee
(In the following  we shall set $|\tau|=1$ in the relevant expressions
to avoid lengthy notations.)
\ba \label{PSCounterTerms}
(2W^{(2)})^{-1} =
  \frac{1}{2k } \l(1 +\frac{4 \nu^2 - 1}{8 x^2}
   + \frac{3 (4 \nu ^2-1) (4 \nu ^2-9)}{128   x^4}
   + \frac{5  (4 \nu ^2-9 )^2 (4 \nu ^2- 17 )}{1024  x^6}
  + ... \r)  .
\ea
The first two terms   in (\ref{PSCounterTerms})
just cancel the first two divergent terms of  (\ref{vksqtps}),
so that the regularized spectrum becomes UV convergent.
The third term $\frac{3 (4 \nu ^2-1 ) (4 \nu ^2-9  )}{128 x^4}$
of (\ref{PSCounterTerms}) happens to cancel
the convergent third  term of (\ref{vksqtps}),
and,  as a  result,  the regularized power spectrum is dominated by
the fourth term  at high $k$,
\ba \label{ParkerPSreghk}
\Delta ^2 _{k\,  reg} & \simeq &
   \frac{ H^2 x^{2}}{4 \pi^2   }
         \frac{5 (9- 4 \nu ^2)}{8 x^6}
       =   \frac{5}{8\pi^2 x^4} \frac{m^2}{H^2}
        \propto k^{-4}
\ea
which is positive and UV   convergent.
As a check, when $ \Delta ^2 _{k\, reg}$ is multiplied by $k^2$,
it is found that  $k^2 \Delta ^2 _{k\, reg}$ is   UV convergent.
This property has an important implication to a proper prescription
for regularization of the stress tensor.
Equivalently, the high $k$ behavior
can be also given    in terms of $\omega$,
by (\ref{vkxi0t1}) at $|\tau|=1$,
\be\label{Deltxi0}
\Delta^2_k = \frac{ k^{3}}{2  \pi^2 a^2 }  \Big(
\frac{1}{2 \bar \omega } +\frac{1}{2 \bar \omega ^3}
-\frac{3 \left(4 \nu ^2-9\right)}{32 \bar \omega ^5}
-\frac{5 \left(4 \nu ^2-9\right) \left(4 \nu ^2+7\right) }{256 \bar \omega ^7}
+\frac{35 \left(176 \nu ^4-472 \nu ^2+171\right)}{1024 \bar \omega ^9}
       + ... \Big)
\ee
where  $\bar \omega \equiv  (k^2 +\frac{m^2}{H^2})^{1/2}$,
and by (\ref{W2xi}),
\be\label{w2t1}
(2W^{(2)})^{-1} =
\frac{1}{2 \bar \omega}
+\frac{1}{2 \bar \omega^3 }+\frac{3 \left(\frac{9}{4}-\nu ^2\right)}{8 \bar \omega ^5 }
-\frac{5 \left(\frac{9}{4}-\nu ^2\right)^2}{ 16 \bar \omega ^7 } ,
\ee
so the $\omega^{-1}$,  $\omega^{-3}$, $\omega^{-5}$ terms are canceled,
yielding
\be\label{PSregxi0}
\Delta ^2 _{k\,  reg} \simeq \frac{ H^2 x^{3}}{ 2 \pi^2   }
 \frac{5 (9- 4 \nu ^2) }{16 \bar \omega ^7}
\ee
which is equivalent to (\ref{ParkerPSreghk}).

The several points should be mentioned.
First,  the regularized power spectrum has a factor
$(9- 4 \nu ^2  )\propto  m^2 $, so that it is zero at $m=0$.
Second, for the massive field,
$\omega=\frac{m}{H}$ at $k=0$,
the subtraction term  (\ref{w2t1})
based on $1/\omega$ expansion is IR finite and  smaller than $|v_k|^2$,
so that regularization does not alter the convergence pattern of
the  power spectrum at low $k$, as shown in Figure \ref{Figure1}(c).
Third,  in regard to inflation cosmology,
the effects upon the power spectrum of massive field by regularization is small
in low $k$ range of the observations of CMB anisotropies
 ($k|\tau_1|\simeq  10^{-28} \sim 10^{-25}$).
This is in contrast to a massless field,
like RGW and scalar metric perturbations,
for which the IR distortions by  regularization
fall into the range of cosmological observations
\cite{WangZhangChen2016,ZhangWangJCAP2018}.

\begin{figure}[htb]
\centering
\subcaptionbox{}
    {%
        \includegraphics[width = .48\linewidth]{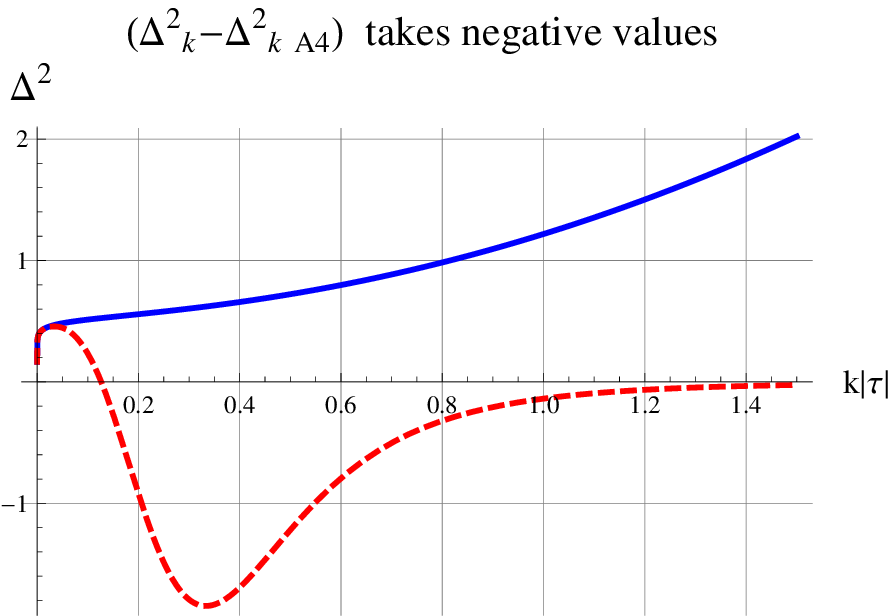}
}
\subcaptionbox{}
    {%
        \includegraphics[width = .48\linewidth]{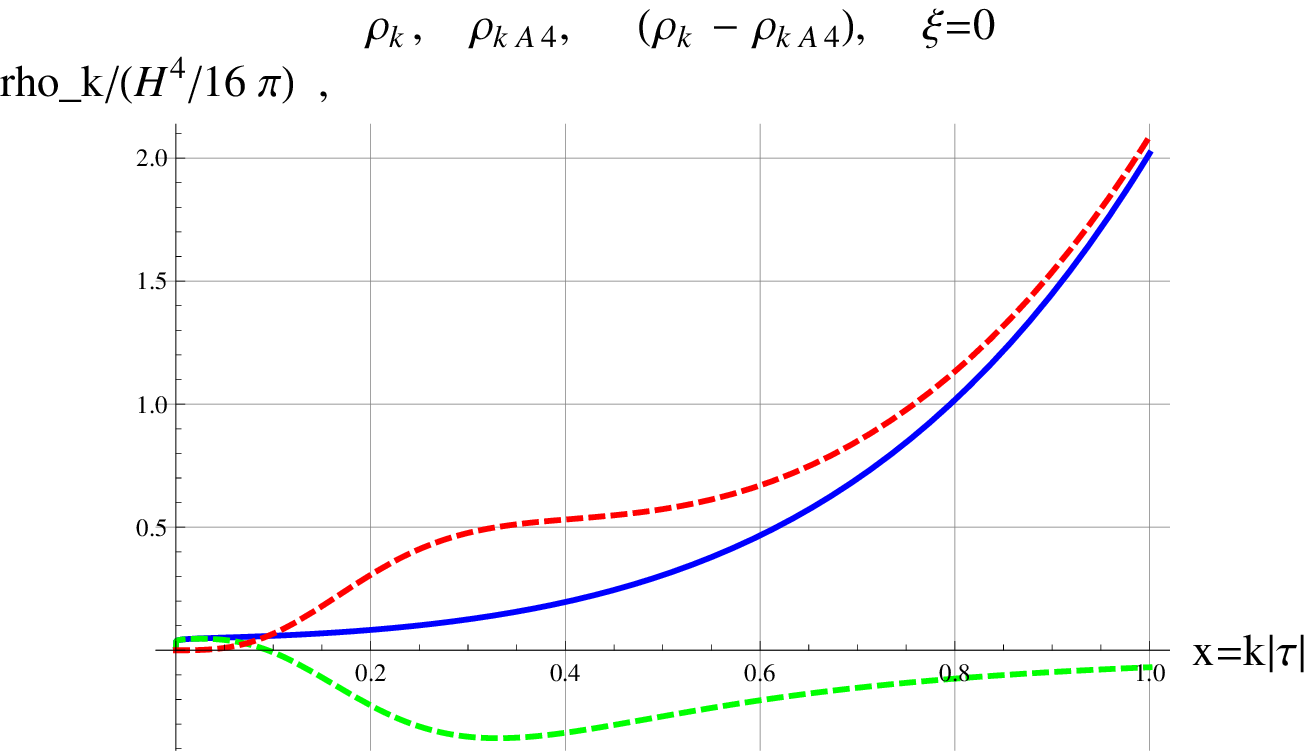}}
\subcaptionbox{}
    {%
        \includegraphics[width = .48\linewidth]{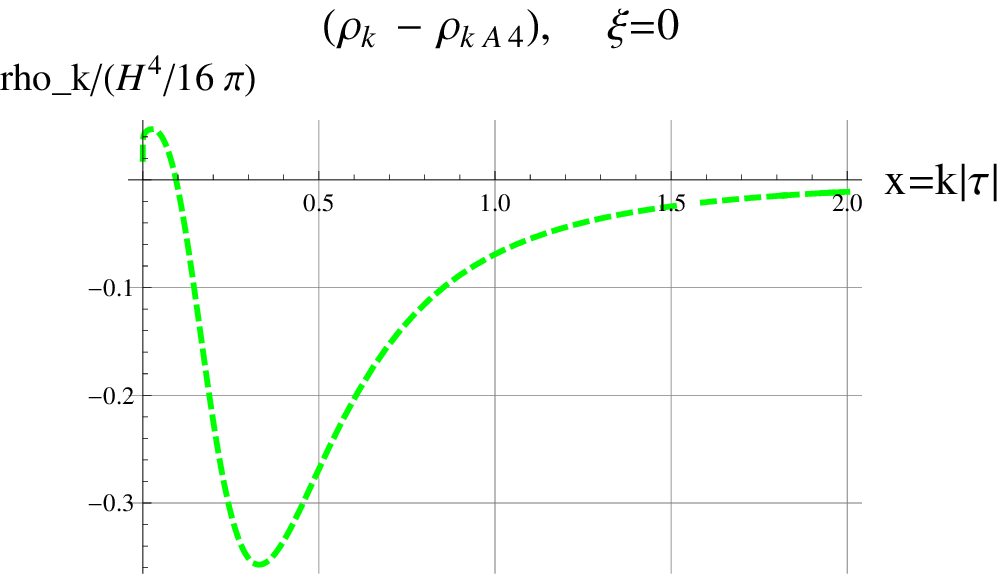}
    }
\subcaptionbox{}
    {%
        \includegraphics[width = .48\linewidth]{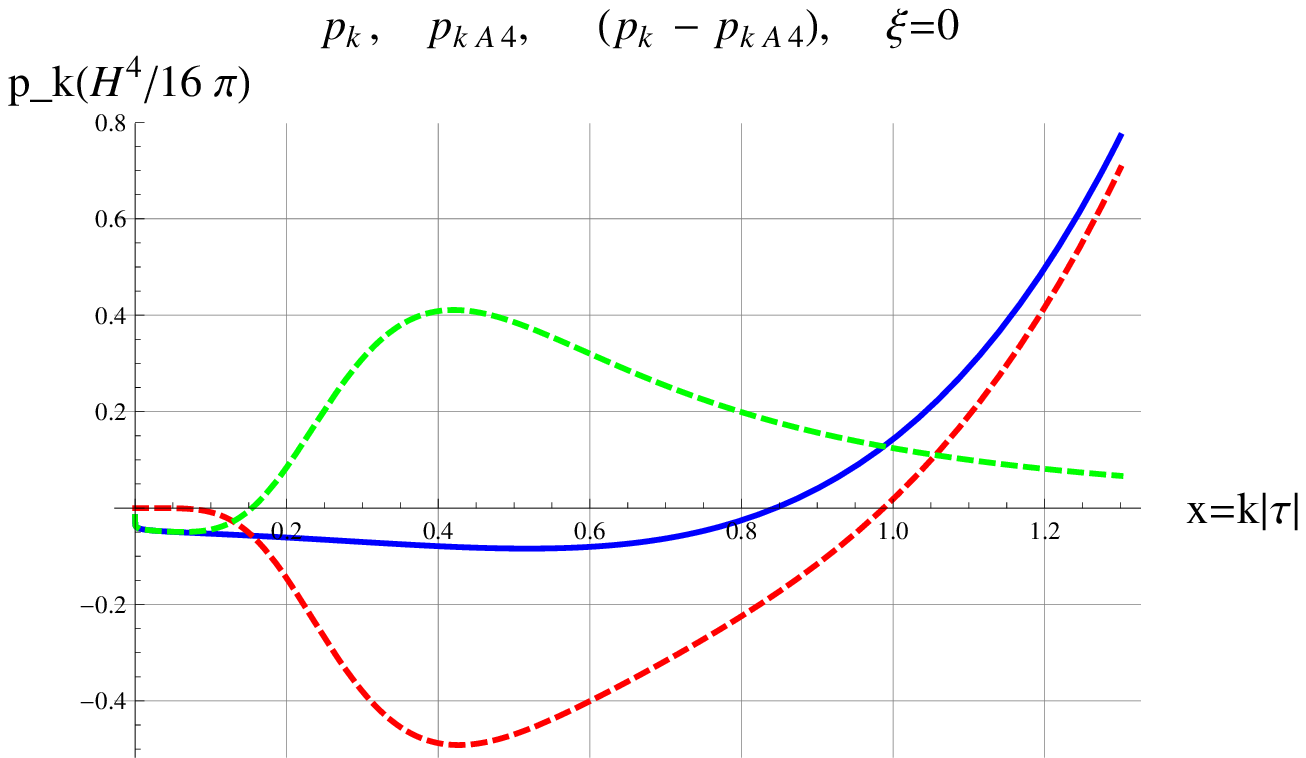}
        }
\caption{
   The model $\xi=0$.
   (a)  Blue: the unregularized $\Delta^2_{k}$.
        Red Dash: the 4th-order regularized power spectrum is  negative.
    (b)     Blue: the unregularized $\rho_{k }$.
Red:   $\rho_{k\, A\, 4 }$.
Green: $(\rho_k- \rho_{k\, A 4})$ which is negative for all  $k|\tau| \gtrsim 0.1$
       because  $\rho_{k\, A\, 4}$ is higher than $\rho_{k }$.
    (c)  $(\rho_k- \rho_{k\, A 4})$,   enlarged from (b).
    (d) Blue: the  pressure  $p_{k }$.
        Red:    $p_{k\, A4}$.
        Green:    $(p_k- p_{k\, A 4})$.
    }
     \label{Figure2}
\end{figure}

If we try to regularize the power spectrum for $\xi=0$
to  the 0th adiabatic order, we would get
\ba
 \frac{ k^{3}}{2  \pi^2 a^2 }
      \Big( |v_k |^2 -|v_k^{(0)} |^2 \Big)
 =  \frac{ k^{3}}{2  \pi^2 a^2 }
      \Big( |v_k |^2 -\frac{1}{2 \omega } \Big)
 \simeq   \frac{ H^2 x^{3}}{2  \pi^2}     \frac{1}{2 x^3}
  = \frac{ H^2  }{4  \pi^2}   \propto k^0 ,
    \nn
\ea
so that the logarithmic divergence  still exists.
If we    try the 4th order regularization,
by  (\ref{W4gen})  (\ref{v2sub}),
\ba \label{ParkerPS4th}
  \frac{ k^{3}}{2  \pi^2 a^2 }
      \Big( |v_k |^2 -|v_k^{(4)} |^2 \Big)
      =\frac{ k^{3}}{2  \pi^2 a^2 }
      \Big( |v_k |^2 -(2W^{(4)})^{-1} \Big)  , \nn
\ea
the resulting spectrum will take negative values,
as shown by the dashed line in Figure \ref{Figure2}(a).
This is also checked at high $k$ as follows.
With  $|v_k |^2 $ in (\ref{vksq}) and $(2W^{(4)})^{-1}$ in (\ref{W4gen})
\ba
(2W^{(4)})^{-1}
& \simeq &
    \frac{1}{2 k}
    \Big( 1 +\frac{4\nu^2-9}{8 x^2}
    +\frac{3  (4 \nu ^2-9) (4 \nu ^2-1 )}{128 x^4}
    +\frac{5  (4 \nu ^2-9) (4 \nu ^2-1 )  (4 \nu ^2-25  )}{1024  x^6} \nn \\
&& +\frac{35 (4 \nu ^2-9 )^2  (16 \nu ^4-264 \nu ^2+705 ) }{32768  x^8}
   +... \Big) ,
\ea
the difference is
\be
\frac{ k^{3}}{2  \pi^2 a^2 }
      \Big( |v_k |^2 -(2W^{(4)})^{-1} \Big)
\simeq  \frac{k^{3}}{2\pi^2 a^2 } \frac{1}{2k}
\frac{175 \left(4 \nu^2-9\right)}{32  x^8}
  =- \frac{ H^2  }{2\pi^2} \frac{175   \frac{m^2}{H^2} }{64 x^6}
 \propto  - k^{-6} <0 .
\ee
Not only the $k^{2}$, $k^{0}$ divergent terms,
but also the $k^{-2}$, $k^{-4}$ convergent terms have been  removed.
Thus, the 4th order regularization of power spectrum
is an  incorrect  prescription,
as it subtracts more than necessary and violates the minimal subtraction rule.
The negativeness will
also occur for  the 4th order regularization of stress tensor in the following.

Now consider the adiabatic regularization of spectral energy density and spectral pressure.
First we try the 4th-order regularization.
The  unregularized $\rho_k$  for $\xi=0$ during de Sitter inflation
and its asymptotic high-$k$ and high-$\bar\omega$ expansions are the following
{\allowdisplaybreaks
\ba \label{rhoExpanxi0}
\rho_{k}  & = &  \frac{H^4}{16 \pi  }
 \Bigg(  x^2  \l|\frac{d}{dx} \big(x^{3/2} H^{(2)}_{\nu} (x)\big) \r|^2
       + x^5 |H^{(2)}_{\nu} (x)|^2
      + (\frac{9}{4}-\nu^2)  x^3 |H^{(2)}_{\nu} (x)|^2 \Bigg)  \nn \\
& \simeq  &  \frac{H^4}{16 \pi}
\Bigg(
\frac{4 x^4}{\pi }+\frac{\left(13-4 \nu ^2\right) x^2}{2 \pi }
-\frac{ (4 \nu ^2-9 ) (4 \nu ^2-1 )}{32 \pi }
-\frac{(4 \nu ^2-9 ) (4 \nu ^2-1) (4 \nu ^2+11)}{256 \pi  x^2} \nn \\
&&   -\frac{5 (4 \nu ^2-9 ) (4 \nu ^2-1)(4 \nu ^2-25 ) (4 \nu ^2+47)}{8192 \pi  x^4} \nn \\
&&  -\frac{7  (4 \nu ^2-9  )  (4 \nu ^2-1 ) (4 \nu ^2-25  )
    (4 \nu ^2-49  ) (4 \nu ^2+99 ) }{65536 \pi  x^6} \nn \\
&&  -\frac{21  (4 \nu ^2-9  )  (4 \nu ^2-1 ) (4 \nu ^2-25  )
    (4 \nu ^2-49  )(4 \nu ^2-81  ) (4 \nu ^2+167 ) }{1048576  \pi  x^8}
    + ...
           \Bigg) \\
& \simeq  & \frac{H^4}{16 \pi}
 \frac{4 x^3}{\pi }
 \Bigg[ \bar\omega +\frac{1}{2 \bar \omega }
 +\frac{\frac{9}{4}-\nu^2}{2 \bar\omega ^3}
 -\frac{(\frac{9}{4}-\nu^2) \left(4 \nu ^2-29\right)}{32 \bar \omega ^5}
 -\frac{5 (\frac{9}{4}-\nu^2) \left(76 \nu ^2-59\right)}{128 \bar \omega ^7} \nn \\
&& - \frac{7 (\frac{9}{4}-\nu^2) \left(32 \nu ^4-1108 \nu ^2+171\right)}{256 \bar \omega ^9} \nn \\
&& + \frac{21 (\frac{9}{4}-\nu^2)(320 \nu^6+44176 \nu ^4 -366724 \nu^2+90963 )}
       {32768 \bar \omega ^{11}}
          +... \Bigg] . \label{rhoExpkXi0}
\ea
}
The first term, either in terms of $k$ or $\bar \omega$,
is quartic divergent, analogous to that in the Minkowski spacetime.
The first three terms, either in terms of $\bar \omega$ or in terms of $k$,
 are divergent,
and the remaining are convergent.
The subtraction term for spectral energy density with general $m$ and $\xi$
is listed in  (\ref{countrh1})  in Appendix \ref{sectionC},
and, for the de Sitter inflation and  $\xi=0$,  it  is
{\allowdisplaybreaks
\ba\label{Regenergykmass}
 \rho_{k\, A 4}
& = & \frac{H^4}{16 \pi}
\frac{4 x^3}{\pi } \Bigg[ \bar \omega  +\frac{1}{2\bar\omega }
    +\frac{\frac{9}{4}-\nu ^2}{2 \bar\omega ^3}
    + \frac{(\frac{9}{4}-\nu^2)  (\frac{29}{4}-  \nu ^2  )}{8 \bar\omega ^5}
   + \frac{ 95   ( \frac94 -\nu^2)^2 }{32 \bar\omega^7} \nn \\
&& ~~~  -\frac{7  (\frac94  -\nu ^2 )^3}{8\bar\omega^9}
    -\frac{105 (\frac{9}{4}-\nu ^2 )^4}{128 \bar\omega ^{11}} \Bigg]  \label{rhosub4om}
     \\
& \simeq &  \frac{H^4}{16 \pi}\Bigg(
 \frac{4 x^4}{\pi }
+\frac{ (13-4 \nu ^2 ) x^2}{2 \pi }
-\frac{(4 \nu ^2-9 ) (4 \nu ^2-1  )}{32 \pi }
-\frac{(4 \nu ^2-9 ) (4 \nu ^2-1 ) (4 \nu ^2+11)}{256 \pi  x^2} \nn \\
&&  - \frac{5 (4 \nu ^2-9 )^2  (16 \nu ^4+120 \nu ^2-927 )}{8192 \pi  x^4}
-\frac{7 (4 \nu ^2-9 )^3 \left(16 \nu ^4+168 \nu ^2-5451\right)}{65536 \pi  x^6} \nn \\
&& -\frac{21 (4 \nu ^2-9 )^4 \left(16 \nu ^4+152 \nu ^2-17895\right)}{1048576 \pi  x^8}
    + ...  \Bigg)  . \label{rhosub4k}
\ea
}
The 4th-order regularized spectral energy density
\ba\label{Regenergykmass2}
  \rho_{k} -\rho_{k\, A 4 }
\ea
is negative at high $k$
as shown in  Figure \ref{Figure2}(b) and Figure \ref{Figure2}(c)
for $\frac{m^2}{H^2}=0.1$.
We  analytically  check this  at high $k$ as the following.
By comparing  (\ref{rhoExpkXi0}) and (\ref{Regenergykmass}),
the first four  terms of $\rho_k$ are canceled
by $\rho_{k\, A 4 }$,
including its convergent $\bar \omega^{-5}$ term,
and the remaining is dominated by the $\bar \omega^{-7}$ term,
\be \label{subrho2om}
\rho_{k} -\rho_{k\, A 4}
\simeq   \frac{H^4}{16 \pi}  \frac{4 x^3}{\pi }
\frac{35 \left(4 \nu ^2-9\right)}{32 \bar\omega ^7}<0
\ee
which is    negative.
Or, equivalently,
comparing (\ref{rhoExpanxi0}) and  (\ref{rhosub4k}) in terms of $k$,
the first four terms ($\propto x^{4}, x^2, x^0, x^{-2}$) of $\rho_{k}$ are canceled,
leaving
\be\label{4thnegat}
\rho_{k} -\rho_{k\, A 4}
\simeq   \frac{H^4}{16 \pi} \frac{35 \left(4 \nu ^2-9\right)}{8 \pi  x^4}
 =  - \frac{H^4}{16 \pi} \frac{35   \frac{m^2}{H^2} }{2 \pi x^4} \propto - k^{-4}<0
\ee
which is   negative, and  equal to (\ref{subrho2om}) up to the order $k^{-4}$.
This negativeness
is a difficulty of the conventional 4th-order regularization
of spectral energy density  for a massive scalar field.
By the 4th-order prescription  (\ref{rhosub4om}),
the subtraction term  $\rho_{k\, A 4}$
 contains the terms from $\omega $ up to  $\omega^{-11}$,
which are more than necessary for removing the UV divergences.
This is also seen in the $k$ expansion of
$\rho_{k\, A 4}$ of  (\ref{rhosub4k}),
among which the $ k^4, k^2, k^0$  terms are sufficient
for removing the UV divergences.
So the 4th-order subtraction   removes too much
and violates the minimal subtraction rule \cite{ParkerFulling1974},
causing a negative regularized spectral energy density.

We also give the 4th-order regularization of the spectral pressure for  $\xi=0$,
{\allowdisplaybreaks
\ba\label{pkex}
p_k & = &   \frac{H^4}{16 \pi}
 \Bigg(  x^2  \l|\frac{d}{dx} \big(x^{3/2} H^{(2)}_{\nu} (x)\big) \r|^2
       -\frac13  x^5 |H^{(2)}_{\nu} (x)|^2
       - (\frac{9}{4}-\nu^2)  x^3 |H^{(2)}_{\nu} (x)|^2 \Bigg) \nn \\
&= &  \frac{H^4}{16 \pi}
\Bigg(
 \frac{4 x^4}{3 \pi }+\frac{\left(4 \nu ^2-13\right) x^2}{6 \pi }
 +\frac{\left(4 \nu ^2-9 \right) \left(4 \nu ^2-1 \right)}{32 \pi }
 +\frac{5 \left(4 \nu ^2-9 \right) \left(4 \nu ^2-1 \right) \left(4 \nu ^2+11\right) }{768 \pi  x^2} \nn \\
&&  +\frac{35 \left(4 \nu ^2-9 \right) \left(4 \nu ^2-1 \right)
   \left(4 \nu ^2-25 \right)\left(4 \nu ^2+47\right) }{24576 \pi  x^4} \nn \\
&& + \frac{21 \left(4 \nu ^2-9 \right)
\left(4 \nu ^2-1 \right)\left(4 \nu ^2-25 \right)\left(4 \nu ^2-49 \right)\left(4 \nu ^2+99\right)}{65536 \pi  x^6} \nn \\
&& +\frac{77 \left(4 \nu ^2-9 \right)\left(4 \nu ^2-1\right)
\left(4 \nu ^2-25 \right) \left(4 \nu ^2-49 \right)
 (4 \nu ^2-81  ) \left(4 \nu ^2+167\right)}{1048576 \pi  x^8} \nn \\
&& +\frac{143 (4 \nu ^2-1) (4 \nu ^2-9) (4 \nu ^2-25)
  (4 \nu ^2-49)(4 \nu ^2-81 ) (4 \nu ^2-121) (4 \nu ^2+251) }{8388608 \pi  x^{10}}
      +... \Bigg)     \\
& = & \frac{H^4}{16 \pi}  \frac{4 x^3}{\pi} \frac13
\Bigg[ \bar \omega  +\frac{\nu ^2-\frac{11}{4}}{ \bar \omega }
+\frac{\nu ^2-\frac{9}{4}}{ \bar \omega ^3}
+\frac{ (  \nu ^2-\frac94 )  (28 \nu ^2+37 )}{32 \bar \omega ^5}
-\frac{5 (4 \nu ^2-9 )  (4 \nu ^4+95 \nu ^2-38 )}{128 \bar \omega ^7} \nn \\
&& -\frac{7  (4 \nu ^2-9 )  (2096 \nu ^4-24544 \nu ^2+5733 ) }{2048 \bar \omega ^9} \nn \\
&& +\frac{21  (4 \nu ^2-9 ) (448 \nu ^6+599216 \nu ^4
  -4289708 \nu ^2+1037529 )}{32768 \bar \omega ^{11}} \nn \\
&& +\frac{33 (4 \nu ^2-9) (8960 \nu ^8+2471424 \nu ^6
 -88933920 \nu ^4+480721088 \nu ^2-114648777 )}{131072 \bar \omega ^{13}}
  + ... \Bigg] , \nn \\
\ea
}
and  by (\ref{countpress}) the 4th-order subtraction term for pressure is
{\allowdisplaybreaks
\ba
p_{k\, A4} & = & \frac{H^4}{16 \pi}   \frac{4 x^3}{\pi }
\frac{1}{3} \Bigg[\bar \omega  +\frac{4 \nu ^2-11}{4 \bar \omega }
+\frac{4 \nu ^2-9}{4 \bar \omega ^3}
-\frac{-112 \nu ^4+104 \nu ^2+333}{128 \bar \omega ^5}
 \nn \\
&& -\frac{5  (9-4 \nu ^2 )^2  (\nu ^2+26 )}{128 \bar \omega ^7}
-\frac{917  (4 \nu ^2-9 )^3}{2048 \bar \omega ^9}
+\frac{147  (9-4 \nu ^2 )^4}{32768 \bar \omega ^{11}}
+\frac{1155  (4 \nu ^2-9 )^5}{131072 \bar \omega ^{13}} \Bigg]   \\
& = & \frac{H^4}{16 \pi}
\Bigg(\frac{4 x^4}{3 \pi }+\frac{\left(4 \nu ^2-13\right) x^2}{6 \pi }
+\frac{ (4 \nu ^2-9  ) (4 \nu ^2-1  )}{32 \pi }
+\frac{5 (4 \nu ^2-1 )(4 \nu ^2-9 ) \left(4 \nu ^2+11\right)}{768 \pi  x^2}\nn \\
&& +\frac{35 (4 \nu ^2 -9)^2  (16 \nu ^4+120 \nu ^2-927 )}{24576 \pi  x^4}
 +\frac{21 (4 \nu ^2-9)^3 (16 \nu ^4+168 \nu ^2-5451)}{65536 \pi  x^6} \nn \\
&& +\frac{77 (4 \nu ^2-9)^4 \left(16 \nu ^4+152 \nu ^2-17895\right)}
   {1048576 \pi  x^8} +... \Bigg)   ,
\ea
}
so  the 4th-order  regularized spectral pressure  for $\xi=0$ is given by
\be \label{pressre}
 p_{k} -p_{k\, A 4} \simeq  \frac{H^4}{16 \pi}   \frac{4 x^3}{\pi }
     \frac{- 245 \left(4 \nu ^2-9\right)}{96 \omega ^7} \propto k^{-4}
\ee
which is positive except at $k|\tau| \gtrsim 0.15$,
as shown in  Figure \ref{Figure2}(d).
A negative, or positive  spectral pressure is allowed, from physics point of view.
Thus,  we mainly focus on the  negativeness  of spectral energy density
  in this paper.

\begin{figure}[htb]
\centering
\subcaptionbox{}
    {%
        \includegraphics[width = .48\linewidth]{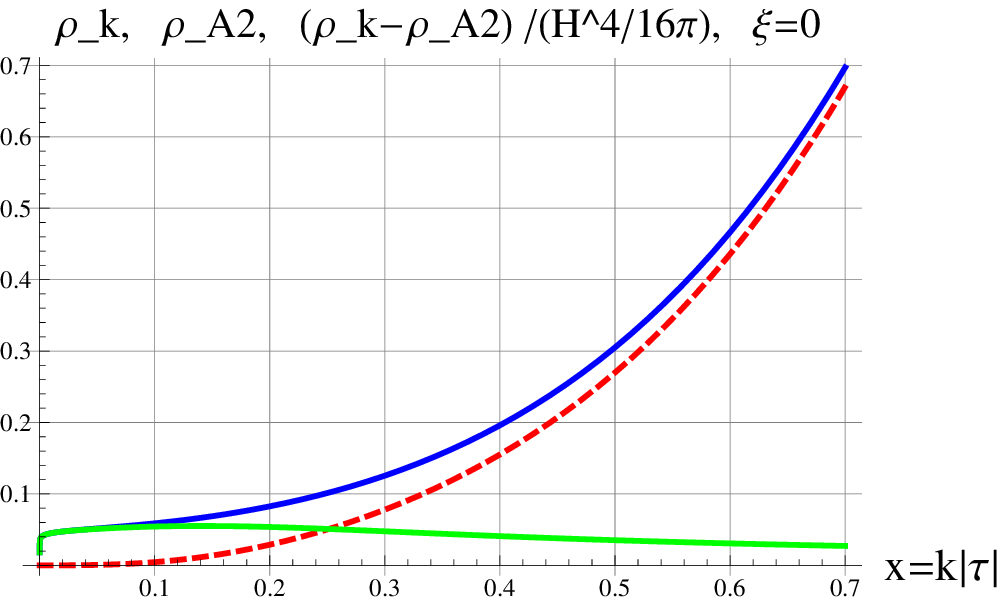}
}
\subcaptionbox{}
    {%
        \includegraphics[width = .48\linewidth]{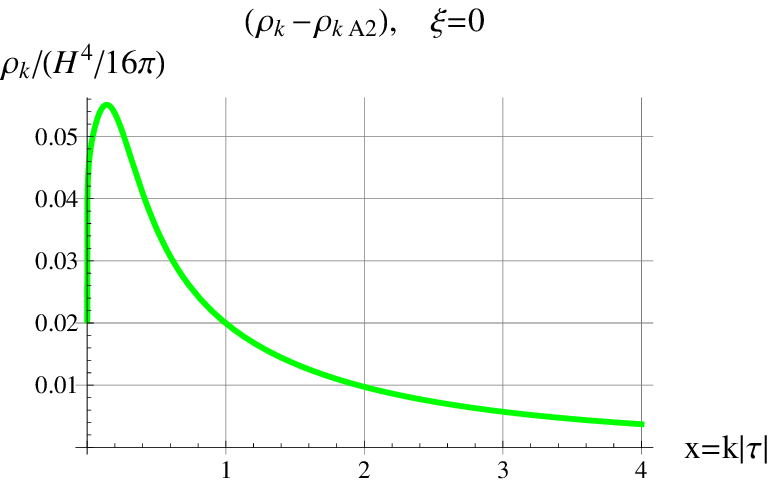}}
\subcaptionbox{}
    {%
        \includegraphics[width = .48\linewidth]{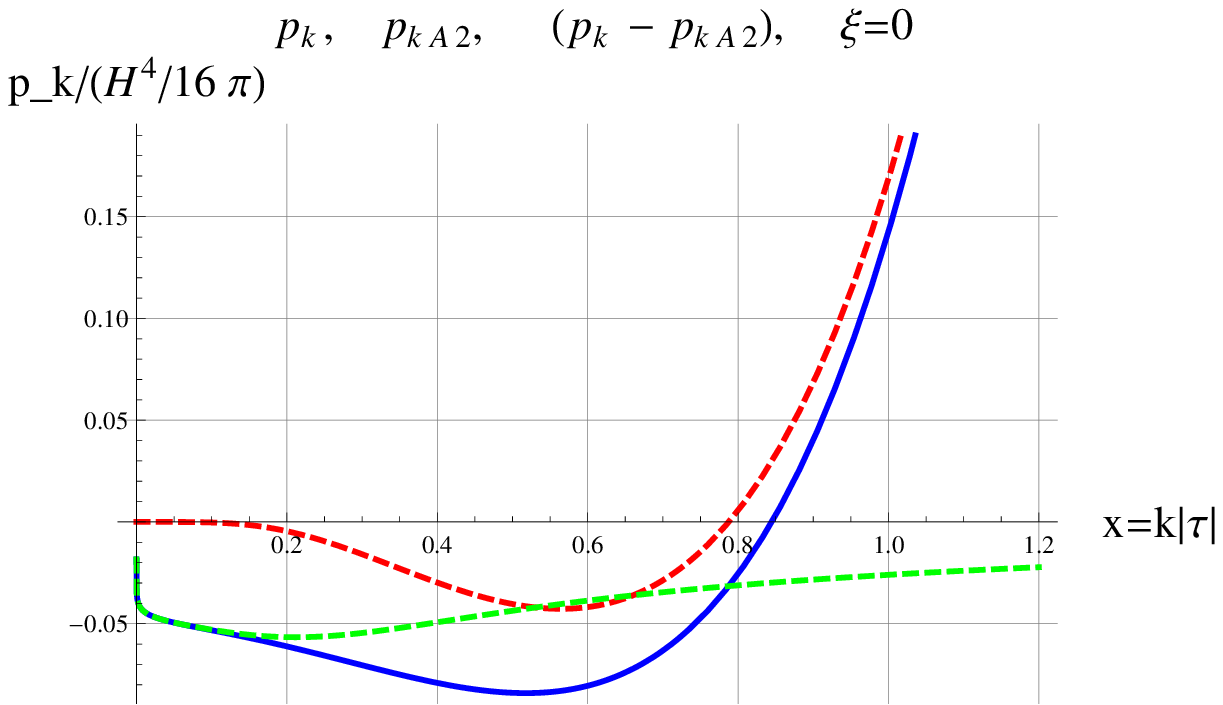}
    }
\subcaptionbox{}
    {%
        \includegraphics[width = .48\linewidth]{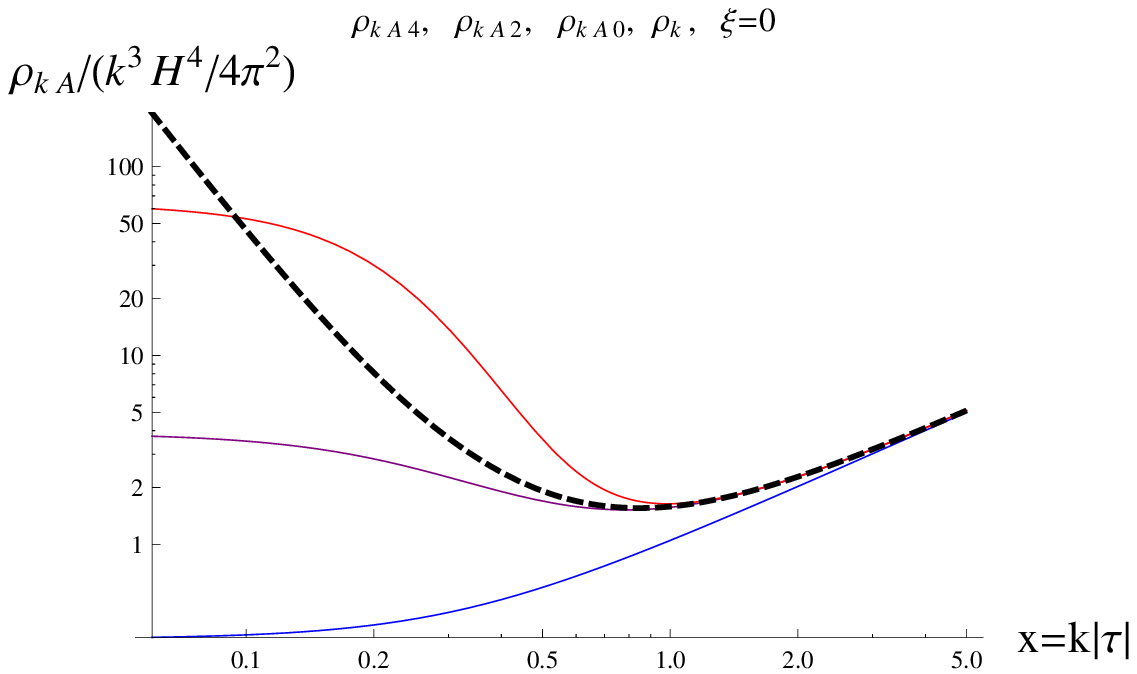}
        }
\caption{
    The model  $\xi=0$.
     (a)  Blue: the unregularized $\rho_{k }$.
Red: the subtraction term  $\rho_{k\,A 2}$.
Green:  the 2nd-order regularized $(\rho_{k }-\rho_{k\,A 2})$.
    (b) $(\rho_{k }-\rho_{k\,A 2})$ is positive and UV-convergent.
         This is enlarged from (a).
    (c)    Blue: the unregularized $p_k$;
Green: the subtraction term  $p_{k\, A2}$;
Red:  the regularized $(p_{k }-p_{k\,A 2})$  becomes UV-convergent
       and is negative for all $k$.
    (d)
 The 4th (Red), 2nd (Purple), 0th (Blue) -order subtraction terms,
   compared with the unregularized $\rho_k$ (Dashed).
    }
     \label{Figure3}
\end{figure}

To avoid the difficulty of the negative spectral energy density,
according to the minimal subtraction rule \cite{ParkerFulling1974},
we propose the 2nd-order  regularization for the stress tensor with $\xi=0$,
i.e.,  the same adiabatic order as we have done upon the power spectrum.
The  2nd-order subtraction term by (\ref{rhoA22count}) for $\xi=0$
is expressed in terms of $\bar\omega$, and  in terms of $k$ as the following
\ba\label{rhoom2mi}
\rho_{k\, A 2}  &  \simeq  &   \frac{H^4}{16 \pi} \frac{4 x^3}{\pi}
 \Bigg( \bar \omega
  +\frac{1}{2 \bar \omega}
  + \frac{\frac{9}{4}-\nu ^2}{2 \bar \omega  ^3}
  + \frac{\left(\frac{9}{4}-\nu ^2\right)^2}{8 \bar \omega^5} \Bigg) \\
& \simeq &
\frac{H^4}{16 \pi} \Bigg(
\frac{4 x^4}{\pi }+\frac{ (13-4 \nu ^2 ) x^2}{2 \pi }
-\frac{ (4 \nu ^2-9  ) (4 \nu ^2-1 )}{32 \pi }
- \frac{ (4 \nu ^2-9 )^2 (4 \nu ^2+19 ) }{256 \pi  x^2} +...
\Bigg) , \label{rhok2mi}  \nn \\
\ea
the 2nd-order  regularized spectral energy density is
\be \label{Regenergy2nd}
\rho_{k\, reg} =  \rho_{k} -\rho_{k\, A 2} .
\ee
The result is  positive and UV convergent,
as shown in Figure \ref{Figure3}(a) and Figure \ref{Figure3}(b),
in contrast to the negative result   by the 4th-order.
It is amazing that although  $\rho_{k\, A 2} $
is defined with only two time derivatives of $a(\tau)$,
it is actually  sufficient to  cancel all the quartic, quadratic,
and logarithmic  divergences  in $\rho_k$.
We analyze how this happens at high $k$.
Comparing  (\ref{rhoExpkXi0}) and (\ref{rhoom2mi}) in $\bar \omega$ expansion,
the first three $\bar \omega$, $\bar \omega^{-1}$,
$\bar \omega^{-3}$ divergent terms of $\rho_k$ are canceled,
nevertheless,  the convergent  $\bar \omega^{-5}$ term remains.
By inspection of (\ref{Regenergykmass}) and (\ref{rhoom2mi}),
the  $\bar \omega^{-5}$ term of  $\rho_{k\, A 4}$
has canceled that same term of  $\rho_{k}$,
but the  $\bar \omega^{-5}$ term of  $\rho_{k\, A 2}$
does not canceled that of  $\rho_{k}$.
This leads  to a big difference
between the 2nd- and 4th-order regularization.
Equivalently,
comparing  (\ref{rhoExpanxi0}) and (\ref{rhok2mi}) in $k$ expansion,
the first three $x^4$, $x^2$, $x^0$  divergent terms of $\rho_k$
are canceled by $\rho_{k\, A 2}$,
but the $x^{-2}$ convergent term remains,
unlike what happens with $\rho_{k\, A 4}$.
Again this is because
the $x^{-2}$ term of $\rho_{k\, A 2}$ has a different coefficient
from that of $\rho_{k\, A 4}$.
Thus,  2nd-order regularized spectral energy density at high $k$,
\be\label{4thrhopos}
\rho_{k\, reg}   \simeq
\frac{H^4}{16 \pi} \frac{4x^3 }{\pi} \frac{5 \left(9-4 \nu ^2\right)}{32 \bar \omega ^5}
  \simeq
\frac{H^4}{16 \pi} \big(\frac{5 \left(9- 4 \nu ^2\right)}{8 \pi  x^2} \big)
    \propto m^2 H^2 k^{-2} >0  ,
\ee
 is positive and UV convergent,  indeed.

The 2nd-order subtraction term (\ref{pA2count}) for  pressure with  $\xi=0$ is
\be\label{pa2xi0}
p_{k\, A 2 }= \frac{H^4}{16 \pi} \frac{4 x^3}{\pi}
\frac{1}{3}\Big(\bar \omega
+\frac{4 \nu ^2-11}{4  \bar \omega }
+\frac{4 \nu ^2-9}{4  \bar \omega ^3}
+\frac{7 \left(9-4 \nu ^2\right)^2}{128 \bar  \omega ^5}
-\frac{5 \left(4 \nu ^2-9\right)^3}{521 \bar  \omega ^7} \Big),
\ee
and   the 2nd-order  regularized spectral  pressure  is
\be \label{press2ndr}
 p_{k\, reg} =  p_{k} -p_{k\, A2 }
\ee
which is UV convergent,
 but is negative for all $k$, as shown in Figure \ref{Figure3}(c).
At high $k$ it behaves as
\ba\label{regPreexi0}
p_{k\, reg} \simeq    \frac{H^4}{16 \pi} \frac{4 x^3}{\pi}
     ( \frac{25 ( 4 \nu ^2-9 )}{96 \bar \omega ^5} )
     \propto - m^2 H^2 k^{-2} <0.
\ea
A negative spectral pressure is allowed,  from point of view of physics.
The trace  at high $k$  is
{\allowdisplaybreaks
\ba\label{traceExpanxi0}
\langle T^\mu\, _\mu \rangle_k
& \simeq &  \frac{H^4}{16 \pi} \Bigg(
\frac{(13-4 \nu ^2 ) x^2}{\pi }
-\frac{ (4 \nu ^2-9 )  (4 \nu ^2-1 )}{8 \pi }
 -\frac{3 (4 \nu ^2-9 ) (4 \nu ^2-1 ) (4 \nu ^2+11 ) }{128 \pi  x^2} \nn \\
& & -\frac{5 (4 \nu ^2-9 ) (4 \nu ^2-1 )(4 \nu ^2-25) (4 \nu ^2+47) }{1024 \pi  x^4} \nn\\
& & -\frac{35 (4 \nu ^2-9 ) (4 \nu ^2-1 )(4 \nu ^2-25)
    (4 \nu ^2-49)(4 \nu ^2+99) }{32768 \pi  x^6}     +...      \Bigg)   \\
& \simeq   & \frac{H^4}{16 \pi}
 \frac{4 x^3}{\pi}
 \Bigg[   \frac{\frac{13}{4}-\nu ^2 }{ \bar \omega }
 + \frac{3(\frac{9}{4}-\nu^2)}{2 \bar\omega ^3}
 -\frac{3  (4 \nu ^2-9 )  (4 \nu ^2+11 ) }{64 \bar\omega ^5} \nn \\
&& +\frac{5  (4 \nu ^2-9 )  (16 \nu ^4+456 \nu ^2-211 )}{512 \bar \omega ^7}\nn\\
&& +\frac{105  (4 \nu ^2-9 )  (144 \nu ^4-1784 \nu ^2+405 )}{2048 \bar \omega ^9}
  +... \Bigg]  ,
\ea
}
and the 2nd-order  regularized spectral trace
at high $k$ is
\ba\label{regtrcxi0}
\langle T^{\beta} \,_{\beta}\rangle_{k\, reg}
& = &         (\rho_{k\, reg} - 3 p_{k\, reg })
  \simeq  \frac{H^4}{16 \pi} \frac{4 x^3}{\pi }
          \frac{15 (9- 4 \nu ^2 )   }{16 \bar \omega ^5}.
\ea

Observe that the  $k$-modes of regularized stress tensor
in the vacuum state  are  not  maximally symmetric,
\be\label{kmtrunmax}
\langle T_{\mu\nu}  \rangle_{k\, reg}
   \ne \frac14 g_{\mu\nu} \langle T^{\beta} \,_{\beta }\rangle_{k\, reg} .
\ee
By numerical  integration over $k$,
the regularized energy density and pressure
for the model $\frac{m^2}{H^2}=0.1$  are
\be\label{rhoreg}
 \rho_{reg} =\int_0^\infty ( \rho_{k} -\rho_{k\, A2})  \frac{dk}{k}
    \simeq   0.895913 \frac{H^4}{16\pi} =89.5913 \frac{m^4}{16\pi} >0,
\ee
\be\label{preg}
p_{reg} = \int_0^\infty ( p_{k} -p_{k\, A 2} ) \frac{dk}{k}
 \simeq    -0.895913  \frac{H^4}{16 \pi} =-89.5913 \frac{m^4}{16\pi} < 0 ,
\ee
and the regularized trace is
\be\label{trint}
\langle T^{\mu} \,_{\mu }\rangle_{reg}
= 4 \rho_{\,reg}  = 3.58365  \frac{H^4}{16 \pi}.
\ee
Therefore,   the regularized stress tensor with $\xi=0$
in the BD vacuum  is maximally symmetric,
\be\label{mtrunmxi0}
\langle T_{\mu\nu}  \rangle_{reg}
  = \frac14 g_{\mu\nu} \langle T^{\beta} \,_{\beta }\rangle_{  reg} .
\ee
(For a model $\frac{m^2}{H^2}=1$,
the numerical integrations give $\rho_{reg} =-p_{reg} \simeq 0.737829 \frac{H^4}{16 \pi}$,
which are of the same magnitude as in the $\frac{m^2}{H^2}=0.1$ model.)
(\ref{mtrunmxi0}) and (\ref{kmtrunmax})
tells that the regularized stress tensor is maximally symmetric,
but it is distributed nonuniformly  in the $k$-modes.
This is an interesting feature of
the structure of the massive scalar field in the vacuum during de Sitter inflation.
(\ref{rhoreg}) and (\ref{preg}) give
an equation of state in the vacuum $w\equiv p_{reg}/\rho_{reg} =-1$,
which is the same as that of the cosmological constant.
Thus, the regularized  $\langle T_{\mu\nu}  \rangle_{reg}$ in the vacuum state
may be used as the background stress tensor that drives the de Sitter inflation.
In connection with inflation cosmology,
for models with $H^2\lesssim G^{-1}$,
the regularized   $\rho_{reg}$ in (\ref{rhoreg})
will be smaller than the inflation energy density, $3 H^2/(8\pi G)$.
However, if the magnitude of $\rho_{reg}$
is comparable to the inflation energy density,
it will be able to play the role to drive the inflation and
the resulting inflation expansion rate will be $H \sim  G^{-1/2}$
which is generally allowed.
Thus, the massive scalar field in the vacuum can play a double role:
its regularized $\langle T_{\mu\nu} \rangle_{reg}$ drives the inflation,
while its $k$-modes $\phi_k$ together with other metric perturbations
constitute the primordial fluctuations during inflation
which leave imprints upon
CMB anisotropies and polarization \cite{ZhaoZhang2006}.
We have also  checked that the four-divergence of the 2nd-order subtraction terms
for the stress tensor is zero,
so that the 2nd-order regularized spectral stress tensor
respects the covariant conservation.
(See  (\ref{consvrhoA2}) (\ref{consvrhoreg2}) in Appendix \ref{sectionC}.)
Hence,    the 2nd-order adiabatic regularization
yields a positive, UV-convergent, covariantly-conserved
spectral energy density
for the minimally-coupling, massive  scalar field.
It is very satisfying that the same  2nd-order regularization
works  simultaneously for
the power spectrum and the spectral stress tensor with $\xi=0$.

We also examine the 0th-order regularization.
The subtraction term for spectral energy density (\ref{rhoA0}) is
\ba\label{rhok0}
\rho_{k\, A 0} = \frac{k^3}{4\pi^2 a^4}\omega
 \simeq   \frac{H^4}{16 \pi}
 \Bigg(  \frac{4 x^4}{\pi }+\frac{ \left(9-4\nu ^2\right) x^2}
 {2 \pi }-\frac{\left(\frac{9}{4}-\nu ^2\right)^2}{2 \pi }
 +\frac{\left(\frac{9}{4}-\nu ^2\right)^3}{4 \pi  x^2} +... \Bigg)
\ea
which  removes only the quartic divergence of $\rho_k$ in (\ref{rhoExpanxi0}),
but  the quadratic and  logarithmic divergences remain.
So the 0th-order regularization fails.

For comparison, we plot the subtraction terms
$\rho_{k\, A 0}$, $\rho_{k\, A 2}$,
$\rho_{k\, A 4}$, and the unregularized $\rho_{k}$
in one graph in Figure \ref{Figure3}(d).
It reveals  that
 $\rho_{k\, A 4}$ is higher than  $\rho_k$  around $k|\tau| \gtrsim 0.1$,
leading to a negative  spectral energy density,
$\rho_{k\, A 2}$ is lower than $\rho_k$ and gives a positive spectral energy density,
$\rho_{k\, A 0}$ is too low.

With regard to renormalization,
since the 2nd-order subtraction terms $\rho_{k\, A 2}$ and $p_{k\, A 2}$
involve only up to the second order time derivatives of metric,
one does not need the fourth order time derivative counter terms
$^{(1)}H_{\mu\nu}$, $^{(2)}H_{\mu\nu}$
in renormalization  \cite{UtiyamaDeWitt1962,DowkerCritchley1976,Bunch1980}.
The 0th-order divergent terms after $k$-integration
\be \label{renormal}
\langle T^0\,_0\rangle^{(0)}
=  \frac{1}{4\pi^2 a^4}  \int_0^\infty dk  k^{2}  \omega,
~~~~ \langle T^1\,_1\rangle^{(0)}
=  - \frac{1}{12 \pi^2 a^4} \int_0^\infty dk  k^{2}
   \Big[ \omega -\frac{m^2 a^2}{\omega} \Big],
\ee
will be absorbed by renormalizing the cosmological constant $\Lambda$,
and the 2nd-order divergent terms after the $k$ integration
\be
\langle T^0\,_0\rangle^{(2)}
=  \frac{1}{4\pi^2 a^4}  \int_0^\infty dk  k^{2}
   (- \frac16)  \Big[ - \frac{3}{\omega} \frac{a'\,^2}{a^2}
                    - \frac{3m^2 a'\, ^2}{\omega^3}  \Big],
\ee
\be
\langle T^1\,_1\rangle^{(2)}
= - \frac{1}{12 \pi^2 a^4}  \int_0^\infty dk  k^{2}
( -\frac16) \Big[ \frac{1}{\omega} (6\frac{a''}{a}- 9\frac{a'\,^2}{a^2})
       +\frac{6m^2 a^2}{\omega^3}(\frac{a''}{a}-\frac{a'\,^2}{a^2})  \Big]
\ee
will be absorbed by  renormalizing the gravitation constant $G$,
in the same manner as described in Refs.\cite{UtiyamaDeWitt1962,Bunch1980}.

Now we study the  massless  minimally-coupling scalar field.
The exact solution for  $m=0=\xi$ is
\be\label{modexi0}
v_k (\tau ) = - \sqrt{\frac{\pi}{2}}\sqrt{\frac{x}{2k}}
      H^{(1)}_{\frac32} ( x)
      =- \frac{1}{ \sqrt{2 k}} \Big( 1- \frac{i}{x} \Big)  e^{i x},
\ee
the unregularized stress tensor (\ref{energykxi0}) (\ref{pk}) reduces to
\ba\label{rho2nd}
\rho_k = \frac{ k^3}{4\pi^2 a^4}
         \Big ( k + \frac{1}{2 k \tau^2} \Big ) ,
~~~~
p_k  =\frac{ k^3}{12\pi^2 a^4}    \Big( k - \frac{1}{2 k \tau^2} \Big)  .
\ea
The 2nd-order subtraction terms (\ref{rhoA22count}) (\ref{pA2count})
 reduce to
\be \label{countrh0m162}
\rho_{k\,A 2}
= \frac{k^3}{4\pi^2 a^4} \Big( k + \frac{1}{2 k \tau^2}   \Big) ,
~~~~
p_{k\,A 2} = \frac{ k^3}{12\pi^2 a^4}    \Big( k - \frac{1}{2 k \tau^2} \Big) ,
\ee
which are  just  equal to
the unregularized    (\ref{rho2nd}).
Thus, we arrive at a zero regularized stress tensor
\be\label{tmunuregxi0}
\langle T^{\mu\nu} \rangle _{k\, reg}=0   .
\ee
This vanishing  result  also follows directly
from the massless limit of (\ref{4thrhopos}) (\ref{regPreexi0}).
(A zero regularized stress tensor also occurs for RGW during de Sitter inflation
  \cite{WangZhangChen2016,ZhangWangJCAP2018},
whose equation is similar to a  massless minimally-coupling scalar field.)
The unregularized power spectrum  for $m=0=\xi$ contains only two terms,
\be \label{psxi16m0}
\Delta^2_{k } =  \frac{ k^{3}}{2  \pi^2 a^2 }       |v_k|^2
=  \frac{ k^{3}}{2  \pi^2 a^2 }   \frac{1}{2 k }  (1+\frac{1}{x^2} )
\ee
where the first term is  UV divergent,
and the second term is both IR and UV log divergent.
Both terms are to be removed.
And after adiabatic subtraction the regularized spectrum  is also zero
 \cite{Parker2007}
\be \label{PSxi0m0}
\Delta^2_{k\, reg } =  \frac{ k^{3}}{2  \pi^2 a^2 }
      \Big( |v_k|^2 -|v_k^{(2)}|^2 \Big)
=  \frac{ k^{3}}{2  \pi^2 a^2 }
   \Big( \frac{1}{2 k }  (1+\frac{1}{x^2} ) -
      (\frac{1}{2 k} +  \frac{1}{2 k  x^2} ) \Big) =0
\ee
where $|v_k^{(2)}|^2$ is given by  (\ref{W2xi}).
The result (\ref{PSxi0m0})
also follows from  the massless limit of
 (\ref{ParkerPSreghk}) or (\ref{PSregxi0})  of the massive scalar field.
Hence, for the minimally-coupling massless scalar field in de Sitter space,
the 2nd-order regularization yields
a zero stress tensor and a zero power spectrum.
We also observe that in this case
the 4th-order regularization is actually equivalent to the 2nd-order one,
\be\label{rhopxi0m0}
\rho_{k\,A4} = \rho_{k\,A2},
~~~~ p_{k\,A4} = p_{k\,A2} ,
\ee
as is seen from the 4th-order subtraction terms
(\ref{countrh1}) (\ref{countpress}),
together with the relations (\ref{desi}) (\ref{desi2}) in de Sitter space.
We like to mention that
the   zero result (\ref{tmunuregxi0})
and (\ref{PSxi0m0}) of the massless field
may not be true in a general RW spacetime.
For instance,
 RGW has nonzero regularized power spectrum and stress tensor
during   the quasi de Sitter inflation
\cite{WangZhangChen2016,ZhangWangJCAP2018}.

The lesson of the $\xi=0$ case is that,
the 2nd-order adiabatic mode $v_k^{(2)}$
regularizes simultaneously  both the power spectrum
and the stress tensor in a consistent manner.
That is, when the power spectrum $( \propto |v_k|^2)$  is correctly regularized by
the 2nd-order subtraction term $|v_k^{(2)}|^2$,
the  spectral energy density
$( \propto   |(\frac{v_k}{a})' |^2  +\frac{k^2}{a^2} |v_k|^2
 +m^2 |v_k |^2 )$
 will  be regularized also correctly by the 2nd-order  subtraction term
$|(\frac{v_k^{(2)}}{a})'|^2 + \frac{k^2}{a^2} | v_k^{(2)} |^2
+m^2  |v_k^{(2)}|^2$.
This  outcome can be further analyzed term by term as the following.
When $|v_k|^2$ is regularized correctly by $|v_k^{(2)}|^2$,
giving a positive, UV convergent result,
 it is obvious that  $m^2|v_k|^2$ after regularization
is also positive and UV convergent.
Next the $\frac{k^2}{a^2}|v_k|^2$ term which has an extra factor $k^2$.
Since the regularized power spectrum $\propto k^{-4}$
as indicated by  (\ref{ParkerPSreghk}),
so multiplying it by the factor $k^2$ will
give a regularized  $\frac{k^2}{a^2}|v_k|^2$ term
which is still $k^{-2}$ convergent.
Finally the time derivative term $|(\frac{v_k}{a})'|^2$.
Since $v_k/a $ and  $v_k^{(2)}/a$  are
 $ \propto e^{-ik\tau}$ at high $k$,
the effect of time differentiation is to bring a factor $-ik$,
so that $(|(\frac{v_k}{a})'|^2 - |(\frac{v_k^{(2)}}{a})'|^2)
 \propto k^2 (|(\frac{v_k}{a})|^2-|(\frac{v_k^{(2)}}{a})|^2)$,
which is also $k^{-2}$ convergent.

In a general curved spacetime,
the possible scheme of adiabatic regularization
may be different from the 2nd  order.
One needs to study concretely
and to see what is the appropriate scheme.
As pointed out in Refs.\cite{Markkanen2018,MarkkanenTranberg2013},
a prescription of adiabatic regularization may be not unique
from perspective of  renormalization,
because the infinities to be absorbed into the bare constants
can always carry along a finite term,
and each different finite term will correspond to
a different scheme  of regularization.

\section{The  regularization of scalar field with $\xi=\frac16$}\label{section4}

Now we consider the   massive scalar field with $\xi=\frac16$,
i.e.,   $\nu=(\frac14-\frac{m^2}{H^2})^{1/2}$.
This is also an interesting case,
and the field equation (\ref{equvk}) in massless limit
reduces to that in   Minkowski spacetime.
As in the last section,
we also explore the 0th-, 2nd-, 4th-order regularization,  respectively.

\begin{figure}[htb]
\centering
\subcaptionbox{}
    {%
        \includegraphics[width = .48\linewidth]{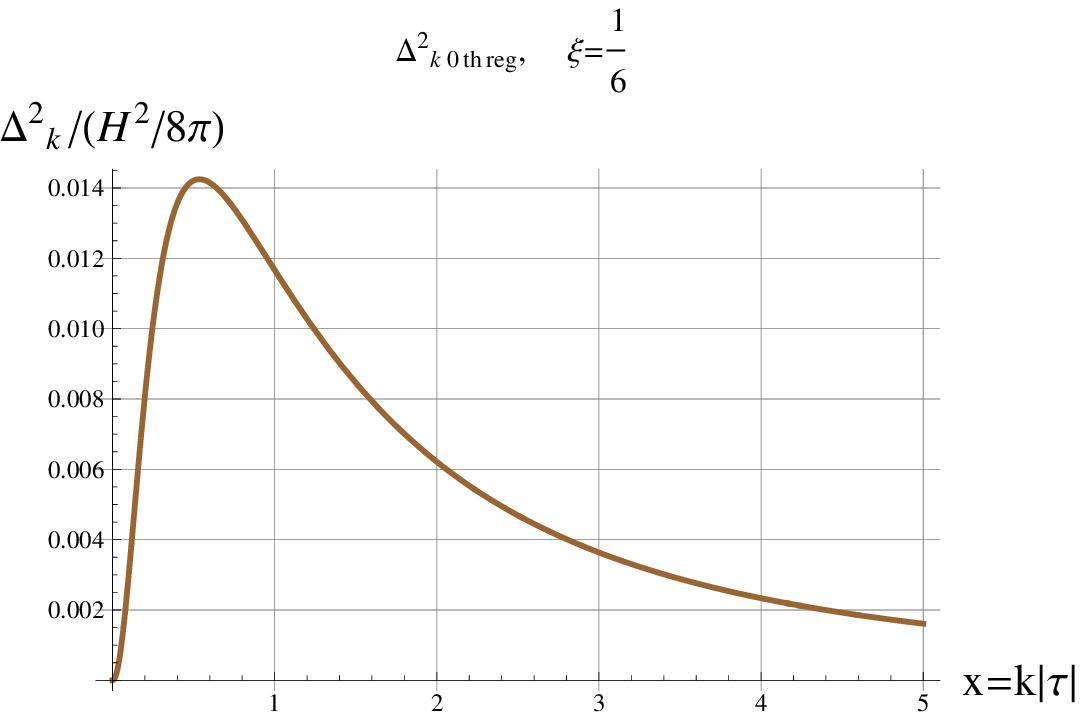}
}
\subcaptionbox{}
    {%
        \includegraphics[width = .48\linewidth]{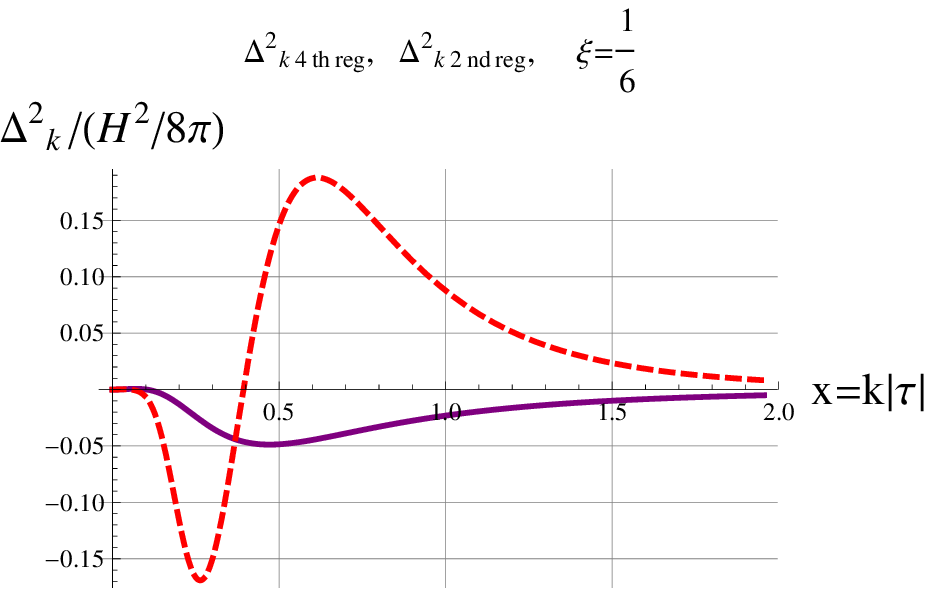}}
\subcaptionbox{}
    {%
        \includegraphics[width = .48\linewidth]{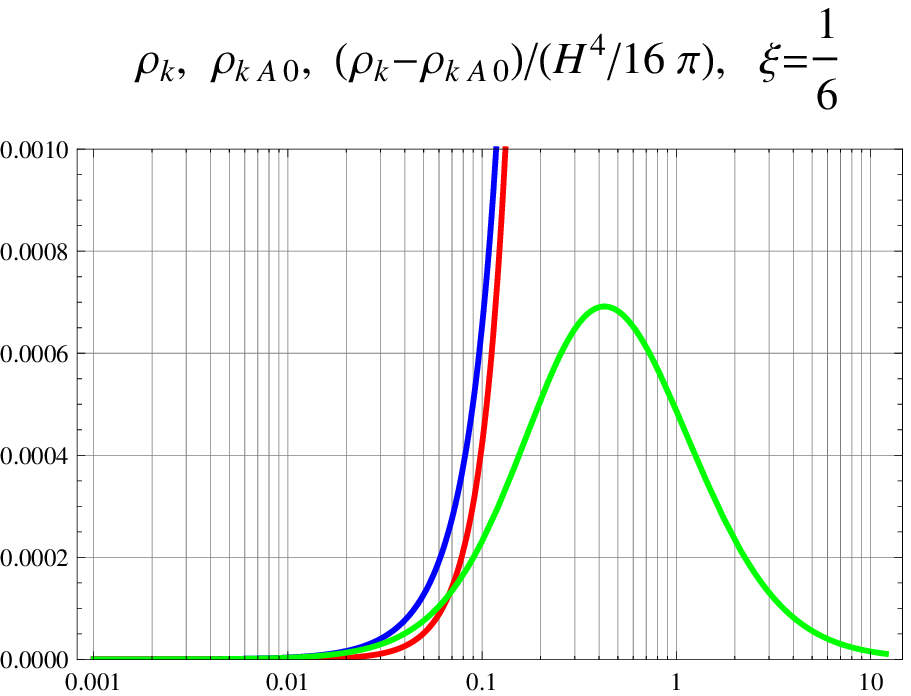}
    }
\subcaptionbox{}
    {%
        \includegraphics[width = .48\linewidth]{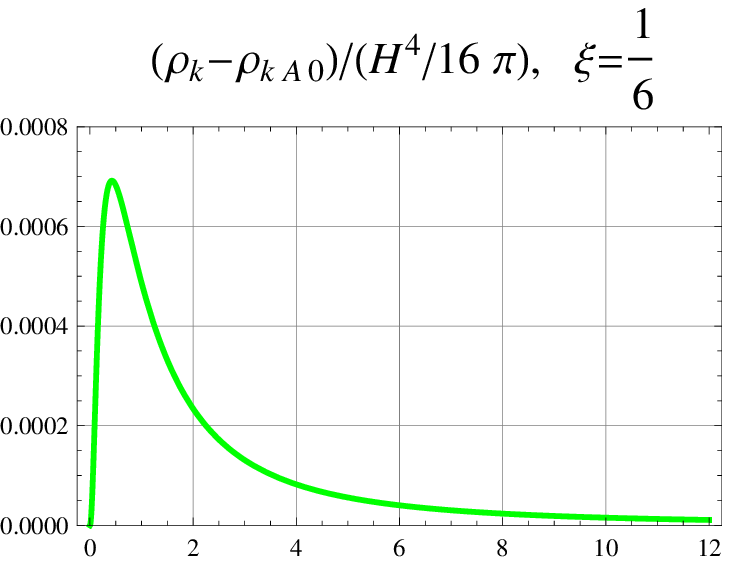}
        }
\caption{
     The model  $\xi=\frac16 $.
    (a) The 0th-order regularized power spectrum $\Delta^2_{k\, reg}$
     is positive and UV convergent.
    (b)   Red Dashed: the 4th-order regularized power spectrum.
          Purple: the 2nd-order.
                Both have negative values.
    (c)   Blue: the unregularized $\rho_k$.
        Red: the subtraction term  $\rho_{k\, A 0 }$.
           Green:  $(\rho_{k }-\rho_{k\,A 0})$  is positive,
           IR- and UV-convergent.
    (d)  $(\rho_{k }-\rho_{k\,A 0})$, enlarged from (c).
    }
     \label{Figure4}
\end{figure}

First   the power spectrum.
The 0th-order regularization is given by
\ba \label{PSxi160}
\Delta^2_{k\, reg  } & = & \frac{ k^{3}}{2  \pi^2 a^2 }
      \Big( |v_k |^2 -|v_k^{(0)} |^2 \Big)
    =  \frac{ k^{3}}{2  \pi^2 a^2 }
      \Big( |v_k |^2 -\frac{1}{2\omega} \Big) .
\ea
The result is  a positive, UV-convergent power spectrum,
as shown in Figure \ref{Figure4}(a).
This  can be checked at high $k$.
Expanding in  terms of $\bar \omega$, using $|v_k |^2$ of (\ref{vexpxi16}),
we have
\be\label{vps16}
\Delta^2_{k}= \frac{ H^2 x^{3}}{2\pi^2} \Big( \frac{1}{2 \bar \omega}
+\frac{3 (1-4 \nu ^2 )}{32   \bar \omega ^5}
+\frac{5  (1-4 \nu ^2 ) (4 \nu ^2-25 )}{256   \bar \omega ^7}
+ ... \Big) .
\ee
Only the first term is UV divergent,
 and it is removed by the subtraction term,
so the regularized power spectrum at high $\omega$ is
\be\label{psregxi}
\Delta^2_{k\, reg}  \simeq  \frac{ H^2 x^{3}}{2\pi^2}
\Big( \frac{3 (1-4 \nu ^2 )}{32   \bar \omega ^5}
+\frac{5  (1-4 \nu ^2 ) (4 \nu ^2-25 )}{256   \bar \omega ^7}
+ ... \Big)
 \propto k^{-2}  >0
\ee
which is positive.
Or,  in terms of $k$,  by
\be
(2\omega)^{-1} \simeq
|\tau| \Big(\frac{1}{2 x} +\frac{4 \nu ^2-1}{16 x^3}
    +\frac{3 \left(4 \nu ^2-1\right)^2\text{  }}{256 x^5}
    +\frac{5\left(4 \nu ^2-1\right)^3\text{  }}{2048 x^7}  +...\Big) ,
\ee
and  by   (\ref{vksq}),
the regularized power spectrum at high $k$ is
\be
\Delta^2_{k\, reg  }
\simeq   \frac{H^2 x^{3}}{2\pi^2}
  \Big(\frac{3(1-4 \nu ^2)}{32 x^5}
  +\frac{5 \left(1-4 \nu ^2\right) \left(4 \nu ^2-7\right)}{64 x^7} + ...\Big)
             \propto k^{-2}  >0
\ee
which is   equivalent to  (\ref{psregxi}).
The regularized power spectrum
has a   factor $(1-4 \nu ^2 )=4\frac{m^2}{H^2}$,
so that it will be zero for  $m=0$.
We also find that the  2nd- and 4th-order regularization
fail to give a positive-definite power spectrum for $\xi=\frac16$,
as shown in Figure \ref{Figure4}(b).

Next  the  spectral stress tensor.
The unregularized $\rho_k$ at high $k$  and high-$\bar\omega$ is  the following
{\allowdisplaybreaks
\ba \label{rhoExpanxi16}
\rho_k &=&  \frac{H^4}{16 \pi  }
 \Big(  x^4  \l|\frac{d}{dx} \big(x^{1/2} H^{(1)}_{\nu} (x)\big) \r|^2
       + x^5 |H^{(1)}_{\nu} (x)|^2
      + (\frac{1}{4}-\nu^2)  x^3 |H^{(1)}_{\nu} (x)|^2  \Big) \nn \\
& \simeq &  \frac{H^4}{16 \pi} \Big(
 \frac{4 x^4}{\pi }
 +\frac{(1-4 \nu ^2 ) x^2}{2 \pi }
 -\frac{\left(1-4 \nu ^2\right)^2}{32 \pi }
  -\frac{ (1-4 \nu ^2 )^2  (4 \nu ^2-9 )}{256 \pi  x^2} \nn \\
&& -\frac{5  (1-4 \nu ^2 )^2  (4 \nu ^2-9  ) (4 \nu ^2-25  )}{8192 \pi  x^4}
   -\frac{7  (1-4 \nu ^2 )^2  (4 \nu ^2-9  ) (4 \nu ^2-25  ) (4 \nu ^2-49  )}{65536 \pi  x^6}\nn \\
&& -\frac{21 (1-4 \nu ^2 )^2 (4 \nu ^2-9  )  (4 \nu ^2-25  ) (4 \nu ^2-49  )
      (4 \nu ^2-81  ) }{1048576 \pi  x^8}
             +...  \Big)   \\
& \simeq &  \frac{H^4}{16 \pi} \frac{4 x^3}{\pi }
\Bigg[
\bar \omega
+\frac{ (1-4 \nu ^2 )^2}{128 \bar \omega ^5}
-\frac{15  (1-4 \nu ^2 )^2}{512 \bar \omega ^7}
-\frac{21  (1-4 \nu ^2 )^2  (4 \nu ^2-13 )}{1024 \bar \omega ^9} \nn  \\
&& -\frac{105  (1-4 \nu ^2 )^2  (16 \nu ^4-584 \nu ^2+1297 )}{32768 \bar \omega ^{11}}
  + ... \Bigg] .  \label{rhoExpanxi16omg}
\ea
}
In terms of powers of $\bar \omega$,
(\ref{rhoExpanxi16omg}) does not contain
$\bar \omega^{-1}$ nor $\bar \omega^{-3}$ terms,
in contrast to (\ref{rhoExpkXi0}) of $\xi=0$.
The first term in (\ref{rhoExpanxi16omg}) is quartic divergent,
and the remaining terms   are convergent.
So,  to achieve  UV convergence,
we need only to remove the $\bar \omega$ term  in (\ref{rhoExpanxi16omg}).
In terms of powers of $k$, the  first three  terms in (\ref{rhoExpanxi16})
 are  $k^4, k^2,k^0$ divergent respectively.
The spectral pressure  for $\xi=\frac16 $ at high $k$   and high-$\bar\omega$ is
{\allowdisplaybreaks
\ba \label{pkxi16omx}
p_k &= & \frac{H^4}{16 \pi  }
 \Big(  x^4  \l|\frac{d}{dx} \big(x^{1/2} H^{(1)}_{\nu} (x)\big) \r|^2
       +    x^5 |H^{(1)}_{\nu} (x)|^2
       -    (\frac{1}{4}-\nu^2)  x^3 |H^{(1)}_{\nu} (x)|^2  \Big)  \nn \\
& \simeq   & \frac{H^4}{16 \pi}
\Bigg(
\frac{4 x^4}{3 \pi }+\frac{\left(4 \nu ^2-1\right) x^2}{6 \pi }
+\frac{\left(1-4 \nu ^2\right)^2}{32 \pi }
+\frac{5 \left(1-4 \nu ^2\right)^2  (4 \nu ^2-9 )}{768 \pi  x^2} \nn \\
&& +\frac{35 \left(1-4 \nu ^2\right)^2  (4 \nu ^2-9 ) (4 \nu ^2-25  ) }{24576 \pi  x^4}
   +\frac{21 (1-4 \nu ^2 )^2  (4 \nu ^2-9) (4 \nu ^2-25)(4 \nu ^2-49)}{65536 \pi  x^6} \nn \\
&&+\frac{77  (1-4 \nu ^2 )^2  (4 \nu ^2-9 ) (4 \nu ^2-25) (4 \nu ^2-49 )\left(4 \nu ^2-81 \right)}{1048576 \pi  x^8} \nn \\
&& +\frac{143 (1-4 \nu ^2)^2 (4 \nu ^2-9) (4 \nu ^2-25)(4 \nu ^2-49)(4 \nu ^2-81)\left(4 \nu ^2-121 \right)}{8388608 \pi  x^{10}}
      + ... \Bigg) \nn \\
&\simeq & \frac{H^4}{16 \pi} \frac{4 x^3}{\pi }
\Bigg[\frac{\bar \omega }{3}+\frac{4 \nu ^2-1}{12 \bar \omega }
 -\frac{5 \left(1-4 \nu ^2\right)^2}{384 \bar \omega ^5}
 -\frac{5 \left(1-4 \nu ^2\right)^2 \left(2 \nu ^2-11\right)}{768 \bar \omega ^7}   \\
&& +\frac{7  (1-4 \nu ^2 )^2  (92 \nu ^2-239 )}{2048 \bar \omega ^9}
 +\frac{7 (1-4 \nu ^2)^2 (2032 \nu ^4-36152 \nu ^2+72271)}{32768 \bar \omega ^{11}} \nn \\
&&  +\frac{55  (1-4 \nu ^2 )^2 (448 \nu ^6-62640 \nu ^4+602628 \nu ^2-1045309)}{131072 \bar \omega ^{13}}
        +... \Bigg] ,  \label{pkxi16om}
\ea
}
and the spectral  trace for $\xi=\frac16 $  is
{\allowdisplaybreaks
\ba \label{tracekxi16}
 \rho_k -3p_k   & \simeq & \frac{H^4}{16 \pi}
\Bigg(
\frac{ (1-4 \nu ^2 ) x^2}{\pi }
-\frac{ (1-4 \nu ^2 )^2}{8 \pi }
-\frac{3  (1-4 \nu ^2 )^2  (4 \nu ^2-9 )}{128 \pi  x^2}  \nn \\
&& -\frac{5 (1-4 \nu ^2 )^2  (4 \nu ^2-9  ) (4 \nu ^2-25  ) }{1024 \pi  x^4}
  -\frac{35 (1-4 \nu ^2 )^2  (4 \nu ^2-9  ) (4 \nu ^2-25  )
   (4 \nu ^2-49  )}{32768 \pi  x^6} \nn \\
&& -\frac{63  (1-4 \nu ^2 )^2 (4 \nu ^2-9 ) (4 \nu ^2-25 )
   (4 \nu ^2-49 )\left(4 \nu ^2-81 \right)}{262144 \pi  x^8} \nn \\
&& -\frac{231  (1-4 \nu ^2 )^2  (4 \nu ^2-9  )
 (4 \nu ^2-25  ) (4 \nu ^2-49  )
 (4 \nu ^2-81  ) (4 \nu ^2-121  )}{4194304 \pi  x^{10}}
 + ...\Bigg)   \\
& \simeq & \frac{H^4}{16 \pi} \frac{4 x^3}{\pi }
\Bigg[\frac{\frac{1}{4}-\nu ^2}{\bar \omega }
+\frac{3 (1-4 \nu ^2 )^2   }{64 \bar \omega ^5}
  + \frac{5 (1-4 \nu ^2 )^2 (  4  \nu ^2 -25)}{512 \, \bar \omega ^7} \nn \\
&& -\frac{105 (1-4 \nu ^2 )^2  (20\nu ^2 -53)}{2048 \, \bar  \omega ^9}
   -\frac{63(1-4 \nu ^2)^2 (176 \nu ^4-3256 \nu ^2+6563)}{8192\,\bar \omega ^{11}} \nn \\
&& +\frac{1155  (1-4 \nu ^2 )^2  (64 \nu ^6-9456 \nu ^4+92268 \nu ^2-160717)}
   {131072 \bar  \, \omega ^{13}}
              + ...    \Bigg] . \label{tracexi16}
\ea
}
In the $\bar \omega$ expansion
the spectral  trace   has only an  $\bar \omega^{-1}$ divergent term,
and,  in the $k$ expansion,  it has    only $k^2,k^0$ divergences.
There is a common factor  $(1-4 \nu ^2 )=4 \frac{m^2}{H^2}$,
so  in the limit $m=0$ the trace is zero,
consistent with its definition (\ref{traceTmunu}).

\begin{figure}[htb]
\centering
\subcaptionbox{}
    {%
        \includegraphics[width = .47\linewidth]{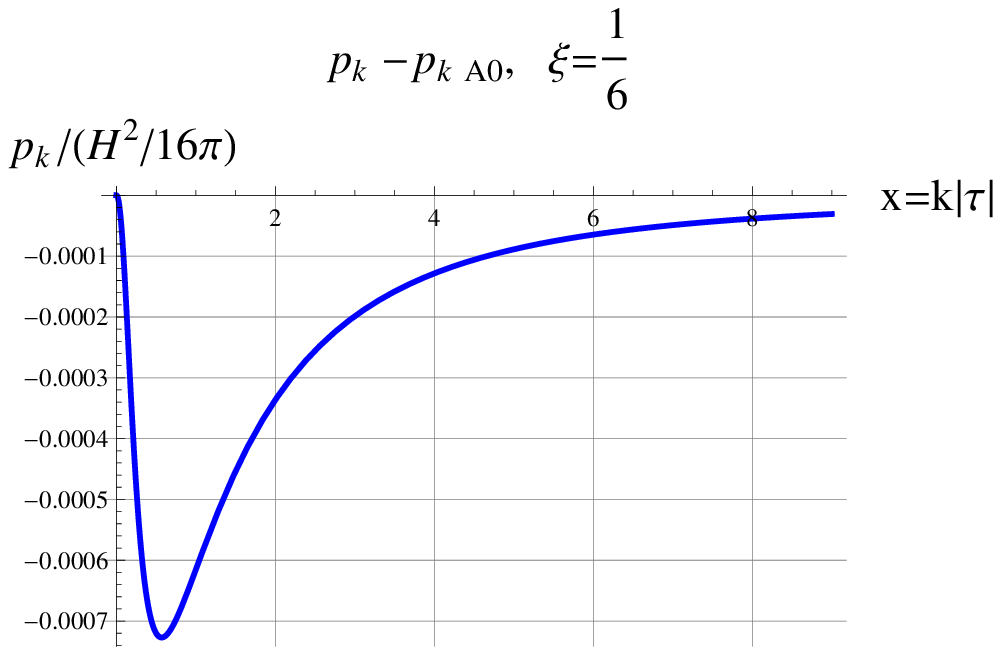}
}
\subcaptionbox{}
    {%
        \includegraphics[width = .47\linewidth]{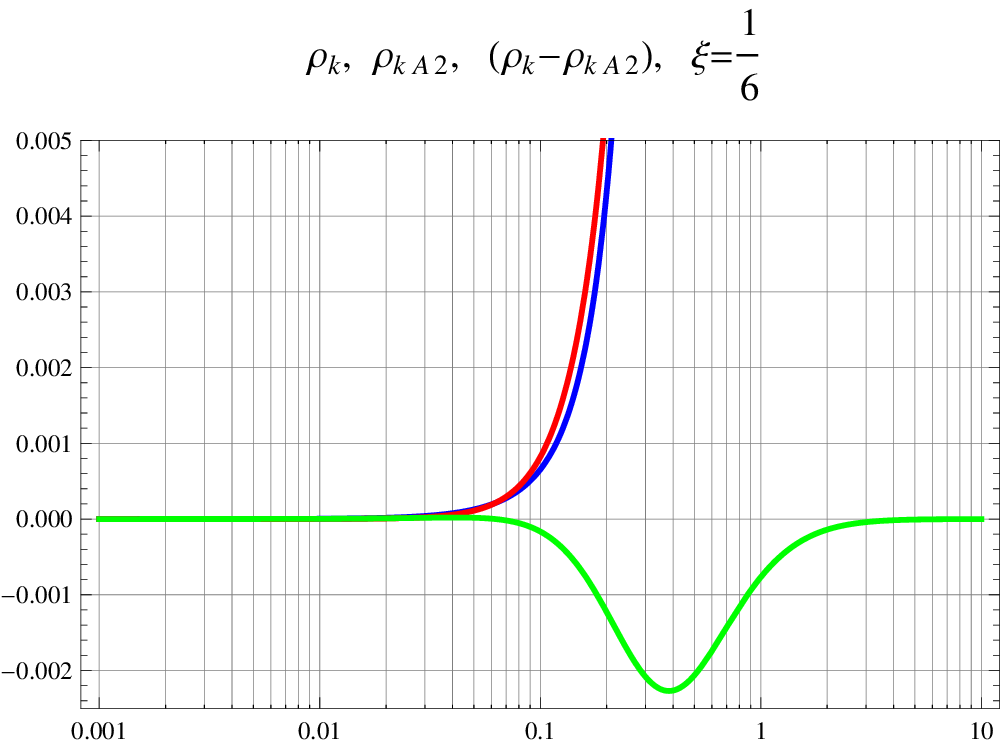}
   }
\subcaptionbox{}
    {%
        \includegraphics[width = .47\linewidth]{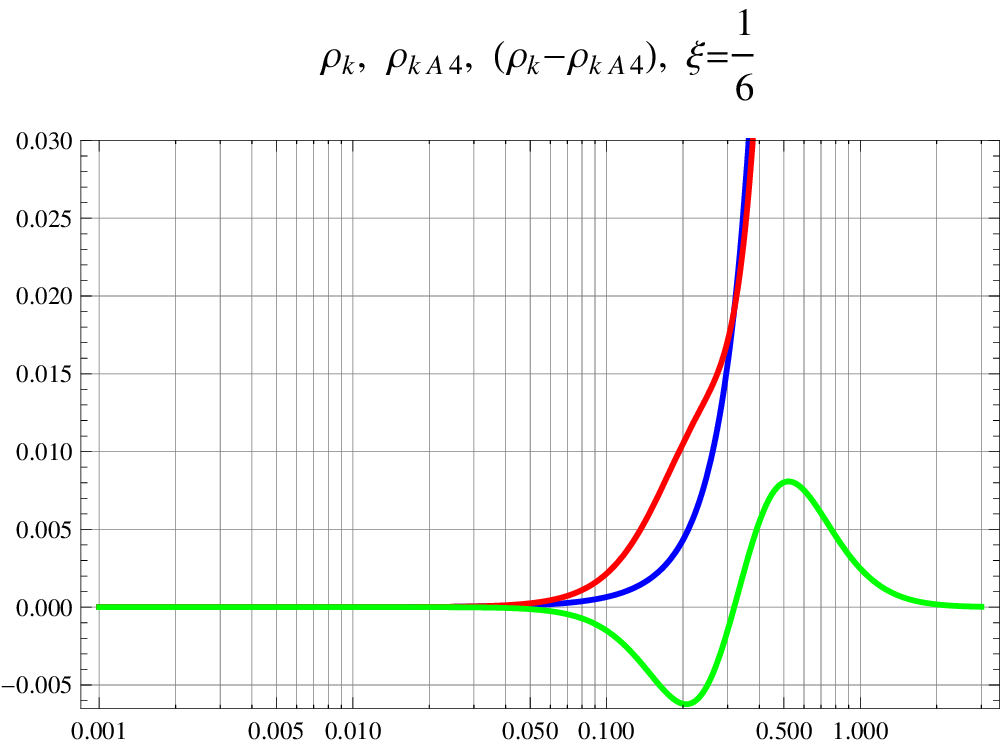}
    }
\subcaptionbox{}
    {%
        \includegraphics[width = .47\linewidth]{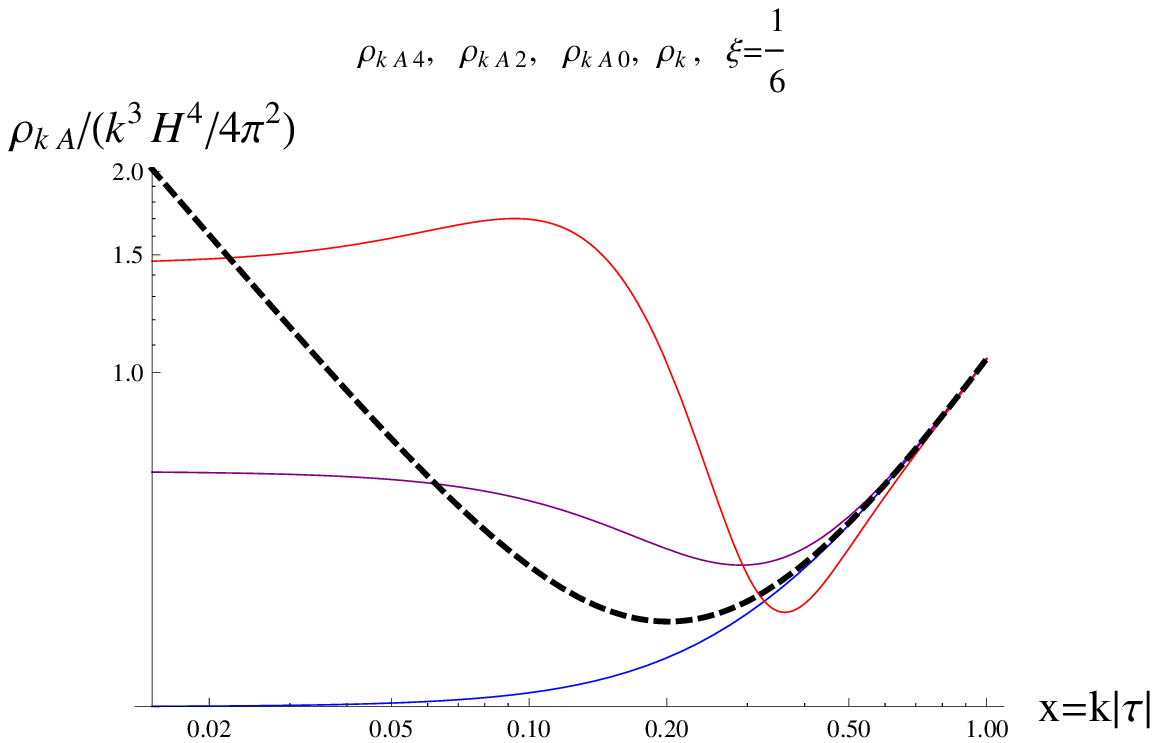}
        }
\caption{
    The model $\xi=\frac16 $.
    (a) The 0th-order regularized spectral pressure $(p_k-p_{k\, A 0})$.
    (b) Blue: the unregularized  $\rho_k$.
        Red: the subtraction term  $\rho_{k\, A 2 }$.
        Green:  $(\rho_{k }-\rho_{k\,A 2})$ is negative for all $k$.
    (c)    Blue: the unregularized  $\rho_k$.
          Red: the subtraction term  $\rho_{k\, A 4 }$ of (\ref{countrh1xi16omg}).
          Green: $(\rho_{k }-\rho_{k\,A 4})$ is negative for  $k|\tau| \lesssim 0.3$.
    (d)  The 4th (Red), 2nd (Purple), 0th (Blue) -order subtraction terms,
         compared with the unregularized $\rho_k$ (Dashed).
    }
     \label{Figure5}
\end{figure}

The 0th-order  subtraction term for spectral energy density
(\ref{rhoA0}) is written as
\ba\label{xi16rhok0}
\rho_{k\, A 0} & = & \frac{H^4}{16 \pi} \frac{4 x^3}{\pi } \bar \omega   \\
&  \simeq &  \frac{H^4}{16 \pi}\Bigg(
   \frac{4 x^4}{\pi }  +\frac{\left(1-4 \nu ^2\right) x^2}{2 \pi }
   -\frac{\left(1-4 \nu ^2\right)^2}{32 \pi }
   +\frac{\left(1-4 \nu ^2\right)^3}{256 \pi  x^2} +... \Bigg) \, .
\ea
In terms of $\bar \omega$, it cancels the only $\bar \omega$ divergent term
of $\rho_{k}$ in (\ref{rhoExpanxi16omg}),
or in terms of $k$,
it cancels all the $k^4, k^2, k^0$ divergent terms
 of $\rho_{k}$ in (\ref{rhoExpanxi16}).
The 0th-order regularized spectral energy density
\be
\rho_{k\, reg} =\rho_{k} -\rho_{k\, A 0}
\ee
is positive, UV convergent, as shown in Figure \ref{Figure4}(c) and Figure \ref{Figure4}(d).
At high $k$, it is dominated by the 4th term
\ba\label{rhrexi}
\rho_{k\, reg}
& \simeq & \frac{H^4}{16 \pi}\frac{4 x^3}{\pi }
\Bigg[ \frac{ (1-4 \nu ^2 )^2}{128 \bar \omega ^5}
-\frac{15  (1-4 \nu ^2 )^2}{512 \bar \omega ^7}
-\frac{21  (1-4 \nu ^2 )^2  (4 \nu ^2-13 )}{1024 \bar \omega ^9} \nn  \\
&& -\frac{105  (1-4 \nu ^2 )^2  (16 \nu ^4-584 \nu ^2+1297 )}{32768 \bar \omega ^{11}}
  + ... \Bigg] \\
& \simeq & \frac{H^4}{16 \pi} \frac{(1-4 \nu ^2)^2}{32 \pi x^2} \propto k^{-2} >0
\ea
which is  $k^{-2}$ convergent at high $k$.
For the spectral pressure,
the 0th-order subtraction term  is given by  (\ref{pkA0})
and is  written as
\be\label{psubt0th}
p_{k\, A 0} =    \frac{H^4}{16\pi}  \frac{4 x^3}{\pi }
    \Big[ \frac{\bar \omega}{3} -\frac{(1-4\nu^2)}{12 \bar \omega} \Big] \, ,
\ee
and the  regularized pressure is given by
\be
p_{k\, reg} = p_{k} -p_{k\, A 0}
\ee
which is negative for the whole range of  $k$,
as shown in Figure \ref{Figure5}(a).
This is checked at high $k$,
\ba \label{preg0thxi16}
p_{k } -p_{k\, A 0}
&\simeq & \frac{H^4}{16 \pi} \frac{4 x^3}{\pi }
\Bigg[
 -\frac{5 \left(1-4 \nu ^2\right)^2}{384 \bar \omega ^5}
 -\frac{5 \left(1-4 \nu ^2\right)^2 \left(2 \nu ^2-11\right)}{768 \bar \omega ^7}
 +\frac{7  (1-4 \nu ^2 )^2  (92 \nu ^2-239 )}{2048 \bar \omega ^9} \nn \\
&& +\frac{7 (1-4 \nu ^2)^2 (2032 \nu ^4-36152 \nu ^2+72271)}{32768 \bar \omega ^{11}}
        +... \Bigg]   \\
& \simeq &  - \frac{H^4}{16 \pi} \frac{4 x^3}{\pi }
       \frac{5 \left(1-4 \nu ^2\right)^2}{384 \bar \omega ^5} <0
\ea
which is negative.  Again a negative pressure is allowed.
The  0th-order subtraction term for the  trace  is
\be\label{trace0th}
\rho_{k\,A  0} - 3 p_{k\, A 0}
       = \frac{H^4}{16\pi}  \frac{4 x^3}{\pi }
         \frac{(1-4\nu^2)}{4 \bar \omega}
\ee
which is proportional to $m^2$.
The difference between (\ref{tracexi16}) and (\ref{trace0th})
gives the 0th-order regularized spectral trace
\ba\label{regtrc}
\langle T^{\mu} \,_{\mu}\rangle_{k\, reg}
& \simeq &
\frac{H^4}{16 \pi} \frac{4 x^3}{\pi }
\Bigg[\frac{3 (1-4 \nu ^2 )^2  }{64\bar \omega ^5}
  + \frac{5 (1-4 \nu ^2 )^2 (  4  \nu ^2 -25)}{512 \, \bar \omega ^7}
  -\frac{105 (1-4 \nu ^2 )^2  (20\nu ^2 -53)}{2048 \, \bar  \omega ^9}
  + ... \Bigg]
  \nn  \\
\ea
which is UV convergent.
The regularized trace has a factor $(1-4 \nu ^2 )^2 = 16 \frac{m^4}{H^4}$,
and  vanishes in the limit $m=0$,
and   no trace anomaly exists.
There is a relation between the regularized spectral trace (\ref{regtrc})
and the regularized power spectrum (\ref{psregxi}) for $\xi=\frac16$
as follows
\ba \label{trcdel}
\langle T^{\beta} \,_{\beta}\rangle_{k\, reg}
& = & H^2 (\frac14- \nu ^2 ) \Delta^2_{k\, reg \, 0}
  = m^2  \Delta^2_{k\, reg \, 0}
\ea
which, after $k$-integration, yields
\be\label{regTG}
\langle T^{\beta} \,_{\beta}\rangle_{  reg} = m^2 G(0)_{reg}
\ee
where $G(0)_{reg} $ is the regularized, finite  auto-correlation function.
The result (\ref{regTG}) is consistent with
the definition relation (\ref{traceTmunu16}) for $\xi=\frac16$.
(See Figure \ref{Figure10}(a) for the value of $G(0)_{reg}$
and  Sect.\ref{section6} for the Green's function.)
After  numerical  integration over $k$,
the 0th-order regularized energy density and pressure
 (for the model   $\frac{m^2}{H^2}=0.1$)
  are the following
\be\label{regrhoxi16}
\rho_{\,reg} =  \int_0^\infty ( \rho_{k} -\rho_{k\, A 0})  \frac{dk}{k}
  \simeq    0.001786 \frac{H^4}{16 \pi} =0.1786    \frac{m^4}{16 \pi}  >0,
\ee
\be \label{regpxi16}
p_{\, reg}  \int_0^\infty ( p_{k} -p_{k\, A 0})  \frac{dk}{k}
  \simeq    - 0.001786 \frac{H^4}{16 \pi} = -0.1786   \frac{m^4}{16 \pi} < 0 ,
\ee
and the regularized trace
\be\label{trv7}
\langle T^{\mu} \,_{\mu }\rangle_{reg}
= 4 \rho_{\,reg}=0.007143 \frac{H^4}{16 \pi}= 0.7143  \frac{m^4}{16 \pi}.
\ee
So the regularized stress tensor with $\xi=\frac16$
in the  vacuum is also maximally symmetric,
\be\label{mtrunmxi16}
\langle T_{\mu\nu}  \rangle_{  reg}
  = \frac14 g_{\mu\nu} \langle T^{\beta} \,_{\beta }\rangle_{  reg} ,
\ee
similar to  Eq.(\ref{mtrunmxi0}) for $\xi=0$,
although  it is nonuniformly distributed among the  $k$-modes.
In regard to the magnitude of vacuum energy density
  (\ref{regrhoxi16}) for  $\xi=\frac16$
is about three orders lower than  (\ref{rhoreg}) for $\xi=0$.
For the regularized stress tensor in the vacuum to drive inflation,
one would have a very high rate expansion $H\sim 50 G^{-1/2}$
for a model $\frac{m^2}{H^2}=0.1$ and $\xi=\frac16$.
Its $k$-modes $\phi_k$ still can be part of the primordial fluctuations.
As (\ref{rhoA0cons}) (\ref{rho0thconsv}) show,
the four-divergence of the 0th-order subtraction terms
for stress tensor is zero,
so that the 0th-order regularized spectral stress tensor
respects the covariant conservation.
Hence,  the 0th-order adiabatic regularization
yields  a positive, UV-convergent, covariantly-conserved
spectral energy density
for the conformally-coupling, massive scalar field.
Thus,  the   0th-order   regularization
works simultaneously  for both the power spectrum
and the spectral stress tensor with  $\xi=\frac16$.

We next examine what happens for the 2nd-order regularization.
From (\ref{rhoA22count}) for $\xi=\frac16$
the 2nd-order subtraction term  is
\ba
\rho_{k\, A 2} & = &
\frac{H^4}{16 \pi} \frac{4 x^3}{\pi }
\Big[   \bar \omega + \frac{(1-4 \nu ^2 )^2}{128 \bar \omega^5}   \Big] \nn \\
& \simeq &  \frac{H^4}{16 \pi}
\Bigg(
 \frac{4 x^4}{\pi }
 +\frac{\left(1-4 \nu ^2\right) x^2}{2 \pi }
 -\frac{\left(1-4\nu ^2\right)^2}{32 \pi}
 -\frac{\left(1-4 \nu ^2\right)^2 \left(4 \nu ^2-9\right)}{256 \pi  x^2} \nn \\
&&  ~~~~ -\frac{5 \left(4 \nu ^2-33\right) \left(4 \nu ^2-1\right)^3}{8192 \pi  x^4}
 +...\Bigg) .
\ea
This will  subtract too much,  removing  not only the $\bar \omega$ term,
but also the $\bar \omega^{-5}$ term of (\ref{rhoExpanxi16omg}),
or in terms of $k$,
removing  the first four terms of $\rho_{k}$  of (\ref{rhoExpanxi16}).
So,  for a conformally-coupling massive scalar field,
the  2nd-order regularization violates the minimal subtraction rule,
and is an incorrect  prescription,
leading to  a negative spectral energy density, as shown in Figure \ref{Figure5}(b).

We then examine  the 4th-order regularization.
From (\ref{countrh1})  for $\xi=\frac16$
the 4th-order subtraction term   is
\ba\label{countrh1xi16omg}
\rho_{k\,A 4}
& =& \frac{H^4}{16\pi} \frac{4 x^3}{\pi}
\Big[
\bar \omega +\frac{ (\frac{1}{4}-\nu ^2 )^2}{8 \bar \omega ^5}
-\frac{15 \left(\frac{1}{4}-\nu ^2\right)^2}{32 \bar \omega ^7}
+\frac{21 \left(\frac{1}{4}-\nu ^2\right)^3}{16 \bar \omega ^9}
-\frac{105 \left(\frac{1}{4}-\nu ^2\right)^4}{128 \bar \omega ^{11}}
\Big]
   \\
& \simeq &  \frac{H^4}{16 \pi}
\Bigg(
 \frac{4 x^4}{\pi }
 +\frac{\left(1-4 \nu ^2\right) x^2}{2 \pi }
 -\frac{\left(1-4\nu ^2\right)^2}{32 \pi}
 -\frac{\left(1-4 \nu ^2\right)^2 \left(4 \nu ^2-9\right)}{256 \pi  x^2} \nn \\
&& -\frac{5  (1-4 \nu ^2 )^2 (4 \nu ^2-9 )(4 \nu ^2-25 )}{8192 \pi  x^4}
      -\frac{7  (4 \nu ^2-1 )^3  (16 \nu ^4-328 \nu ^2+1809  )}{65536 \pi  x^6}
 +... \Bigg) .  \label{countrh1xxi16} \nn \\
\ea
This also subtracts off too much,
canceling not only the divergent $\bar \omega$ term
and but also the convergent $\bar \omega^{-5}$, $\bar \omega^{-7}$  terms
in (\ref{countrh1xi16omg}),
thus violating the minimal subtraction rule.
Or, in terms $k$,
it removes  not only the $k^{4}, k^{2}, k^0$ divergent terms
but also the  $k^{-2}, k^{-4}$ convergent terms of $\rho_{k}$  in (\ref{rhoExpanxi16}).
The resulting spectral energy density $(\rho_{k} -\rho_{k\, A 4})$
is negative in the low $k$ range,
as shown in Figure \ref{Figure5}(c).
Thus,
the 4th-order regularization is   an incorrect  prescription
for a conformally-coupling massive scalar field.
We plot
$\rho_{k\, A 0}$, $\rho_{k\, A 2}$, $\rho_{k\, A 4}$, $\rho_{k}$
together in Figure \ref{Figure5}(d),
which reveals that only $\rho_{k\, A 0}$ is lower than $\rho_k$
and yields a positive $\rho_{k \, reg}$,
whereas $\rho_{k\, A 2}$ and $\rho_{k\, A 4}$
are higher than $\rho_k$.

When the power spectrum $( \propto |v_k|^2)$  is correctly regularized by
the 0th-order subtraction term $|v_k^{(0)}|^2$,
the  spectral energy density
$( \propto   | v_k ' |^2  + k^2 |v_k|^2
 +m^2 a^2 | v_k |^2 ) \propto
 \bar \omega +\frac{\left(1-4 \nu ^2\right)^2}{128\bar \omega^5}$
will be also correctly regularized by the 0th-order  subtraction term
$| v_k^{(0)} \, '|^2 + k^2  | v_k^{(0)} |^2
+m^2 a^2 | v_k^{(0)} |^2 \propto \bar \omega$.

With regard to renormalization
for the conformally-coupling  massive scalar field,
the divergent terms of  $\langle T^\mu\,_\nu \rangle$
in (\ref{rhoExpanxi16omg}) and (\ref{pkxi16om})
are of the 0th-order   (no time derivatives of $a(\tau)$),
and will be absorbed by renormalizing the cosmological constant $\Lambda$.
Neither the gravitation constant nor the counter terms
$^{(1)}H_{\mu\nu}$, $^{(2)}H_{\mu\nu}$ are involved
in the renormalization  \cite{CandelasRaine1975}.

Now we investigate  the  conformally-coupling massless scalar field.
With $m=0$ and  $\xi=\frac16$,
the exact  mode (\ref{u}) reduces to
\be  \label{u016}
v_k (\tau )  \equiv  -i \sqrt{\frac{\pi}{2}}\sqrt{\frac{x}{2k}}
  H^{(2)}_{1/2} ( x)
  =    \frac{1}{\sqrt{2k} }   e^{-i k\tau} ,
\ee
Eqs.(\ref{rhok})  (\ref{sprectpressure})  reduce to
\ba \label{rhok16m}
\rho_k  = \frac{ k^3}{4\pi^2 a^4}  ( |v_k'|^2 + k^2  |v_k|^2 )
= \frac{ k^3}{4\pi^2 a^4} k ,
~~~ ~  p_k = \frac13 \rho_k ,
\ea
and the trace   $\langle T^\mu\, _\mu \rangle_k=0$.
The spectral energy density and pressure in (\ref{rhok16m}) have only one term,
like in the Minkowski spacetime.
The 0th-order subtraction terms  (\ref{rhoA0}) (\ref{pkA0})
for $m=0$ (independent of $\xi$) are give by
\be\label{rho0thsb}
\rho_{k\,A 0}=\frac{ k^3}{4\pi^2 a^4} k ,
~~~ ~  p_{k\,A 0}  =\frac{ k^3}{12\pi^2 a^4} k  ,
\ee
 just equal to the unregularized stress tensor (\ref{rhok16m}).
Thus,  the regularized stress tensor is zero, so is the regularized trace,
\be\label{zerostress}
\langle T^{\mu\nu} \rangle _{k\, reg}=0,
~~~~ \langle T^{\beta}\, _\beta \rangle _{k\, reg}=0 .
\ee
Hence, there is no trace anomaly
for the conformally-coupling massless scalar field,
and  the 0th-order regularization respects the conformal symmetry.
The power spectrum has only one term,
\be\label{m0Xi16PS}
\Delta^2_{k} = \frac{ k^{3}}{2  \pi^2 a^2 }
      \Big( |v_k(\tau)|^2  \Big)
  =  \frac{ k^{3}}{2  \pi^2 a^2}  \frac{1}{2 k} ,
\ee
which  is  UV divergent,
the  0th-order regularization is also sufficient,
giving a  vanishing regularized spectrum
\ba\label{ps16m0}
\Delta^2_{k\, reg} & = & \frac{ k^{3}}{2  \pi^2 a^2 }
      \Big( |v_k(\tau)|^2 -|v_k^{(0)}(\tau)|^2 \Big)
      =  \frac{ k^{3}}{2  \pi^2 a^2 }
      \Big( \frac{1}{2 k} -\frac{1}{2 k} \Big) =0.
\ea
The vanishing results (\ref{zerostress}) (\ref{ps16m0})
tell us  that the 0th-order adiabatic regularization of
the conformally-coupling massless scalar field in de Sitter space
is essentially the same as the conventional regularization (by normal ordering)
of the massless scalar field in Minkowski spacetime.

It is an interesting observation that,
for the  case $m=0$ and $\xi=\frac16$,
the 2nd- and 4th-order subtraction terms
(\ref{rhoA22count}) (\ref{pA2count}) (\ref{countrh1}) (\ref{countpress})
  are actually
 equal to the 0th-order terms,
\be\label{rhoA024}
\rho_{k\,A4}= \rho_{k\,A2}=\rho_{k\,A0},
~~~~ p_{k\,A4}= p_{k\,A2}=p_{k\,A0} ,
\ee
and  consequently, the 0th-, 2nd-, and 4th-order regularization are the same.
This is true not only for de Sitter space,
and but also for a general flat RW spacetime,
as is seen from the expressions
(\ref{rhoA22count}) (\ref{pA2count}) (\ref{countrh1}) (\ref{countpress}).
The relations    (\ref{rhoA024})
can be also rechecked   as follows.
For $m=0$ and $\xi=\frac16$,
the 0th-order effective frequency $W^{(0)}=k$,
and the 0th-order WKB solution
\be\label{v(0)m0}
v_k^{(0)}= \frac{1}{\sqrt { 2 W^{(0)}} } e^{-i\int W^{(0)}d \tau}
=  \frac{1}{\sqrt{2 k} } e^{-i k \tau} ,
\ee
which is equal to the exact mode  $v_k$ in (\ref{u016}).
The 2nd-order  by (\ref{W2expression})
is   $W^{(2)}= k$,
and the 4th-order $W^{(4)}= k$ by (\ref{Wk4}),
so that
\be \label{v1024}
v_k^{(2)}= v_k^{(4)} = v_k^{(0)}.
\ee
Thus,
for the conformally-coupling massless scalar field in a general RW spacetime,
the WKB approximate solution (\ref{v(0)m0}) is the exact solution,
as is evident from the wave equation (\ref{equvk}),
the regularized spectral stress tensor and power spectrum will be always
zero under adiabatic regularization.

The zero trace  (\ref{zerostress}) can also follow
from taking the massless limit of 0th-order regularized trace of a massive field.
In contrast,  the so-called conformal trace anomaly in literature was made up
from the massless limit of 4th-order regularized trace of a massive field.
The stress tensor (\ref{rhok}) (\ref{sprectpressure}) (\ref{tracekgen})
of the massive scalar field
contains the mass $m$ as a smooth parameter.
Starting with  $m\ne 0$,
the unregularized trace  (\ref{traceTmunu16})    (\ref{tracekgen})
with $\xi=\frac16$   give
\be \label{trace16}
\langle T^\mu\, _\mu \rangle_{k}
=\frac{ k^3}{2 \pi^2 a^2}      m^2  |v_k|^2  ,
\ee
the 0th-order subtraction term for the  trace in (\ref{trcsb0}) is
\be\label{tracesub}
\langle T^\mu\, _\mu \rangle_{k\, A0}
  =\frac{m^2 k^3}{2\pi^2 a^2} \frac{1}{2\omega } ,
\ee
so, the 0th-order regularized trace  is
\ba\label{0thtr}
\langle T^\mu\, _\mu \rangle_{k}
 -\langle T^\mu\, _\mu \rangle_{k\, A0}
 = \frac{m^2 k^3}{2 \pi^2 a^2}  \Big(|v_k|^2 -\frac{1}{2\omega} \Big) ,
\ea
which holds  for $\xi= \frac16$ and a general $m$.
Now taking  the limit $m=0$ of  (\ref{0thtr}),
using   $\omega=k$ and $|v_k|^2 =\frac{1}{2k}$ by (\ref{u016}),
one obtains
\ba\label{0tr}
\langle T^\mu\, _\mu \rangle_{k}
 -\langle T^\mu\, _\mu \rangle_{k\, A0}
 = \lim_{m\rightarrow 0} \,
 \frac{m^2 k^3}{2 \pi^2 a^2}  \Big(\frac{1 }{2k} -\frac{1 }{2k} \Big)  =0  ,
\ea
and its  $k$-integration  is also zero,
$\langle T^{\mu} \,_{\mu}\rangle_{reg}=0$.
In Sect.\ref{section6}, we shall rederive the results (\ref{zerostress})  (\ref{ps16m0})
by the method of regularization of Green's function.
The trace anomaly in adiabatic regularization
\cite{Bunch1980,BunchParker1979,AndersonParker1987}
occurred in the finite part of
4th-order adiabatic subtraction term (\ref{trcsub}) for  $\xi=\frac16$.
Since  the 4th-order regularization leads to a negative spectral energy density,
the trace anomaly is a consequence  of the improper  prescription.

\section{The  regularization of  scalar field with a general $\xi$}\label{section5}

For a massive scalar field with a general coupling $\xi$,
the situation of adiabatic regularization is more complicated than
the $\xi=0$ and $\xi=\frac16$ cases in Sects. \ref{section3}, \ref{section4}.
In this section  we consider
the value of $\xi$ in the open interval $(0,\, \frac16)$
 in de Sitter space.
We briefly report the calculation  results.

First we find that
the 0th-order regularization fails to  remove UV divergences
in the spectral energy density and power spectrum.

\begin{figure}[htb]
\centering
\subcaptionbox{}
    {%
        \includegraphics[width = .48\linewidth]{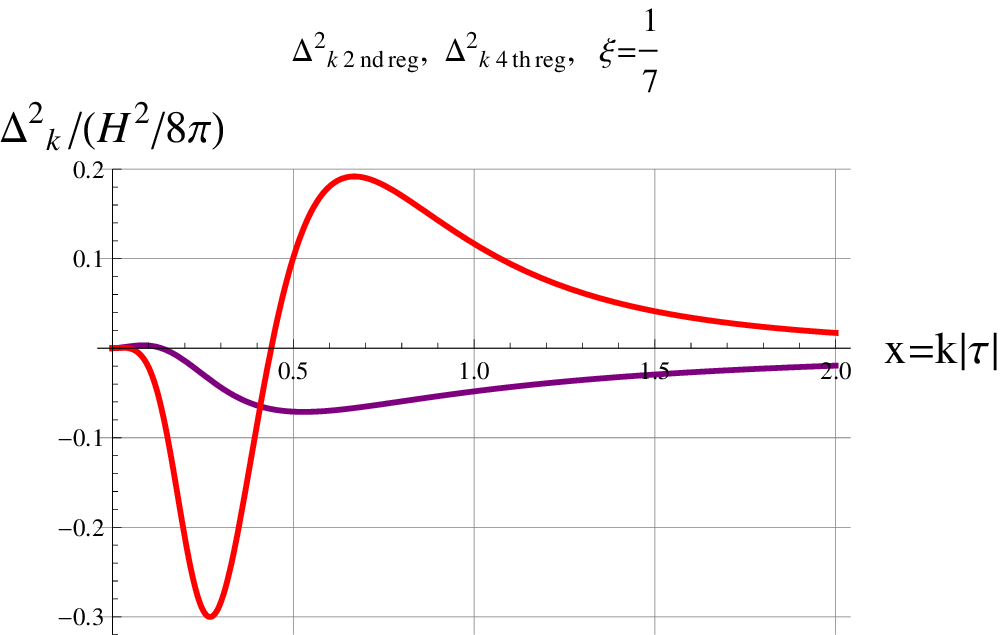}
}
\subcaptionbox{}
    {%
        \includegraphics[width = .48\linewidth]{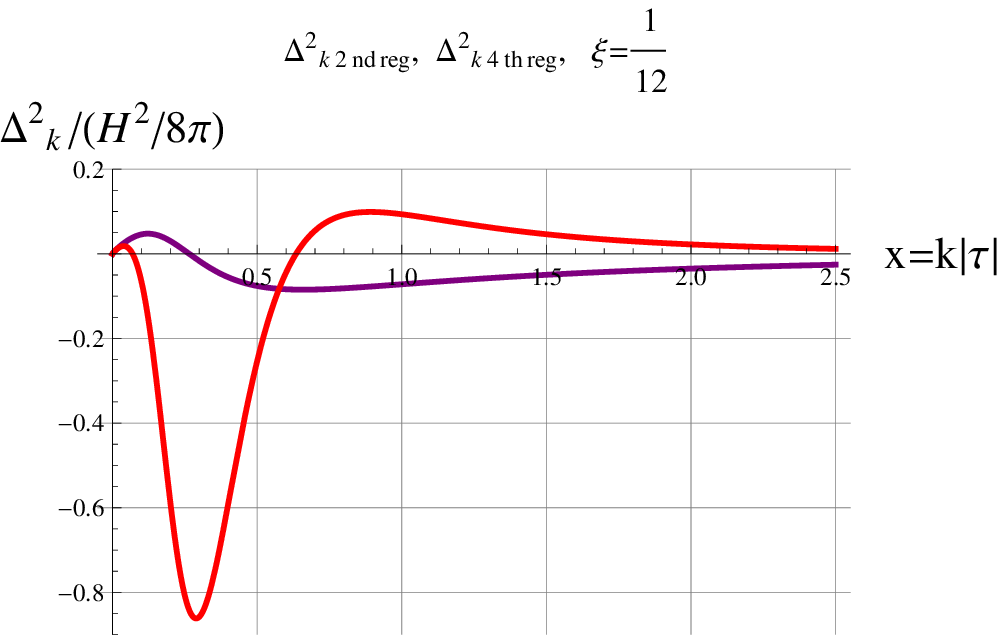}
   }
\subcaptionbox{}
    {%
        \includegraphics[width = .48\linewidth]{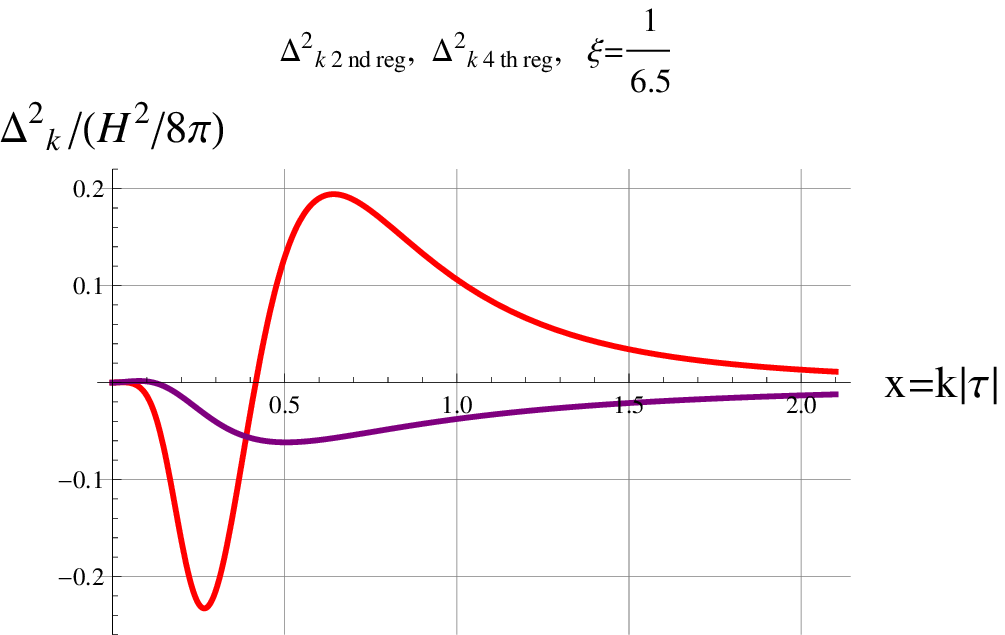}
    }
\caption{
    (a) Red: the 4th-order regularized power spectrum.
        Purple: the 2nd-order.
        Both have negative values for    $\xi=\frac{1}{7}$.
    (b)
   Red: the 4th-order regularized power spectrum.
    Purple: the 2nd-order.
    Both have negative values for   $\xi=\frac{1}{12}$.
    (c)   Red: the 4th-order regularized power spectrum.
          Purple: the 2nd-order.
             Both have negative values for    $\xi=\frac{1}{6.5}$.
    }
     \label{Figure6}
\end{figure}

Next we find that the power spectrum is negative
for $\xi$ in $(0,\, \frac16)$ and for a general $m$,
by either  the 2nd- or 4th-order regularization,
as shown in Figure \ref{Figure6}(a), Figure \ref{Figure6}(b),  Figure \ref{Figure6}(c).
Comparatively, the 4th-order is more interesting in that
it  gives a power spectrum which is positive at high $k$,
but only has some distortions
with negative values at low $k$ around $k|\tau| \lesssim 0.6$.

\begin{figure}[htb]
\centering
\subcaptionbox{}
    {%
        \includegraphics[width = .48\linewidth]{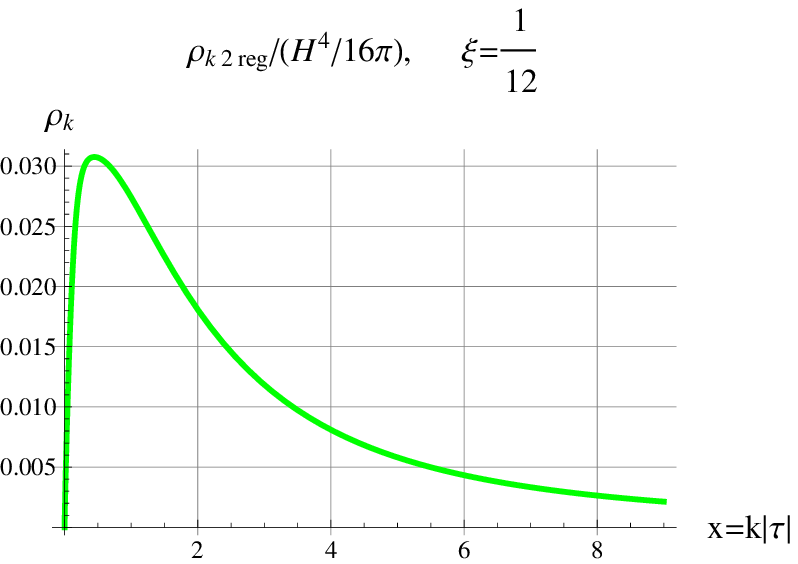}
}
\subcaptionbox{}
    {%
        \includegraphics[width = .48\linewidth]{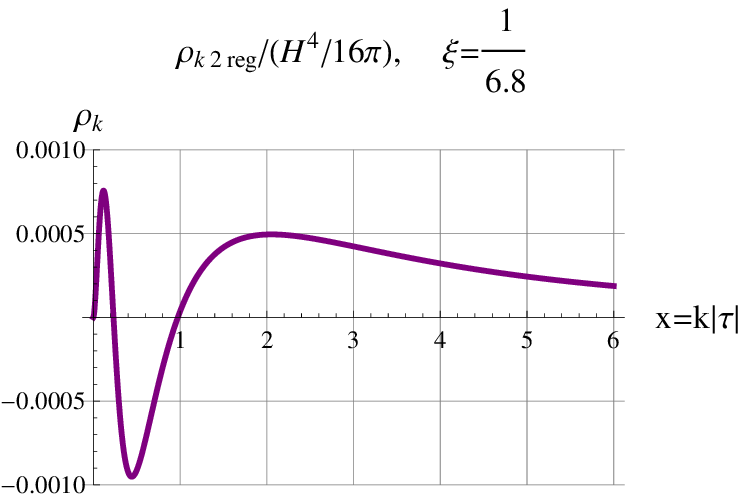}
   }
\subcaptionbox{}
    {%
        \includegraphics[width = .48\linewidth]{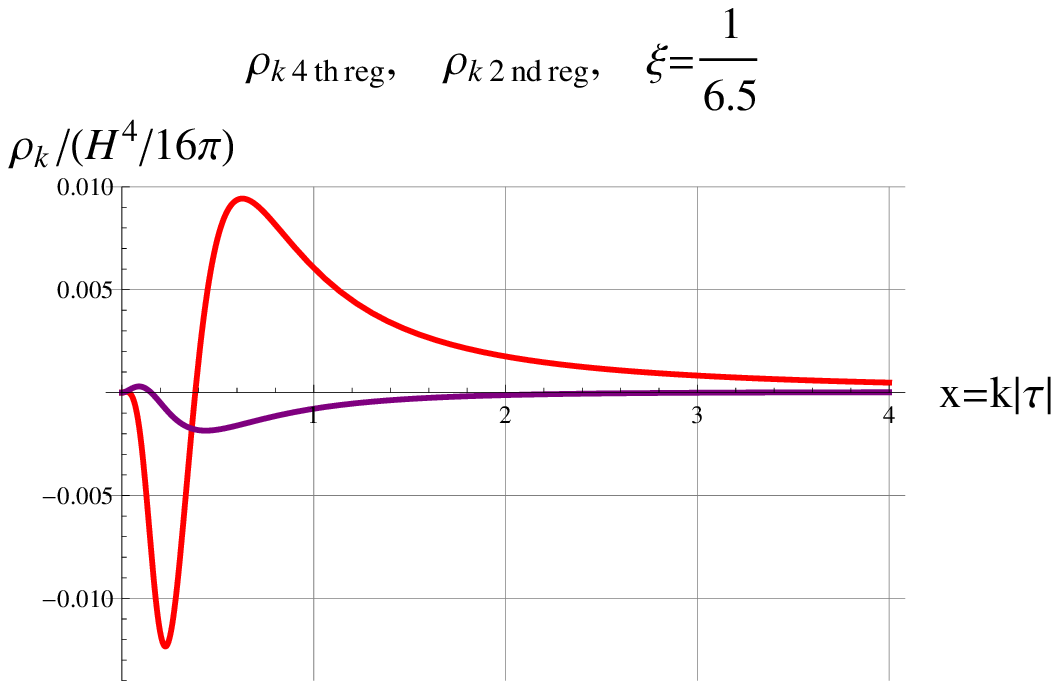}
    }
\subcaptionbox{}
    {%
        \includegraphics[width = .48\linewidth]{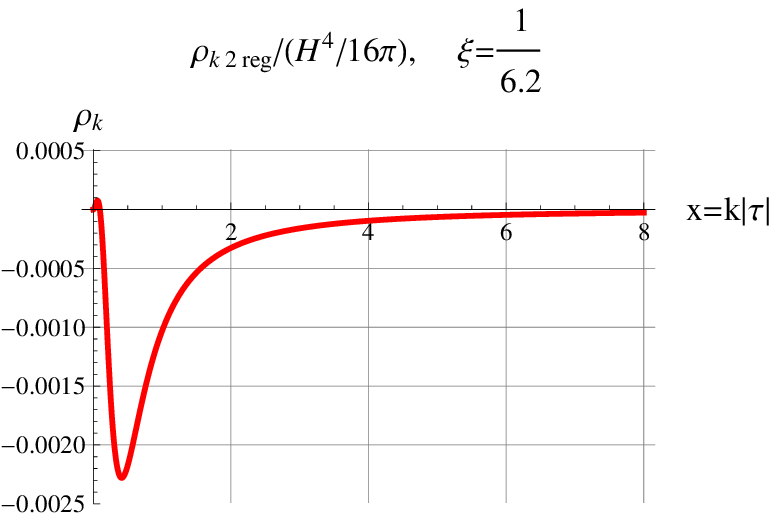}
    }
\caption{
    (a)  The 2nd-order regularized $(\rho_{k }-\rho_{k\,A 2})$
    is positive definite
   for  $\xi=\frac{1}{12}$.
    (b)   The  2nd-order   $(\rho_{k }-\rho_{k\,A 2})$
is positive at high $k$ and negative at low $k$ for  $\xi=\frac{1}{6.8}$.
    (c)  The regularized spectral energy densities have negative values
         for  $\xi=\frac{1}{6.5}$.
        Red: the 4th-order;
       Purple: the 2nd-order.
   (d)  The  2nd-order  $(\rho_{k }-\rho_{k\,A 2})$  is negative at high $k$
        for   $\xi=\frac{1}{6.2}$.
    }
     \label{Figure7}
\end{figure}

\begin{figure}[htb]
\centering
\subcaptionbox{}
    {%
        \includegraphics[width = .48\linewidth]{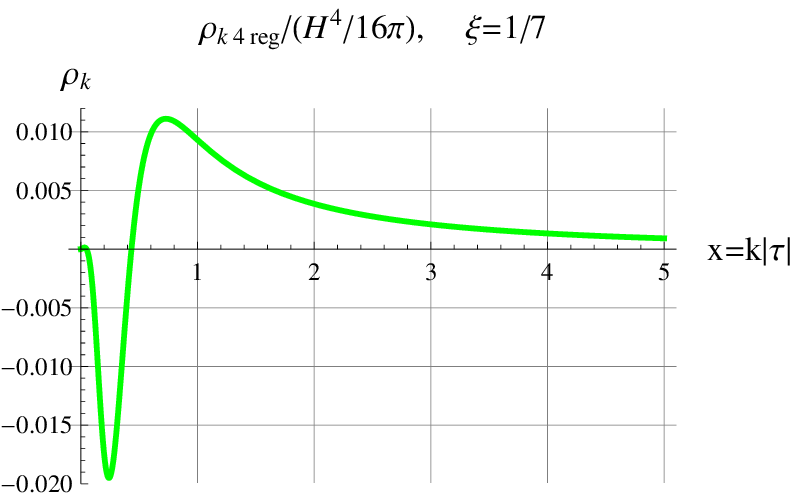}
}
\subcaptionbox{}
    {%
        \includegraphics[width = .48\linewidth]{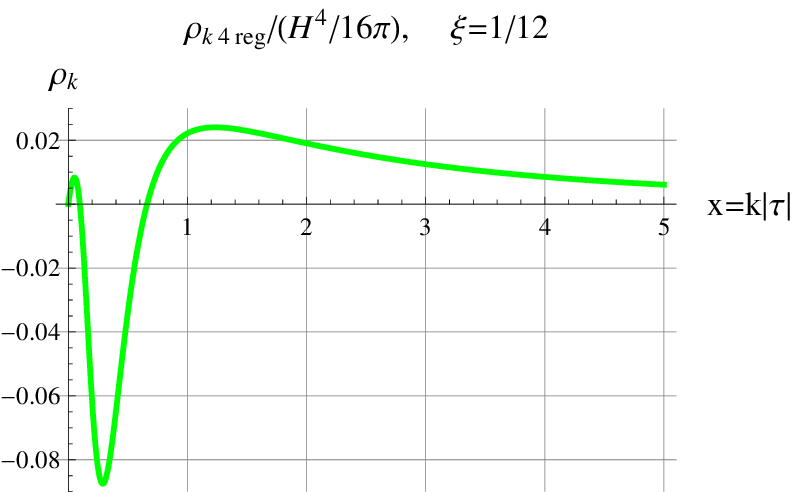}
   }
\subcaptionbox{}
    {%
        \includegraphics[width = .48\linewidth]{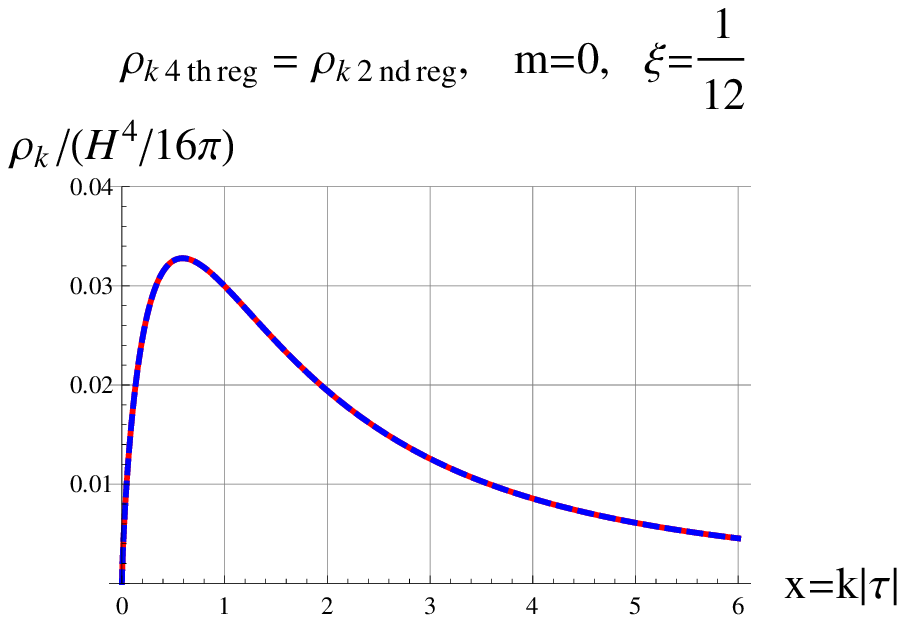}
    }
\subcaptionbox{}
    {%
        \includegraphics[width = .48\linewidth]{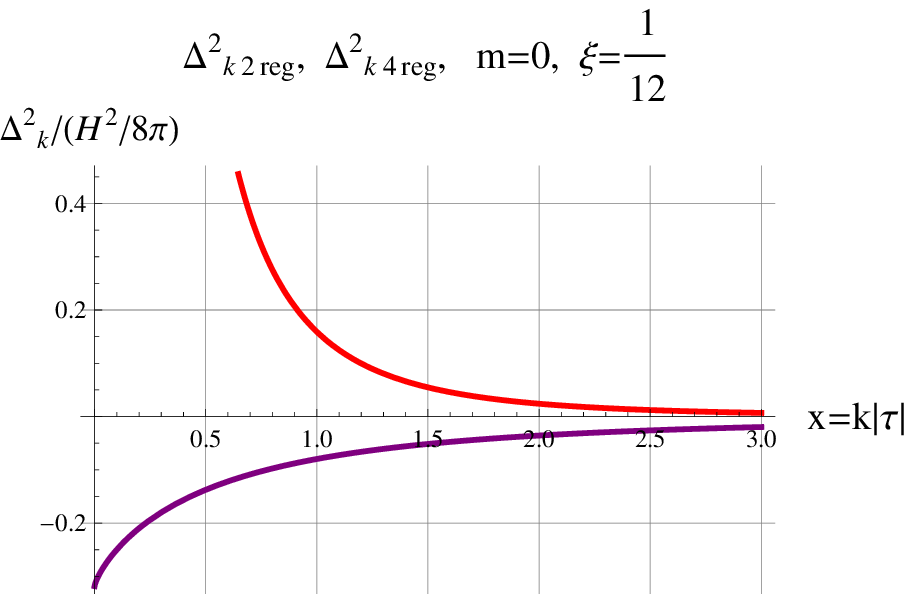}
    }
\caption{
    (a)  The 4th-order  $(\rho_{k }-\rho_{k\,A 4})$
is positive at high $k$ and negative at low  $k$
for $\xi=\frac{1}{7}$.
    (b)
   The 4th-order $(\rho_{k }-\rho_{k\,A 4})$
is positive at high $k$ and negative at low  $k$
for   $\xi=\frac{1}{12}$.
    (c)
  The 2nd- and 4th-order regularized spectral energy densities  are overlapped,
       both positive and UV-convergent
        for   $m=0$ and  $\xi=\frac{1}{12}$.
 (d)
 Purple: the 2nd-order is negative;
 Red: the 4th-order regularized power spectrum
   is positive and UV-convergent, but is IR divergent
 for   $m=0$ and $\xi=\frac{1}{12}$.
    }
     \label{Figure8}
\end{figure}

Then for the spectral energy density,
 we find that the  2nd-order regularized one
is positive  for $\xi \in (0, \frac{1}{12}]$
as shown in  Figure \ref{Figure7}(a),
and is positive at high $k$ and negative at low $k$
for $\xi \in (\frac{1}{12},\frac{1}{6.5})$
as shown in  Figure \ref{Figure7}(b),
and is negative at high $k$ for $\xi \in (\frac{1}{6.5},\frac{1}{6})$,
as shown in Figure \ref{Figure7}(c) and Figure \ref{Figure7}(d).
The 4th-order regularization
leads to a spectral energy density which is
positive at high $k$,
and has some distortions with negative values
at low  $k$ around $k|\tau| \lesssim   0.6$,
as shown in  Figure \ref{Figure7}(c), Figure \ref{Figure8}(a),  Figure \ref{Figure8}(b).

The above results for  $\xi \in  (0,\frac{1}{6})$ with $m\ne 0$
tells  that the conventional adiabatic regularization
can not give  simultaneously positive spectral energy density
and positive power spectrum by the same adiabatic order.
The results also suggest that
only the 4th-order regularization is hopeful
in that it gives positive spectra at high $k$,
while the distortions and negative values
occur only at low $k$ outside the horizon of inflation.
In order to get positive spectra without low $k$ distortions,
some modifications may be made to the 4th-order regularization
for a general $\xi$,
like what we have done for RGW \cite{ZhangWangJCAP2018}.

For a massless scalar field with  $\xi \in (0,\frac{1}{6})$ in de Sitter space,
we find that  the 0th-order regularization again fails
to remove UV divergences.
The 2nd-order  subtraction terms (\ref{rhoA22count}) (\ref{pA2count})
for the stress tensor
turn out to be  equal to the 4th-order (\ref{countrh1}) (\ref{countpress}),
\be\label{countrh0mxi}
\rho_{k\,A 4} = \rho_{k\,A2}
= \frac{k^3}{4\pi^2 a^4}
\Big[ k + (\xi-\frac16) \frac{-3}{k \tau^2}
        \Big] ,
\ee
\be\label{countrh0mxip}
p_{k\,A 4} = p_{k\,A2} =
 \frac{k^3}{12\pi^2 a^4}
\Big[ k
    +  (\xi-\frac16) \frac{1}{k} \frac{3}{\tau^2}
    \Big],
\ee
and the regularized spectral energy density is positive and UV convergent,
as shown in Figure \ref{Figure8}(c) for  $\xi=\frac{1}{12}$ as an illustration.

However,  for the power spectrum,
the 2nd- and 4th-order subtraction terms
are not equal, $|v^{(2)}_k|^2 \ne |v^{(4)}_k|^2$ (see (\ref{W2xi}) and (\ref{W4gen})).
The 2nd-order regularized power spectrum is negative,
and the  4th-order one is positive and UV convergent
but suffers from  IR divergence, as shown in Figure \ref{Figure8}(d).
This IR divergence is due to the zero mass,
analogous to that of RGW \cite{ZhangWangJCAP2018}.
Some modifications at low $k$
are  needed to save the 4th-order   regularization.
Overall,
 the issue of adiabatic regularization for a general $\xi\in (0,\frac16)$
is unsettled.

\section{The Green's function regularization with $\xi=0$ and $ \frac16$ }\label{section6}

\begin{figure}[htb]
\centering
\subcaptionbox{}
    {%
        \includegraphics[width = .48\linewidth]{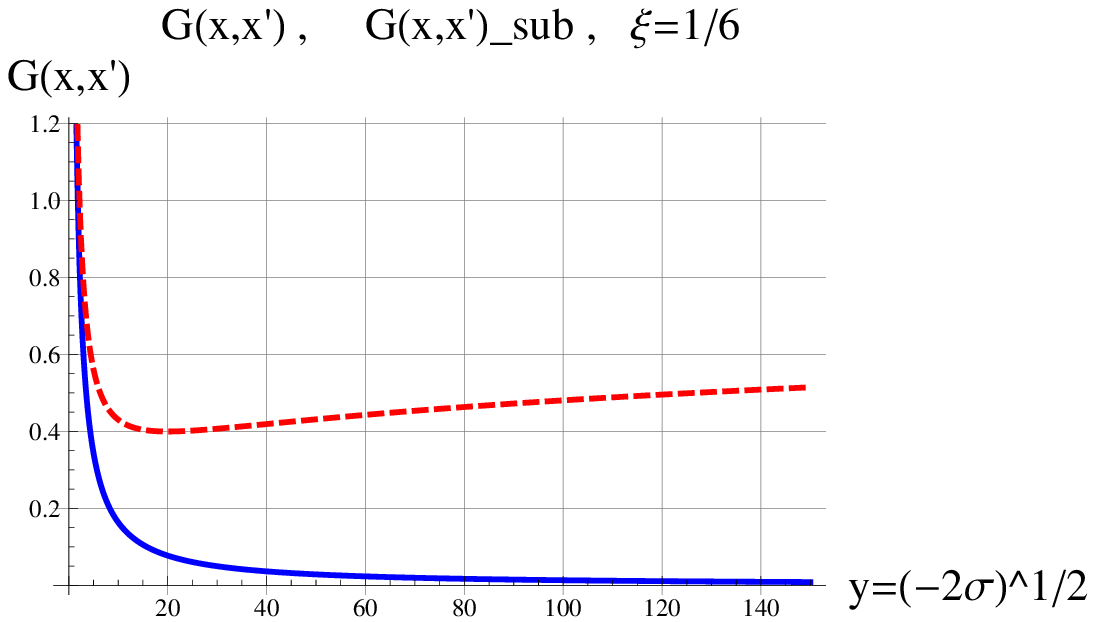}
}
\subcaptionbox{}
    {%
        \includegraphics[width = .48\linewidth]{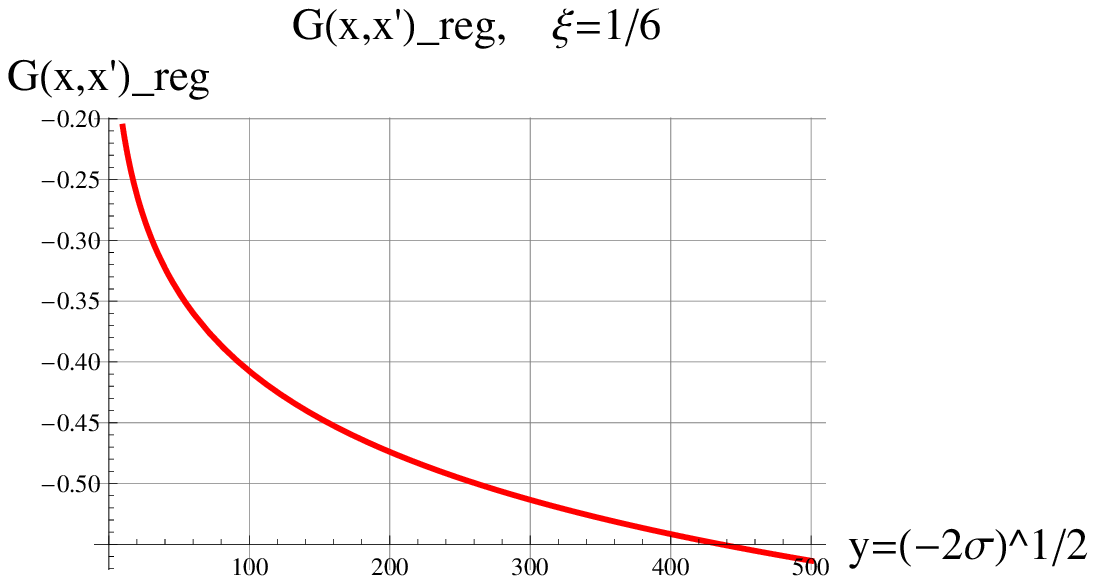}
   }
\subcaptionbox{}
    {%
        \includegraphics[width = .48\linewidth]{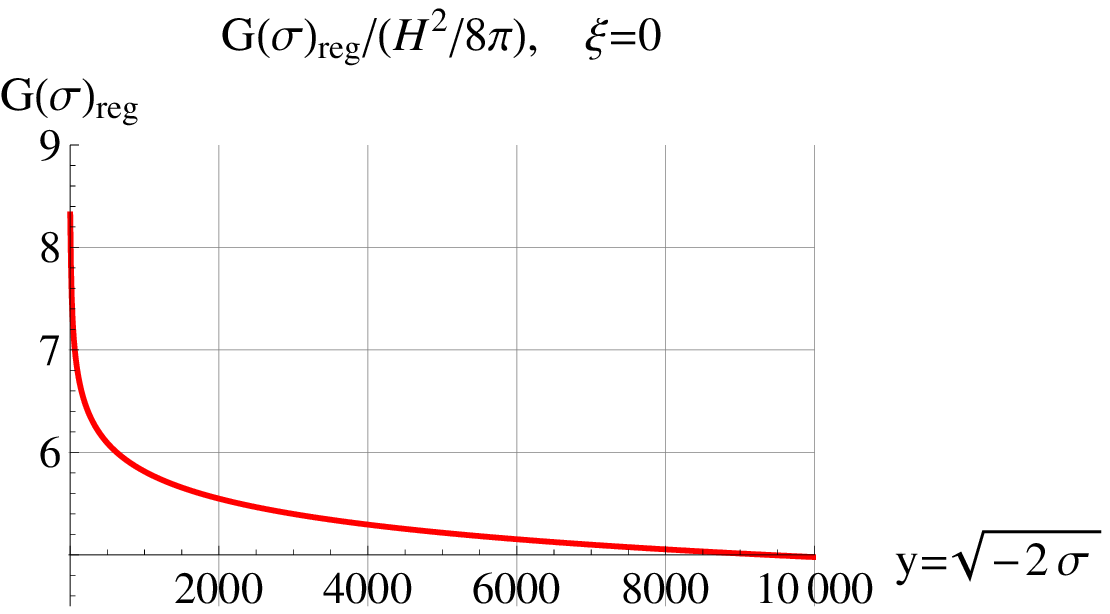}
    }
\subcaptionbox{}
    {%
        \includegraphics[width = .48\linewidth]{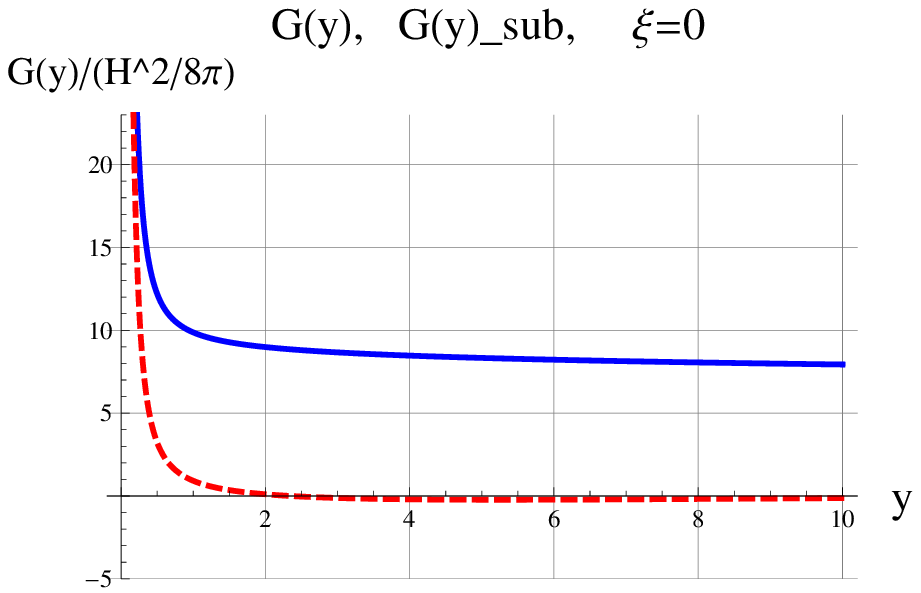}
    }
\caption{
    (a)  The model   $\xi=\frac{1}{6}$.
    Blue: the unregularized $G(y)$  in (\ref{GreeHyper}).
  Red Dashed: the subtraction term $G(y)_{sub}$ in (\ref{subGreens}).
    (b)
 The model   $\xi=\frac{1}{6}$.
  The regularized Green's function prescribed by  (\ref{regGreen}),
  suffers from the log IR divergence at $\sigma\rightarrow \infty$
 caused by the subtraction term   (\ref{subGreens}).
    (c)   The model  $\xi=0$.
    The 2nd-order regularized $G(y)_{reg }$
      given by  (\ref{regGreenf0})  is  IR- and UV-convergent.
 (d)   The model  $\xi=0$.
   Blue: the unregularized  $G(y)$;
   Red Dashed:  the subtraction in Eq.(\ref{sungrxi0}).
    }
     \label{Figure9}
\end{figure}

The 2-point Green's function and the power spectrum are
the Fourier transformation  of each other,
and they contain the same information and are complementary to each other.
In de Sitter space with the mode solution (\ref{u}),
the  unregularized  Green's function   in the  vacuum for a general $\nu$ is
\ba \label{GreenInteg}
G(x^\mu, x'\, ^\mu) & = &
   \frac{1}{(2\pi)^3} \int d^3k \, e^{i \bf k \cdot (r-r')}
           \phi_k(\tau) \phi_{k'}^* (\tau ')  \nn \\
& = & \frac{|\tau|^{1/2} |\tau'|^{1/2} }{8 \pi  a(\tau)a(\tau')}
  \frac{1}{|r-r'|} \int_0^\infty  dk  k  \sin( k|r-r'|)
  H^{(1)}_{\nu} (k\tau ) H^{(2)}_{\nu} (k\tau') .
\ea
The  integration (\ref{GreenInteg})
can be carried out \cite{BunchDavies1978,Watson1958},
and the result can be written in terms of a hypergeometric function
as the following
\be\label{GreeHyper}
G(\sigma)= \frac{H^2}{16 \pi^2}
\Gamma \big( \frac{3}{2}-\nu \big) \Gamma \big( \nu +\frac{3}{2} \big)
 \, _2 F_1  \left[\frac{3}{2}+\nu ,\frac{3}{2}-\nu ,2, ~  1 + \frac{\sigma}{2}
    \right]
\ee
where  $\sigma=  [(\tau-\tau')^2-(r-r')^2 ]/(2 \tau \tau')$.
(We have checked the result (\ref{GreeHyper}) by integration,
and noticed that there should be a factor 2 in Eq.(3.10) in Ref.\cite{BunchDavies1978}.
See also Refs.\cite{CandelasRaine1975,DowkerCritchley1976} for another derivation.)
We plot $G(\sigma )$  of a massive field in the solid line in Figure \ref{Figure9}(a),
which is  IR convergent but UV divergent.
These asymptotic behaviors can be demonstrated analytically as follows.
At large $\sigma$
the  series expansion of (\ref{GreeHyper}) for a general $\nu$ is
\be\label{lrgsigGreen}
G(\sigma) = \frac{H^2}{16 \pi^2}\Bigg[
\sigma ^{\nu -3/2 }
 \Big(\frac{ 2^{\frac{3}{2}-\nu }\Gamma (\frac{3}{2}-\nu )
  \Gamma (2 \nu)}{\Gamma (\nu +\frac{1}{2} )} \Big)
  +\sigma^{-\nu -3/2}
  \Big( \frac{ 2^{\nu +\frac{3}{2}}
  \Gamma (\frac{3}{2}+\nu ) \Gamma (-2 \nu )}{\Gamma (\frac{1}{2}-\nu )} \Big)
   +... \Bigg]
\ee
which is IR convergent for general cases   $\nu <3/2$.
At small $\sigma $ the series expansion    is
\ba\label{GreenExpdsm}
G(\sigma)& = & \frac{H^2}{16 \pi^2}
\Bigg[- \frac{2}{\sigma} +  \ln  \sigma   (\frac{1}{4}-\nu ^2)
       + (\frac{1}{4}-\nu ^2)
        \Big(\psi ^{(0)} (\frac{3}{2}+\nu) +\psi ^{(0)} (\frac{3}{2}-\nu)
           +i\pi  +2 \gamma -1-\ln 2 \Big)
           \nn  \\
&&  +    \frac{\sigma }{4} (\frac{1}{4}-\nu ^2) ( \nu ^2- \frac94 )
  \Big( \ln \sigma + \psi ^{(0)}(\frac{5}{2}+\nu )+ \psi ^{(0)} (\frac{5}{2}-\nu)
     +i\pi +2 \gamma -\frac52-\ln 2 \Big)   +...\Bigg] \nn \\
\ea
where   $1/\sigma$ and
  $\ln  \sigma$ are  divergent.
The  UV divergences  are to be removed by  regularization.

However, the  problem  is what part of $G(\sigma)$
should be subtracted off.
In literature, one is commonly  guided by
the series expansion $G(\sigma)$ of (\ref{GreenExpdsm}).
But this guide may not insure to work out.
Let us compare (\ref{GreenExpdsm})    with
the series expansion of power spectrum $\Delta^2_{k}$
of (\ref{Deltxi0}) (\ref{vps16}) at high $k$.
Generally,  a power of $\sigma$ in (\ref{GreenExpdsm})
is not in a one-to-one correspondence
to a power of $\omega$ in  $\Delta^2_{k}$.
For instances,
the divergent term $1/\omega$ can generally contribute to
the two divergent terms $-\frac{2}{\sigma}$,
 $\ln \sigma (\frac{1}{4}-\nu^2)$, and many other terms.
On the other hand, the constant term in  (\ref{GreenExpdsm})
can receive contributions from all the convergent
terms $\omega^{-5}$, $\omega^{-7}$, etc,
that is, by terms of all adiabatic orders.
The conventional  thinking is to subtract
 $1/\sigma$, $\ln  \sigma$ and some finite terms,
but this simple subtraction
will   cause some difficulty to
the method of regularization of the Green's function as we shall show soon.

Only in two special cases of the massless scalar field,
we  successfully   carry out the direct regularization of Green's function
in position space as the following.
First, for  $m=0$ and $\xi=\frac16$,
the Green's function  (\ref{GreeHyper}) reduces to
\be\label{m0xi16Green}
G(\sigma)= - \frac{H^2}{16 \pi^2}  \frac{2}{\sigma}   .
\ee
We have also double checked this  result
by direct integration of (\ref{GreenInteg}).
(\ref{m0xi16Green})  has only one term,  and
corresponds to the unregularized power spectrum (\ref{m0Xi16PS})
that also  has only one term.
This is like the Green's function
of a massless scalar field in the Minkowski spacetime.
For the   equal-time $\tau= \tau'$ case
 (\ref{m0xi16Green})   reduces to
\ba
G(\sigma)
    =   \frac{H^2}{4 \pi^2}  \frac{\tau^2}{|r-r'|^2 }.
\ea
Since (\ref{m0xi16Green}) has only one UV divergent term,
the regularized   Green's  function is simply given by
\be\label{regGreenm0Xi16}
G(\sigma )_{reg}  =  G(\sigma)  - G(\sigma)_{sub}  =0
\ee
where the subtraction term
\be\label{subGrm0xi16}
G(\sigma)_{sub} = - \frac{H^2}{16 \pi^2}\frac{2}{\sigma} .
\ee
And the regularized  auto-correlation function is also zero,
$G(0)_{reg}  = \lim_{\sigma\rightarrow 0} G(\sigma)_{reg}=0$,
here  the coincidence limit $\sigma\rightarrow 0$
is taken after the subtraction.
The zero regularized  Green's  function  (\ref{regGreenm0Xi16})
and the zero regularized power spectrum (\ref{ps16m0})
agree    with each other,
as they are the Fourier transformation of each other.
The vanishing result  (\ref{regGreenm0Xi16})
subsequently yields,  by the relation (\ref{regTG}),  a zero regularized trace,
\be\label{trgre}
 \langle T^{\beta }\, _{\beta }  \rangle_{reg}  = m^2 G(0)_{reg} =0\times 0 =0 ,
\ee
and a zero regularized stress tenor,
\be \label{tmunugre}
\langle T_{\mu\nu}  \rangle_{reg} =
 \frac{1}{4}g_{\mu\nu}  \langle T^{\beta }\, _{\beta }  \rangle_{reg} =0 ,
\ee
by the maximal symmetry relation (\ref{mtrunmxi16}).
The results (\ref{trgre}) (\ref{tmunugre}) are consistent
with (\ref{zerostress})  (\ref{0tr})
that has been derived in the $k$-space by adiabatic regularization in Sect.\ref{section4}.
Hence,  for the conformally-coupling massless scalar field,
the regularization of Green's function in the position space,
the 0th-order adiabatic regularization, and the massless limit of massive field,
all these three procedures yield the same result.

Next, for   $m=0$ and $\xi=0$,
the factor $\Gamma \big( \frac{3}{2}-\nu \big)$
of expression  (\ref{GreenExpdsm}) would give a formal infinity,
as its mathematical form is not well presented for $\nu=\frac32$.
We can  directly  integrate  Eq.(\ref{GreenInteg})
to obtain the Green's function
\bl
G(x^\mu, x'\, ^\mu)
&=\frac{1}{(2\pi)^3  } \frac{1}{  a(\tau)a(\tau')}
  \int \frac{1}{k}d^3k \,  e^{i {\bf k\cdot(r-r')}-ik(\tau-\tau')}
   \frac12 \bigg(1+i (\frac{1}{\tau'}-\frac{1}{\tau})\frac{1}{k}
  + \frac{1}{k^2}\frac{1}{\tau\tau' } \bigg) \nn \\
&  = \frac{ 1}{8\pi^2 a(\tau)a(\tau') |\tau\tau'|  }
           \big( - \frac{1}{\sigma} - \ln  \sigma  \big)
\el
which is also valid for the coupling massless scalar field
with $\nu=\frac32$   in a general RW spacetime.
In de Sitter space, it is
\be
   G(\sigma) = \frac{H^2}{8\pi^2 } \big( - \frac{1}{\sigma} - \ln \sigma \big)  .
\ee
This Green's function has two terms just
corresponding to the power spectrum (\ref{psxi16m0}).
The first term is the same as (\ref{m0xi16Green}) of $\xi=\frac16$ case,
and the second term is the  log term.
Both terms are UV divergent at $\sigma \rightarrow 0$,
and, moreover, the  log term
is also IR divergent at $\sigma \rightarrow \infty$,
a behavior  just anticipated  from  the power spectrum  (\ref{psxi16m0}).
Therefore, the appropriate regularization is to remove both terms,
and the subtraction term is taken to be
\be
   G(\sigma)_{sub} = \frac{H^2}{8\pi^2 } \big( - \frac{1}{\sigma} - \ln \sigma \big)  ,
\ee
so that
\be\label{xi0reps}
G(\sigma)_{reg }= G(\sigma)  - G(\sigma)_{sub} =0 \, .
\ee
This zero regularized Green's function
and the zero regularized power spectrum (\ref{PSxi0m0})
agree with each other.
The minimally-coupling massless field is a special case
in which the $\ln \sigma$ term is removed properly.
For the massive scalar field, nevertheless,
the $\ln\sigma$ term appearing in the series expansion  (\ref{GreenExpdsm})
can not be simply subtracted off,
for doing this will cause IR divergence,
as we demonstrate in the following.

For general  $m$ and $\xi$,
we simply do not know what divergent terms to be subtracted
off the Green's function in position space
even when the analytical $G(\sigma)$ is given.
In literature,
looking at  the series expansion  (\ref{GreenExpdsm}),
the following regularization was often  proposed
\cite{CandelasRaine1975,DowkerCritchley1976,
Brown1977,Christensen1976,BunchDavies1978}
\be\label{regGreen}
G(\sigma)_{reg} = G(\sigma)  -G(\sigma)_{sub}
\ee
where the subtraction term was taken to be the Hadamard form as  the following
\ba\label{subGreens}
G(\sigma)_{sub} & = & \frac{H^2}{16 \pi^2}\Bigg[
   - \frac{2}{\sigma} +  \ln  \sigma (\frac{1}{4}-\nu ^2)
    \nn \\
&& +(\frac{1}{4}-\nu ^2)
  \Big(\psi ^{(0)} (\frac{3}{2}-\nu  )  +\psi ^{(0)} (\frac{3}{2}+\nu )
    +2 \gamma -1-\ln  2 \Big) \Bigg] .
\ea
(As we shall see later,  this  subtraction term is actually not correct,
and the correct subtraction terms is
(\ref{sungrxi0})  for $\xi=0$,  and (\ref{grsbxi16}) for $\xi=\frac16$,  respectively.)
By appearance,
$G(\sigma)_{sub}$ contains the same UV divergence
as the series expansion  of $G(\sigma)$ at small $\sigma$,
including a $\ln\sigma $ term which is both UV and IR divergent.
The last term of (\ref{subGreens}) is  constant  and   finite
(it contains the 4th-order adiabatic terms.)
To examine  this conventional  prescription,
we plot  $G(\sigma)_{sub}$ and $G(\sigma)$ together
in Figure \ref{Figure9}(a) for a comparison.
$G(\sigma)$ itself is actually IR convergent  at $\sigma\rightarrow \infty$,
and does not have the log IR divergence.
To show the difficulty,
we plot the regularized  $G(\sigma)_{reg}$ in Figure \ref{Figure9}(b),
which becomes, nevertheless, IR divergent at large $\sigma$
due to the $\ln \sigma$ term  in $G(\sigma)_{sub}$.
Unfortunately, this log IR divergence was neglected
and unaddressed  in literature
 \cite{CandelasRaine1975,DowkerCritchley1976,Brown1977,
Hawking1977,Christensen1976,Christensen1978,BunchDavies1978}.
In fact,
as long as one takes the subtraction term in a form as   Eq.(\ref{subGreens}),
one inevitably faces  this $\ln \sigma$ difficulty
 for general $m$ and $\xi$.

The difficulty of regularization of  the Green's function in position space
 \cite{CandelasRaine1975,DowkerCritchley1976,
 Christensen1976,Hawking1977,BunchDavies1978}
is overcome  by adiabatic regularization
when the latter provides a positive, UV-  and IR-convergent power spectrum.
For de Sitter space,
given a regularized power spectrum,
such as (\ref{ParkerPS}) (\ref{PSxi160}),
we can perform Fourier transformation
and convert it into a corresponding regularized Green's function,
\ba\label{Grnregdf}
G({\bf r}, {\bf r}')_{reg}
 =   \frac{1}{|r-r'|} \int_0^\infty  \frac{\sin( k|r-r'|)}{k^2}
          \Delta^2_k (\tau)_{reg} \, d k
\ea
where the equal-time  case  $|\tau|=|\tau'|=1$ is considered
for illustration.
The definition (\ref{Grnregdf}) is analogous
to the definition  (\ref{Greenunregdf}).

First,   for $\xi=0$,
we substitute the 2nd-order regularized
$\Delta^2_{k\, \, reg}$ of (\ref{ParkerPS})
into   (\ref{Grnregdf}) and have   the following
\ba\label{Greenreg2}
G(y)_{reg}
& = &  \frac{H^2}{8\pi} \frac{1}{y} \int_0^\infty d k
  \,  k \sin( k y) \Big( |H^{(1)}_\nu(k)|^2 -\frac{4}{\pi}\frac{1}{2 W^{(2)}} \Big)
  \nn \\
& = &  G(y)  -G(y)_{sub}
\ea
where  $G(y)$ with $y \equiv  |r-r'|$ is given by (\ref{GreeHyper}),
and  $G(y)_{sub}$ is the subtraction term
and  can be calculated as the following.
By   (\ref{w2t1}),  it involves  the following integrations
\ba
 \int_0^\infty d k
  \,  k \sin( k y)  \frac{1}{ W^{(2)}}
=  \int_0^\infty d k
  \,  k \sin( k y) \Big[    \frac{1}{\bar \omega}
+\frac{1}{\bar \omega^3 }
+\frac{3 \frac{m^2}{H^2}}{4 \bar \omega ^5 }
-\frac{5 \frac{m^4}{H^4}}{  8 \bar \omega ^7 }\Big] .
\ea
By the following integration formulae
\be
\int_0^\infty d k   \,  \frac{ k \sin( k y) }{ (k^2 +\frac{m^2}{H^2})^{n+\frac12}}
= \frac{- \sqrt \pi}{2^n (\frac{m^2}{H^2})^{n/2} \Gamma(n+\frac12)}
    \,  \frac{d}{d y} \Big( y^n K_n(y \frac{m}{H} ) \Big )
\ee
where $K_n$ is the modified Bessel's function
(see 3.773 in Ref.\cite{GradshteynRyzhik1980}),
and by the relations:
$\frac{d}{dx} K_0(x) = -K_1(x)$,
$\frac1x \frac{d}{dx}  (x K_1(x)) =  -  K_0(x)$,
$\frac1x \frac{d}{dx}  (x^2 K_2(x)) = - x K_1(x)$,
$\frac1x \frac{d}{dx}  (x^3 K_3(x)) = - x^2 K_2(x)$ \cite{NISTHandbook2010},
the subtraction term for the Green's function for $\xi=0$ is given by
\bl \label{sungrxi0}
G(y)_{sub}
 = & \frac{H^2}{4\pi^2} \Bigg[
        \frac{m}{H} \, \frac{1}{y} K_1 \Big(  \frac{m}{H} y \Big)
     +  K_0 \Big(  \frac{m}{H} y \Big)  + \frac{1}{4}
    \frac{m}{H} \, y  K_1 \Big(  \frac{m}{H} y \Big)
     - \frac{1}{24}  \frac{m^2}{H^2}
       y^2  K_2 \Big(  \frac{m}{H} y \Big) \Bigg] .
\el
Thus,  we obtain the 2nd-order  regularized analytical Green's function
for $\xi=0$ as the following
\bl\label{regGreenf0}
G(y)_{reg}
= & \frac{H^2}{16 \pi^2}
\Gamma \big( \frac{3}{2}-\nu \big) \Gamma \big( \nu +\frac{3}{2} \big)
 \, _2 F_1  \left[\frac{3}{2}+\nu ,\frac{3}{2}-\nu ,2,
     ~  1 - \frac{ y^2}{4} \right] \nn \\
 & -\frac{H^2}{4\pi^2} \Bigg[
        \frac{m}{H} \, \frac{1}{y} K_1 \Big(  \frac{m}{H} y \Big)
     +  K_0 \Big(  \frac{m}{H} y \Big)  + \frac{1}{4}
    \frac{m}{H} \, y  K_1 \Big(  \frac{m}{H} y \Big)
     - \frac{1}{ 24}  \frac{m^2}{H^2}
       y^2  K_2 \Big(  \frac{m}{H} y \Big)
\Bigg] \, .
\el
We plot the resulting $G(y)_{reg}$   in Figure \ref{Figure9}(c),
which exhibits more clearly the UV finiteness and the IR convergence.
The finite value at $y=0$ is
\be
G(0)_{reg } \simeq  8.369 \frac{H^2}{8\pi}
\ee
for the model $\frac{m^2}{H^2}=0.1$ and $\xi=0$.
For a better understanding of this subtraction, in Figure \ref{Figure9}(d)
we  plot  the subtraction $G(y)_{sub}$ given by (\ref{sungrxi0})
together with the unregularized $G(y)$.
It is seen that $G(y)_{sub}$ approaches to $G(y)$ from below
at small $y$,
and thus removes the UV divergences of $G(y)$.
At large $y$,
$G(y)_{sub}$ is lower and approaches to zero faster than  $G(y)$,
and therefore does not affect the IR convergence of $G(y)$.
This is a desired pattern,
in sharp contrast to the conventionally prescribed  subtraction (\ref{subGreens})
shown in Figure \ref{Figure9}(a) by the  Green's function in position space.

\begin{figure}[htb]
\centering
\subcaptionbox{}
    {%
        \includegraphics[width = .48\linewidth]{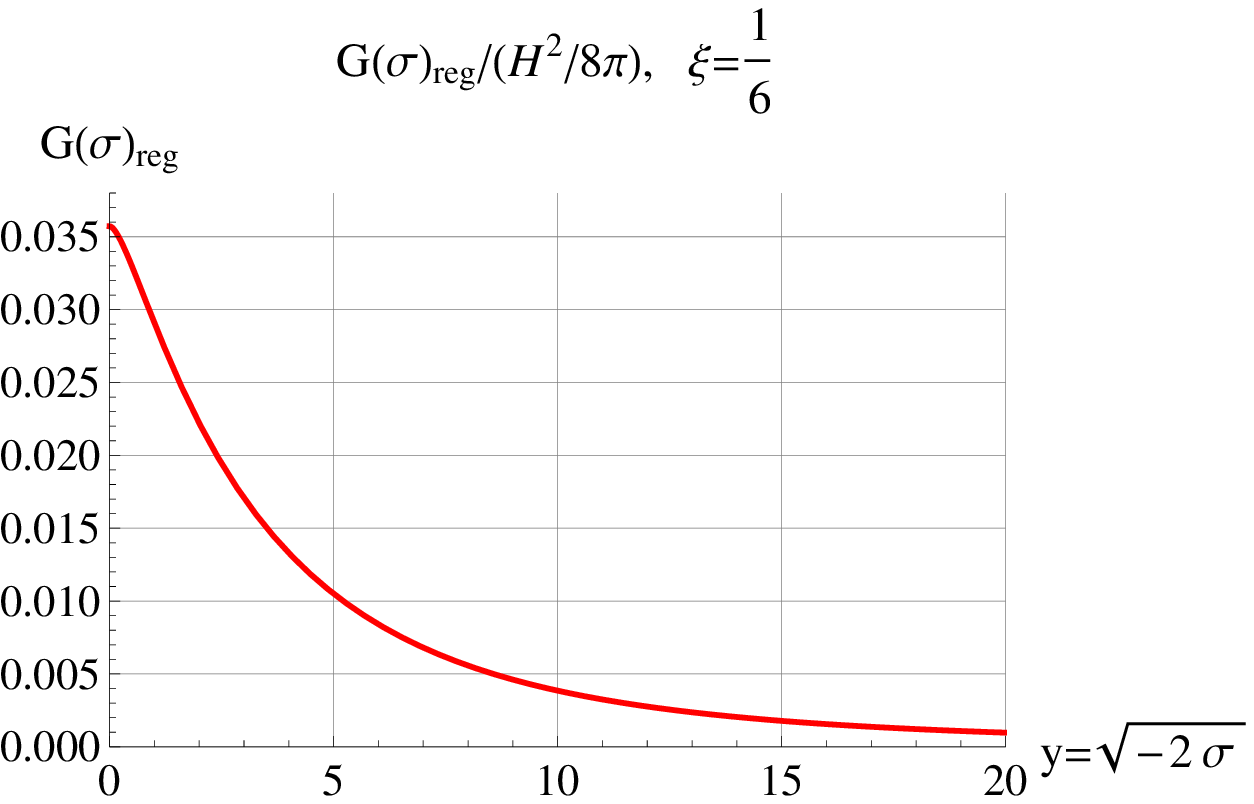}
}
\subcaptionbox{}
    {%
        \includegraphics[width = .48\linewidth]{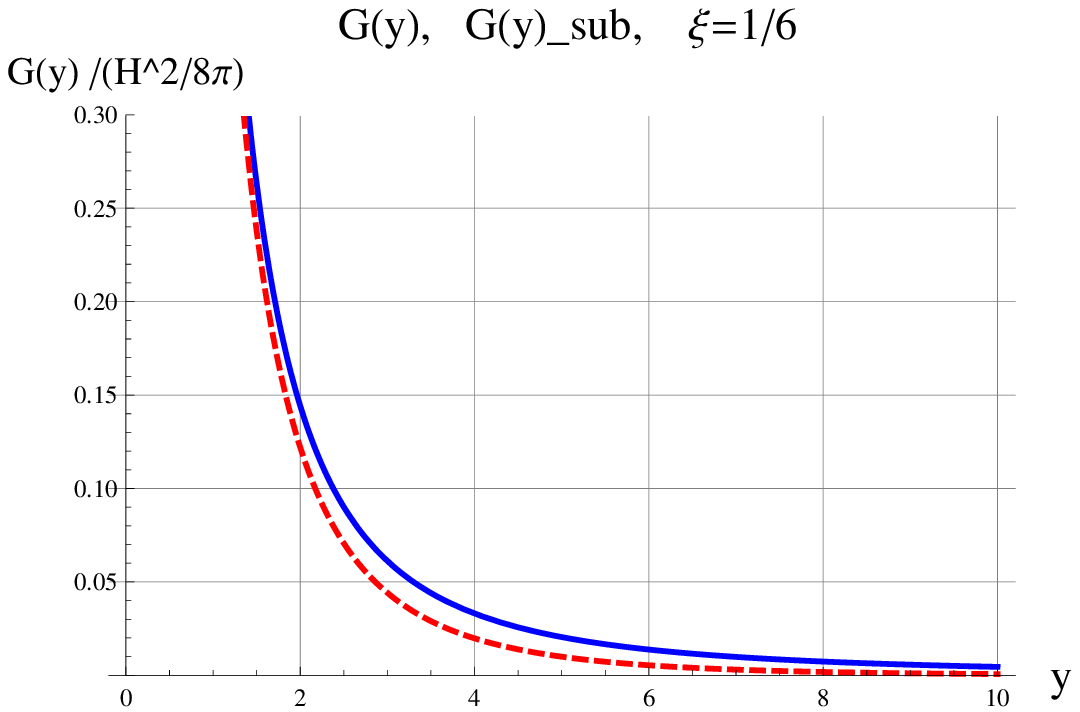}
   }
\caption{
    The model  $\xi=\frac16$.
    (a)    The 0th-order  regularized $G(y)_{reg}$
         in Eq.(\ref{Greenreg16}) is IR- and UV-convergent.
    (b)  Blue: the unregularized  $G(y)$.
         Red Dashed:  the subtraction term $G(y)_{sub}$ in Eq.(\ref{grsbxi16}).
    }
     \label{Figure10}
\end{figure}

Next,   for $\xi=\frac16$,
we substitute  the 0th-order regularized $\Delta^2_{k\, \, reg}$ of  (\ref{PSxi160})
into   (\ref{Grnregdf}) and have the following
\ba \label{Greentrasfreg16}
G(y)_{reg}
& = & \frac{H^2}{8\pi} \frac{1}{y} \int_0^\infty d k
  \,  k \sin( k y) \Big( |H^{(1)}_\nu(k)|^2
   -\frac{4}{\pi}\frac{1}{2 \omega } \Big) \nn \\
& = &  G(y)  -G(y)_{sub}
\ea
where  $G(y)$  is given by (\ref{GreeHyper})
and the subtraction term is
\be\label{grsbxi16}
G(y)_{sub}  = \frac{H^2}{4\pi^2}
  \frac{m}{H} \, \frac{1}{y}   K_1\Big( \frac{m}{H} y \Big) .
\ee
(Ref.\cite{Birrell1978} adopted the 4th-order regularization,
used an approximate subtraction term  valid only at small $y$,
and derived  the asymptotic expression of
$G(y)_{sub}$ to the order of $y^2$.)
The 0th-order regularized analytical  Green's function
for $\xi=\frac16 $ as the following
\bl \label{Greenreg16}
G(y)_{reg}
 = & \frac{H^2}{16 \pi^2}
\Gamma \big( \frac{3}{2}-\nu \big) \Gamma \big( \nu +\frac{3}{2} \big)
 \, _2 F_1  \left[\frac{3}{2}+\nu ,\frac{3}{2}-\nu ,2,
     ~  1 - \frac{ y^2}{4} \right] \nn \\
 & -\frac{H^2}{4\pi^2}
  \frac{m}{H} \, \frac{1}{y}   K_1\Big( \frac{m}{H} y \Big)  \, .
\el
The result is plotted  in Figure \ref{Figure10}(a),
which shows the UV finiteness  and IR convergence.
The finite value at $y=0$ is
\be
G(0)_{reg } =  0.03572  \frac{H^2}{8\pi} ,
\ee
for the model $\frac{m^2}{H^2}=0.1$ and $\xi=\frac16$.
This, together with
the relation $\langle T^{\mu} \,_{\mu }\rangle_{reg }= m^2 G(0)_{reg}$
of (\ref{regTG}),
gives a value of regularized trace
$\langle T^{\mu} \,_{\mu }\rangle_{reg }$
which agrees with that given by (\ref{trv7}).
In Figure \ref{Figure10}(b)
we also plot  the subtraction $G(y)_{sub}$   given by (\ref{grsbxi16})
and the unregularized $G(y)$.
When $m=0$,  (\ref{regTG}) yields a zero trace
$\langle T^{\beta} \,_{\beta}\rangle_{  reg} =0$,
which again agrees with  (\ref{zerostress}) (\ref{0tr})
derived from
the 0th-order adiabatic regularization method in Sect \ref{section4}.

Hence,  for the minimally- and conformally-coupling massive scalar fields
 in de Sitter space,
via the Fourier transformation of regularized power spectra,
we have achieved  the regularized, analytical Green's functions in position space,
which are  UV as well as IR convergent.
We expect that, in general,
when a regularized power spectrum is available,
the Fourier transformation will
provide a viable procedure  to obtain
the corresponding regularized Green's function.

The subtraction terms (\ref{sungrxi0})  (\ref{grsbxi16})
are formed from the modified Bessel's functions
which are  a special combination of
an infinite number of powers of $y$ when expanded at small $y$.
All these terms as a whole are necessary
to ensure the UV convergence and as well as the IR convergence,
also to ensure the covariance of regularized Green's function,
and the covariant  conservation of regularized stress tensor.
This inspection explains why the conventional prescription
based on the Green's function in position space  will not work in general,
because one does not know what the proper substraction term should be exactly.
The conventional prescription is guided by
the series expansion at small $y$,
and proposes  to subtract only the several divergent terms in the  series,
such as $1/y$ and $\ln y$.
And that does not work out.

In the remaining of this section
we examine the issue of trace anomaly
that has been claimed in literature.
The  trace anomaly of the scalar field with $\xi=\frac16$
was first mentioned in Ref.\cite{DowkerCritchley1976}
using the Green's function for a massive scalar field in the limit $m=0$.
In the following we go into details
and investigate its occurrence in the regularization of the Green's function
\cite{DowkerCritchley1976,Christensen1976,Christensen1978,
Brown1977,Hawking1977,Wald1978,BunchDavies1978,
Birrell1978,BunchChristensenFulling1978,Bunch1979,Bunch1980}.
In particular, we shall see how it was brought about by an invalid procedure
of taking massless limit of the 4th-order adiabatic subtraction term.

To  get the trace anomaly  in de Sitter space,
Ref. \cite{BunchDavies1978}
started with  the incorrect subtraction term (\ref{subGreens}),
and expanded its the constant part at large  $m^2/H^2$,
\ba \label{exppsi}
(\frac{1}{4}-\nu ^2)
  \Big(\psi ^{(0)} (\frac{3}{2}-\nu  ) +\psi ^{(0)} (\frac{3}{2}+\nu ) \Big)
   \simeq
   -\frac{H^2}{15 m^2} -\frac{8 H^4}{315 m^4}
  -\frac{ m^2 \ln  \frac{H^2}{m^2}  }{H^2}+\frac{2}{3}
    +... \,  ,
\ea
plugged   (\ref{subGreens})   into (\ref{regGreen})
to  cancel the UV divergent $1/\sigma$ and $\ln  \sigma$ terms,
\bl \label{grex}
G(\sigma)_{reg} = &
\frac{H^2}{16 \pi^2}
 \Gamma \big( \frac{3}{2}-\nu \big) \Gamma \big( \nu +\frac{3}{2} \big)
 \, _2 F_1  \left[\frac{3}{2}+\nu ,\frac{3}{2}-\nu ,2, ~  1 + \frac{\sigma}{2}
    \right] \nn \\
&     -    \frac{H^2}{16 \pi^2}\Big[ - \frac{2}{\sigma} +  \ln  \sigma (\frac{1}{4}-\nu ^2)
     -   \frac{H^2}{15 m^2} + ... \Big]  ,
\el
and, after taking the coincidence limit $\sigma \rightarrow 0$,
got  the regularized stress tensor,
keeping up to the order of $H^4$ (the adiabatic 4th-order),
\be\label{rtace}
\langle T^\mu\, _\mu \rangle_{reg}
 =  m^2 G(\sigma=0)_{reg}
 = m^2  \frac{H^2}{16\pi^2}
   \Big[\frac{H^2}{15 m^2}  +... \Big] ,
\ee
then  took  the limit $m=0$ and claimed  that
the conformally-coupling massless scalar field
has the following nonzero trace (i.e., the trace anomaly) \cite{BunchDavies1978}
\be\label{trvapp}
\lim_{m=0} \langle T^\mu\, _\mu \rangle_{reg}
    = \frac{H^4}{240\pi^2} = \frac{R^2}{34560 \pi^2}.
\ee
Thus the appearance of trace anomaly is caused by
two procedures:
starting with a presumed massive field for a massless field,
and deploying the 4th-order subtraction term.
These have been  commonly adopted in literature which claimed  the trace anomaly.
Inspection of the  above procedures    reveals several problems.
First, if one works directly with a massless scalar field with $\xi=\frac16$,
instead of starting with an artificially  presumed massive field,
 one simply obtains a vanishing $G(\sigma)_{reg}=0$,
and will never run into a trace anomaly,
as demonstrated by the paragraph of (\ref{m0xi16Green}) through (\ref{tmunugre}).
Second,
the subtraction term  (\ref{subGreens}) is incorrect for a massive field,
as it contains the $\ln\sigma$ term
and will generate   the $\ln$ IR divergence,
as we have pointed   earlier.
Third,
 even if the subtraction term  (\ref{subGreens}) were  allowed,
it will generally lead to negative power spectrum and  negative spectral energy density,
which are not acceptable for physics.
Consequently,   for  general $\xi$,
  regularization of Green's function in position space
has not been accomplished yet,
in the sense that the methods have not been able to
yield an explicit expression of UV and IR convergent Green's function $G(\sigma)$
even in  de Sitter space.
In fact, the adequate  subtraction terms of the Green's function
are   given by (\ref{sungrxi0}) for $\xi=0$ ,
and respectively  by (\ref{grsbxi16}) for $\xi=\frac16$,
which have been derived by Fourier transformation
 of the adiabatic subtraction terms for power spectra
 from adiabatic regularization.

For a general curved spacetime,
the analytical solution of Green's function may be not available.
Often in literature,
 the Green's function is written formally as
the DeWitt-Schwinger proper-time integration \cite{Schwinger1951,DeWitt1975}
\bl \label{SchDe}
G(\sigma)  \propto  \sum_{n=0}^{\infty} a_n
 \Big( -\frac{\partial}{\partial m^2} \Big)^n
 \int _0^\infty \frac{d s }{s^2}
 \exp{\Big[- i ( m^2 s -\frac{\sigma}{2s}) \Big]} .
\el
As already pointed in
Refs. \cite{DeWitt1975,Christensen1976,Christensen1978,AdlerLiebermanNg1977},
the $s$-integration  (\ref{SchDe}) is undefined at $m=0$
and is valid only at $m\ne0$.
So, the series expansion of $G(\sigma)$ at $m=0$ is not known
in this formulation.
This is the limitation of  DeWitt-Schwinger proper-time integration,
and that is why   in literature  strangely  a massive field was started with
to get the trace anomaly of a massless field.
(By the way, we like to say  that the DeWitt-Schwinger integration
is not necessary to study the Green's function,
and  that one can study the Green's function of a massless field
in terms of  a Hadamard series.)
At $m\ne 0$,  (\ref{SchDe}) can be expanded as a series
in the following form \cite{DeWitt1975,DowkerCritchley1976,Christensen1976}
\ba \label{GreenChrist}
G(\sigma) & \propto &  \frac{1}{4\pi^2}
\Big\{  a_0 \Big[ \frac{1}{\sigma}
 +m^2 (\gamma+ \frac12 \ln |\frac12 m^2\sigma  |)(1+\frac14 m^2\sigma+...)
  -\frac12 m^2 ...\Big] \nn \\
&&  - a_1 \Big[ (\gamma+ \frac12\ln |\frac12 m^2 \sigma|)(1+\frac12 m^2\sigma+...)
       -\frac12 m^2 \sigma  ...\Big] \nn \\
&&    +a_2 \sigma \Big[ (\gamma+ \frac12\ln |\frac12 m^2 \sigma|)(1+\frac18 m^2\sigma+...)
         -\frac14 - ...\Big] \nn \\
&&    +\frac{1}{2 m^2} \big[a_2 + O(\sigma)...  \big]
    +\frac{1}{2 m^4} \big[a_3  + O(\sigma) ...  \big] + O(m^{-6}) \Big\}
\ea
which has a structure similar to (\ref{GreenExpdsm}) in de Sitter space.
(We must emphasize that this expression is invalid at $m=0$.)
Then it was  commonly proposed to subtract off
some UV divergent terms of (\ref{GreenChrist}),
such as  $a_0/\sigma$, $a_0\frac12 m^2  \ln \sigma$,  $a_2/2 m^2$,  etc.
(These subtraction terms have been commonly adopted
in the methods of regularization in position space,
such  as the dimensional regularization
   \cite{DowkerCritchley1976,Brown1977},
the point-splitting   \cite{Christensen1976,Christensen1978,BunchDavies1978},
the zeta function regularization
   \cite{DowkerCritchley1976,Hawking1977},
and  as well as the path-integral formulation  \cite{Fujikawa1980}.
The zeta function regularization
in Refs. \cite{DowkerCritchley1976,Hawking1977}
and its  path-integral formulation in Ref.  \cite{Fujikawa1980}  are equivalent,
and use the same subtraction terms which are explicitly mentioned
in Refs. \cite{DowkerCritchley1976,Hawking1977}.)
Nevertheless, doing this will  lead to  an  IR divergence
since the term  $a_0 \frac12 m^2 \ln \sigma$ is also IR divergent,
just like in de Sitter space.
This is a vital difficulty of regularization
guided by the series expansion (\ref{GreenChrist}) in position space,
and has not been addressed in literature.
Without solving this difficulty,
one still continued to pick up the subtraction term $a_2/2m^2$ (4th adiabatic order),
multiplied by $m^2$, took the limit $m=0$, and claimed that the trace anomaly is
$\langle T^\mu\, _\mu   \rangle   = \frac{1}{4\pi^2} a_2 $
\cite{DowkerCritchley1976,Christensen1976,Brown1977,Hawking1977,BunchDavies1978}.
But again, this limit-taking procedure is invalid
simply because the series (\ref{GreenChrist}) does  not hold at $m= 0$.
We emphasize that it is mathematically incorrect to pick up a term in (\ref{GreenChrist})
and regard it as a term of $G(\sigma)$ at $m=0$.
For instance, the term $\frac{1}{2 m^4}a_3$
and the subsequent infinitely many terms ($ \propto m^{-6}$, $ m^{-8}$, etc)
will be  divergent at $m=0$.
The series expansion (\ref{GreenChrist})
does not tell what $G(\sigma)$  looks like  at $m=0$,
nor if the $a_2$ term ever exists in  $G(\sigma)$ at $m=0$.
Thus, it is invalid to claim that $a_2$ is the trace anomaly at $m=0$.
The real issue is to find adequate subtraction terms
to the Green's function,
and this has not been solved yet for general curved spacetimes.

\section{Conclusion and discussions}\label{section7}

We have studied, in great details,
adiabatic regularization of a coupling  massive scalar field
  in de Sitter inflation,
and  found that the conventional 4th-order regularization
subtracts off more than necessary and
leads to a negative spectral energy density in a broad range of $k$.
A negative spectral energy density
is unacceptable in the context of cosmology.
For a massive scalar field,
the adiabatic expansion is made in  powers of frequency $\omega$,
whereas the degree of UV divergences is counted in powers of $k$.
When the powers of $\omega$ is further expanded in terms of $k$,
there are infinitely many  powers $k^{-n}$.
This is the origin of unmatch of the adiabatic order
and the degree of UV divergences for a massive scalar field.
So we seek an appropriate prescription
which will give a positive (nonnegative),  IR- and UV-convergent spectral energy density.
As it turns out,
the order of regularization  depends sensitively on the coupling $\xi$.

For the massive field with   $\xi=0$ in Section \ref{section3},
we find that the 0th- and 4th-order prescriptions fail,
and  the 2nd-order regularization
is sufficient to remove quartic, quadratic, and logarithmic
UV divergences of the stress tensor,
and simultaneously
the quadratic, and logarithmic UV divergences of  the power spectrum.
The resulting regularized spectral energy density
and power spectrum are positive, IR- and UV-convergent.
Although nonuniformly distributed  in the $k$-modes,
the regularized stress tensor in the vacuum is maximally symmetric,
and respects the covariant conservation.
Thus, in connection to inflation cosmology,
the massive scalar field can play a double role:
its regularized $\langle T_{\mu\nu} \rangle_{reg}$  drives the inflation,
while its $k$-modes $\phi_k$ are the primordial fluctuations in the  vacuum.
With regard to renormalization,
the 2nd-order regularization involves
renormalization of the cosmological and gravitational constants,
and does not need the fourth order time-derivative of metric
such as $^{(1)}H_{\mu\nu}$, $^{(2)}H_{\mu\nu}$.
In the special case of $m=0=\xi$,
the 2nd-order regularization is the same as  the 4th-order one,
and gives a zero stress tensor  and a zero power spectrum.

For the  massive field with $\xi=\frac16$  in Section \ref{section4},
we find that  the 2nd- and 4th-order prescriptions fail,
and the 0th-order  regularization is sufficient to remove UV divergences
in the stress tensor and the power spectrum,
and the regularized spectra are positive,  and IR- and UV-convergent.
Like the $\xi=0$ case,  the regularized stress tensor
is maximally symmetric,  but nonuniformly distributed  in the $k$-modes.
The 0th-order regularization involves only
  renormalization of the cosmological constant.
In the special case  $\xi=\frac16$ and $m=0$,
we find that the prescriptions of 0th-, 2nd-, and 4th-order regularization
are the same,
all lead to a zero regularized stress tensor and a zero regularized power spectrum,
and no trace anomaly exists.
Even if we start with $m \ne 0$ and $\xi=\frac16$,
the regularized trace is $ \propto m^2$,
so that the  trace is still zero in the limit $m \rightarrow 0$.
Our finding is in contrast to the  literature-reported trace anomaly
which appeared in the 4th-order regularization
of a massive scalar field with $\xi=\frac16$
in  the limit $m \rightarrow 0$.
Since the 4th-order regularization leads to a negative spectral energy density,
it is an improper  prescription,
and the trace anomaly is an artifact.

For the massive   field with a general $\xi \in (0,\frac16)$
 in Section \ref{section5},
the regularized spectra generally have irregular distortions and  negative values.
It seems no regularization of equal order could achieve
positive spectral energy density and power spectrum simultaneously.
We also notice that, by the 4th-order regularization,
both the spectral energy density and power spectrum are positive at high $k$,
whereas  distortions with negative values occur
only at the low $k$  range outside the horizon.
This pattern indicates that one may modify regularization at low $k$
in order to achieve positive spectra,
like what we have done for RGW \cite{ZhangWangJCAP2018}.
Regularization for a general $\xi$ remains unsettled
and more endeavors are needed.

As a complement to the adiabatic regularization,
we have studied the regularization of Green's function in position space
in Section \ref{section6}.
DeWitt \cite{DeWitt1975} suggested that
``after the infinities have been split off,"
the effective Lagrangian  (expressed in terms of the Green's function)
``is both coordinate invariant and conformally invariant,"
however, he did not tell how to achieve this by an explicit demonstration.
In curved spacetimes,
the analytical Green's function   is generally not available,
it is hard to implement a direct regularization.
In de Sitter space,  $G(\sigma)$ for general $m$ and $\xi$
is known as a hypergeometric function,
but its direct regularization is still not easy,
because one does not know its proper subtraction term.
Only in the massless cases $m=0$ with $\xi=\frac16$ or  $\xi=0$,
 $G(\sigma)$  contains one (or two) term which is divergent,
we are able to directly  regularize it to zero,
and no trace anomaly exists.
This outcome agrees with the result of zero spectra
from the 0th- (or 2nd-) order adiabatic regularization.
In the cases $m\ne 0$ with $\xi=\frac16$ or $\xi=0$
when the regularized power spectra are available,
we perform Fourier transformation,
and obtain the regularized Green's function which is UV-finite and IR-convergent.
And, as it turns out,  the subtraction terms  $G(\sigma)_{sub}$
consist of the modified Hankel functions $K_n(\sigma)$
as in Eqs.(\ref{sungrxi0}) and (\ref{grsbxi16}),
not just the commonly-suggested Hadamard form as in  Eq.(\ref{subGreens})
which contains the  $\ln \sigma$ term.
We demonstrate explicitly
how this log IR divergence plagues the Green's function regularization,
which is an issue not been addressed in literature.
When an adequately regularized power spectrum is available,
the Fourier transformation
provides a viable way to obtain the regularized Green's function.
We also analyze, for de Sitter space,
how the literature-reported trace anomaly showed up in the Green's function method,
and actually it was generated by the improper 4th-order subtraction term.
It should be remarked that,
due to the homogeneity and isotropy in de Sitter space,
the fluctuations represented by the scalar field
is a stationary process in terminology of statistics,
so that the Green's function is a function of the separation
of two spacetime points, $G(x^\mu, x^\mu\, ')=G(x^\mu- x^\mu\, ')=G(\sigma)$,
as explicitly seen in Eq.(\ref{GreeHyper}).
In this case,  when a locally   determined part $G^L(\sigma)$
is constructed  to  satisfy
the wave equation in $x$ and in $x'$ \cite{AdlerLiebermanNg1977},
so does  the  boundary-condition-dependent part
$G^B(x^\mu, x^\mu\, ') \equiv  G(\sigma)-G^L(\sigma)$.
So the presumption that $ G^B(x^\mu, x^\mu\, ')$
not satisfy the wave equation in $x'$
 \cite{Wald1978}
does not happen in de Sitter space.
In general curved spacetimes,
the analytical solution  of $G(\sigma)$ is not known,
nor is the appropriate subtraction term,
it is  premature  to claim the trace anomaly.
In retrospect,
the common thing that led to the trace anomaly in those different approaches
consisted of  two procedures:
starting with an artificially-presumed massive field for a massless field,
and adopting the incorrect  4th-order subtraction term.
If one  directly starts with a massless scalar field,
instead of  a  massive field,
one will never run into a trace anomaly,
as we have shown
in the paragraph of (\ref{u016}) through (\ref{ps16m0})
by  adiabatic regularization method,
and in the paragraph of (\ref{m0xi16Green}) through (\ref{tmunugre})
by  the Green's function method.
Even starting with a massive field with $\xi=\frac16$,
if one adopts the 0th-order regularization, instead of the  4th-order,
one will never run into
the trace anomaly nor the negative spectral energy density.

Our study   shows the great advantage of adiabatic regularization
which works particularly  in RW spacetimes.
When the exact solution of field in a RW spacetime is given,
all UV-divergent terms are isolated as powers of $k$,
while the subtraction terms of various orders are always
given explicitly by WKB approximation.
With these in hand, one is able to see
if the spectra are UV- and IR-convergent
and if a spectrum is negative  at certain  order.
Therefore, one is more effectively
guided in choosing a proper regularization scheme.
Furthermore,
the spectra of a massive field are  functions of $\omega$,
adiabatic subtraction can be implemented without causing  the log IR divergence.
These  important features of adiabatic regularization
are not shared by the direct regularization of Green's function in position space.
Finally we remark  that
the regularized stress tensor and Green's function obtained
by adiabatic regularization in this paper are covariant,
and  the corresponding quantities in other coordinates
can be given by coordinate transformations.

\

\textbf{Acknowledgements}

Y. Zhang is supported by
NSFC Grant No. 11421303, 11675165, 11633001,  11961131007.
The authors would like to thank A. Marciano for valuable discussions.

\appendix
\numberwithin{equation}{section}

\section{  High $k$ expansions of exact modes }\label{sectionA}

In this  appendix,  we list some series expansions
following from the analytical solution in de Sitter space
which are  used in the context.

In high   $k $ limit, the  mode $v_k(\tau)$ of (\ref{u})
 approaches to
\ba \label{uinfl}
v_{k } &  \simeq &     \frac{1}{\sqrt{2k}}e^{ix }
     \Big(1 + i\frac{(4\nu^2-1)}{8x} -\frac{(16\nu^4 -40\nu^2 +9)}{128 x^2}
     -i\frac{(64\nu^6 -560\nu^4 +1036\nu^2 -225)}{3072 x^3}  \nn \\
&&¡¡+   \frac{(256\nu^8-5376\nu^6+31584 \nu^4 -51664 \nu^2 +11025 )}{98304 x^4} \nn \\
&&  - i \frac{  \left(-1024 \nu ^{10}+42240 \nu ^8-561792 \nu ^6
     +2764960 \nu ^4-4228884 \nu ^2+893025\right)}{3932160 x^5} + ...
    \Big)   \nn \\
\ea
with $x\equiv k|\tau|$,
where the first term corresponds to the positive-frequency massless mode
in Minkowski spacetime,
and other terms are due to mass and expansion effects.
Notice the pattern that
$\nu^2$ goes with $1/x$ in this high $k$ expansion.
That is,  the high $k$ expansion implies also an  expansion
at low $\frac{m^2}{H^2}$.

\ba\label{vksq}
|v_k|^2
& = &  \frac{1}{2k} \Big( 1 + \frac{4\nu^2-1}{8 x^2}
     + \frac{3(4 \nu ^2 -1) (4 \nu ^2-9)}{128 x^4}   + \frac{5 (4 \nu ^2 -1) (4 \nu ^2-9) (4 \nu ^2-25) }{1024 x^6} \nn \\
&&   + \frac{35  (4 \nu ^2-1 )(4 \nu ^2-9)(4 \nu ^2-25)(4 \nu ^2-49)}{32768 x^8}
        + ...  \Big).   \nn   \\
\ea
It can be also expressed in terms of $\omega$ for $\xi=0$ as follows
\ba \label{vkom}
|v_k|^2
& = & \frac{1}{2 \omega }
+ \frac{-4 \nu ^2 \tau ^2+4 \nu ^2+9 \tau ^2-1}{16 \tau ^2 \omega ^3}
+\frac{3 \left(4 \nu ^2-9\right) \left(4 \nu ^2 \tau ^4
   -8 \nu ^2 \tau ^2+4 \nu ^2-9 \tau ^4+2 \tau ^2-1\right)}{256 \tau ^4 \omega ^5} \nn \\
&&    -\frac{5 \left(4 \nu ^2-9\right) A_\tau}{2048 \tau ^6 \omega ^7 }
+\frac{35 \left(4 \nu ^2-9\right) B_\tau }{65536 \tau ^8 \omega ^9}
\ea
where
\ba
A_\tau & \equiv & 16 \nu ^4 \tau ^6-48 \nu ^4 \tau ^4
  +48 \nu ^4 \tau ^2-16 \nu ^4-72 \nu ^2 \tau ^6
  +120 \nu ^2 \tau ^4-120 \nu ^2 \tau ^2 \nn \\
&& +104 \nu ^2+81 \tau ^6-27 \tau ^4+27 \tau ^2-25
\ea
\ba
B_\tau & \equiv & 64 \nu ^6 \tau ^8-256 \nu ^6 \tau ^6
 +384 \nu ^6 \tau ^4-256 \nu ^6 \tau ^2+64 \nu ^6-432 \nu ^4 \tau ^8
 +1216 \nu ^4 \tau ^6 \nn \\
 && -1824 \nu ^4 \tau ^4
 +2240 \nu ^4 \tau ^2-1200 \nu ^4+972 \nu ^2 \tau ^8 -1584 \nu ^2 \tau ^6
  +2376 \nu ^2 \tau ^4 \nn \\
&& -4144 \nu ^2 \tau ^2+5196 \nu ^2-729 \tau ^8
    +324 \tau ^6-486 \tau ^4+900 \tau ^2-1225 .
\ea
In the context we shall plot all the graphs at a time $|\tau|=1$
 for illustration,
and the above is  simplified  as
\ba\label{vkxi0t1}
|v_k|^2 & = &  |\tau| \Big( \frac{1}{2 \bar \omega } +\frac{1}{2 \bar \omega ^3}
-\frac{3 \left(4 \nu ^2-9\right)}{32 \bar \omega ^5}
-\frac{5 \left(4 \nu ^2-9\right) \left(4 \nu ^2+7\right)}{256 \bar \omega ^7} \nn \\
&& +\frac{35 \left(4 \nu ^2-9\right) \left(44 \nu ^2-19\right)}{1024 \bar \omega ^9}
  + ... \Big)
\ea
with $\bar \omega \equiv
  (k^2 +\frac{m^2}{H^2})^{1/2}$ which equals
  $\omega=(k^2 +\frac{m^2}{H^2\tau^2})^{1/2}$ at $|\tau|=1$.
For $\xi=\frac16$, (\ref{vkxi0t1}) is  replaced by the following
\be\label{vexpxi16}
|v_k |^2=
|\tau| \Big( \frac{1}{2 \bar \omega}
+\frac{3 (1-4 \nu ^2 )}{32   \bar \omega ^5}
+\frac{5  (1-4 \nu ^2 ) (4 \nu ^2-25 )}{256   \bar \omega ^7}
+\frac{105 \left(4 \nu ^2-1\right) \left(20 \nu ^2-53\right)}{1024 \bar\omega ^9}
+ ... \Big).
\ee

The time   derivative  terms  to the 4th adiabatic order are  given by
{\allowdisplaybreaks
\ba\label{vkpr}
v_k' & \simeq & - \sqrt{\frac{k}{2}}e^{i x}
\Big(
  i  -\frac{4\nu^2 -1}{8x} - i\frac{16\nu^4 +24\nu^2 -7}{128 x^2}
 ¡¡+\frac{64\nu^6 + 208\nu^4  -884 \nu^2 +207}{3072 x^3} \nn \\
&& +i \frac{256\nu^8 +768\nu^6 - 22176\nu^4 +47792\nu^2 -10575}{98304 x^4} +...
\Big) ,
\ea
\ba
|v_k'|^2 & = & k \l(\frac{1}{2} - \frac{  ( 4 \nu^2 -1)}{16 x^2}
         - \frac{ (16\nu^4 - 104\nu^2 +25)}{256 x^4}
         -\frac{ (64\nu^6 -2096 \nu^4 +4876\nu^2 -1089)}{2048 x^6}
       + ... .  \r) , \nn \\
\ea
or, at  $|\tau|=1$ for  plotting,
{\allowdisplaybreaks
\ba
|v_k'|^2 =
k \left(\frac{1}{2} +\frac{1-4 \nu ^2}{16 \bar \omega ^2}
+\frac{48 \nu ^4-56 \nu ^2+11 }{256 \bar \omega ^4}
+\frac{-320 \nu ^6+2288 \nu ^4-3900 \nu ^2+837 }{2048 \bar \omega ^6} +... \right) ,\nn \\
\ea
}
{\allowdisplaybreaks
\ba
(\frac{v_k}{a})' & = &  - \frac{ H}{\sqrt {2 k}}e^{i x}
\Bigg( i x   + \frac{(9-4 \nu^4)}{8}
     -i  \frac{(16\nu^4 -40\nu^2 +9)}{128x}
    + \frac{(4\nu^2-9)(1-4\nu^2)^2}{3072 x^2}     \nn \\
&& + i \frac{(2\nu-5)(2\nu-3)(2\nu-1)(2\nu+1) (2\nu+3)(2\nu+5)(4\nu^2+15) }
         {98304 x^3}
    \nn \\
&& - \frac{(2\nu-7)(2\nu-5)(2\nu-3)(2\nu-1)(2\nu+1) (2\nu+3)(2\nu+5)(2\nu+7)(4\nu^2+39)}
{ 3932160 x^4}    + ...
\Bigg)  \nn \\
\ea
}
\ba\label{hikvpri}
|(\frac{v_k}{a})'|^2  & = &
 \frac{\pi H^{2}}{4k} \l|\frac{d}{dx} \big(x^{3/2} H^{(2)}_{\nu} (x)\big) \r|^2 \nn \\
& =& H^2 \Big( \frac{x^2}{2 k}
    + \frac{9- 4 \nu^2}{16 k}
    -\frac{ 16\nu^4 - 40 \nu^2 +9 }{256 k x^2} +\frac{441 -1996 \nu^2
    + 944\nu^4 -64\nu^6 }{2048 k x^4}  + ... \Big) .
       \nn \\
\ea
}
In low $k$  limit,
\be \label{ulow}
v_{k }  \simeq   (\frac{x}{2})^{-\nu + \frac12}
  \frac{  \Gamma(\nu)}{\sqrt{2 \pi k}}
                 e^{i \frac{\pi}{2}(\nu - \frac12) } ,
\ee
\be\label{sqvkm}
|v_k|^2 \simeq    x^{-2\nu } |\tau| \frac{ 2^{2\nu-2} \Gamma(\nu)^2 }{ \pi },
\ee
\be\label{sqvkma}
| \frac{v_k}{a}|^2 \simeq
H^2 x^{-2\nu } |\tau|^3 \frac{ 2^{2\nu-2} \Gamma(\nu)^2 }{ \pi },
\ee
and the time derivatives are
{\allowdisplaybreaks
\be
v_k' \simeq   -(\frac{x}{2})^{-\nu - \frac12}
   k^{1/2}
  \frac{  \Gamma(\nu)-2\Gamma(\nu +1)}{4 \sqrt{2 \pi }}
                 e^{i \frac{\pi}{2}(\nu - \frac12) } ,
\ee
\be
|v_k'|^2  \simeq    (\frac{x}{2})^{-2\nu - 1}
   k
  \frac{  (\Gamma(\nu)-2\Gamma(\nu +1))^2 }{32 \pi},
\ee
\be
(\frac{v_k}{a})'= - H (\frac{x}{2})^{-\nu +\frac12 } k^{-\frac12 }
     \frac{(3 \Gamma(\nu)-2\Gamma(\nu +1)) }{{2 \sqrt {2 \pi}} }  e^{i \frac{\pi}{2}(\nu - \frac12) }   ,
\ee
\ba \label{lwkvpsq}
|(\frac{v_k}{a})'|^2
& =& H^2 x^{-2\nu  } |\tau|  \,
     \frac{2^{2\nu} (3 \Gamma(\nu)-2\Gamma(\nu +1))^2}{16 \pi}.
\ea
}
In these expressions the index $\nu$ is for a general $\xi$.

\section{ The WKB modes}\label{sectionB}

The WKB  approximation \cite{Chakraborty1973,ParkerFulling1974,FullingParkerHu1974,
Bunch1980,AndersonParker1987}
to the solution of (\ref{equvk})  in a flat RW spacetime
is   written as the following
\be\label{vn}
v_k^{(n)}(\tau)
  = (2W_k(\tau))^{-1/2}   \exp \Big[  -i \int^{\tau} W_k(\tau')d\tau' \Big]
\ee
where the effective frequency is
\be\label{equWk}
\big( W_k(\tau) \big )    = \Big[  k^2 + m^2 a^2 + (\xi -\frac16)a^2R
-\frac12 \l( \frac{ W_k  '' }{ W_k}
- \frac32 \big( \frac{W_k  '}{W_k} \big)^2 \r)  \Big]^{1/2}.
\ee
The WKB solution of $W_k$ is obtained by iteratively solving (\ref{equWk})
to a desired adiabatic  order.
Take  the 0th-order effective frequency
\cite{Bunch1980},
\be
W_k^{(0)}=  \omega
          =    \sqrt{k^2+m^2 a^2 } ,
\ee
and the 0th-order adiabatic  mode
\be\label{vkWKB0}
v_k^{(0)}(\tau)
  =(2 \omega )^{-1/2}
    e^{  -i \int^{\tau} \omega d\tau'} .
\ee
The 0th-order quantities that appear in
the 0th-order subtraction terms are
\be
|v^{(0)}_k|^2   =\frac{1}{2 W_k^{(0)}} = \frac{1}{2 \omega} ,
\ee
\be
|v^{(0)}_k \, ' |^2
= \frac12
      \Big(  \frac{ (W_k^{(0)\, '})^2 }{4 (W_k^{(0)})^{3} } +  W_k^{(0)}  \Big)
\simeq  \frac{\omega}{2}  ,
\ee
\[
v_k ^{(0)}\, ' v_k^{(0)}\, ^*  + v_k^{(0)} v_k^{(0)}\, ^*\, ' \simeq 0 .
\]
The 2nd-order adiabatic  mode
\be\label{vkWKB2}
v_k^{(2)}(\tau)
  =(2W_k^{(2)}(\tau))^{-1/2}
    e^{  -i \int^{\tau} W_k^{(2)}(\tau')d\tau'} ,
\ee
The first iteration of (\ref{equWk}) is
\be \label{W2expression}
W_k^{(2)}= \l[ k^2 + m^2 a^2 + (\xi -\frac16)a^2R
-\frac12 \l( \frac{ W^{(0)}_k\,  '' }{ W^{(0)} _k }
- \frac32 \big( \frac{W^{(0)}_k\,  '}{W^{(0)}_k} \big)^2 \r) \r]^{\frac12}
  \, ,
\ee
By
\[
(\xi -\frac16)a^2R
- \frac12  \l( \frac{ W^{(0)}_k\,  '' }{ W^{(0)} _k }
   - \frac32 \big( \frac{W^{(0)}_k\,  '}{W^{(0)}_k} \big)^2 \r)
   =  6 (\xi -\frac16) \frac{a''}{a}
     -\frac{ m^2 \left(a a''+ a'^2\right)}{2 \omega ^2}
     + \frac{5 m^4 a^2 a'^2}{4 \omega ^4},
\]
keeping only up to two time-derivatives of metric,  $a''$ and $a'\, ^2$,
gives  the 2nd-order effective frequency,
\ba\label{Wk2nd}
W_k^{(2)}
& \simeq  &  \omega
 + 3  (\xi-\frac16)\frac{1}{\omega } \frac{a''}{a}
      - \frac{ m^2 (a'' a+ a'\,  ^2 )}{4 \omega^3}
      +  \frac{ 5 m^4 a'\, ^2 a^2 }{8\omega^5 }
        ,
\ea
\be\label{W2xi}
(W_k^{(2)})^{-1}
 \simeq   \frac{1}{\omega}
       - 3  (\xi-\frac16)  \frac{a''/a}{\omega^{3}}
 + \frac{ m^2 (a'' a+ a'\,  ^2 )}{4 \omega^5}
      -  \frac{ 5 m^4 a'\, ^2 a^2 }{8\omega^7 }      .
\ee
It should be observed that
$W_k^{(2)}$  actually contains four terms up to $\omega^{-5}$,
and $(W_k^{(2)})^{-1}$ has up to  $\omega^{-7}$,
even though retaining only two time-derivatives.
The 2nd-order subtraction term for the power spectrum uses
\be\label{v2subsq}
|v^{(2)}_k|^2   =(2 W_k^{(2)})^{-1}
\ee
which is given by (\ref{W2xi}).
The  2nd-order subtraction term for $\rho_k$ has threes terms,
its   $k^2$  and mass terms are $ \propto |v^{(2)}_k|^2 $,
and its  time derivative term is proportional to
\ba\label{energyk}
|(\frac{v^{(2)}_k}{a})' |^2    =
 \frac{1}{a^2} \Big( |v^{(2)}_k \, ' |^2
 - \frac{a'}{a} (v_k ^{(2)}\, ' v_k^{(2)}\, ^*  + v_k^{(2)} v_k^{(2)}\, ^*\, ')
   +(\frac{a'}{a})^2 |v^{(2)}_k|^2     \Big)  .
\ea
By
\ba
v_k^{(2)\, '}(\tau)
&  = & \frac{1}{ \sqrt 2}
  \Big( \frac{- W_k^{(2)\, '} }{  2 (W_k^{(2)} )^{3/2} }
  -i  \frac{W_k^{(2)} }{\sqrt{W_k^{(2)} }} \Big)
   e^{ -i \int^{\tau} W_k^{(2)} d\tau'} ,
\ea
one has up to the 2nd adiabatic order
$W_k^{(2)\, '} \simeq \omega '=     \frac{ m^2 a^2  }{\omega} \frac{a'}{a}$,
\be
|v^{(2)}_k \, ' |^2  = \frac12
      \Big(  \frac{ (W_k^{(2)\, '})^2 }{4 (W_k^{(2)})^{3} } +  W_k^{(2)}  \Big)
   =\frac{\omega}{2} +\frac{7  m^4 a^2 a'^2}{16 \omega ^5}
   -\frac{ m^2 (a a''+   a'^2 )}{8\omega ^3}
   +(\xi-\frac16)\frac{3    a''}{2\omega a } ,
\ee
\ba
v_k ^{(2)}\, ' v_k^{(2)}\, ^*  + v_k^{(2)} v_k^{(2)}\, ^*\, '
&= &  - \Big(  \frac{ W_k^{(2)\, '} }{2 W_k^{(2)} } +i W_k^{(2)}  \Big)
   (2W_k^{(2)} )^{-1}
   - \Big(  \frac{ W_k^{(2)\, '} }{2 W_k^{(2)} } - i W_k^{(2)}  \Big)
   (2W_k^{(2)} )^{-1}  \nn \\
& = & -   \frac{ W_k^{(2)\, '} }{2 ( W_k^{(2)})^2 }
   =-\frac{m^2aa'}{2\omega^3}  .
\ea
Putting together, one has
\ba
|(\frac{v^{(2)}_k}{a})' |^2
& = &    \frac{1}{a^2} \Bigg[\frac{\omega }{2}
+\frac{1}{2  \omega } \frac{a'^2}{ a^2}
+\frac{m^2\left(3a'^2-a a''\right)}{8 \omega ^3}
+\frac{7  m^4 a^2  a'^2}{16 \omega ^5}
+(\xi-\frac16)\frac{3  a''}{2 \omega a } \Bigg] .
\ea
Similarly,  the 4th-order adiabatic  mode
\be\label{vkWKB4}
v_k^{(4)}(\tau)
  =(2W_k^{(4)}(\tau))^{-1/2}
    e^{  -i \int^{\tau} W_k^{(4)}(\tau')d\tau'} ,
\ee
the 4th-order effective frequency is derived  by iteration
\be\label{Wk4}
W_k^{(4)}= \l[ k^2 + m^2 a^2 + (\xi -\frac16)a^2R
-\frac12 \l( \frac{ W^{(2)}_k\,  '' }{ W^{(2)} _k }
- \frac32 \big( \frac{W^{(2)}_k\,  '}{W^{(2)}_k} \big)^2 \r) \r]^{\frac12} .
\ee
Dropping terms of five time derivatives, one obtains
{\allowdisplaybreaks
\ba\label{W4s}
W_k^{(4)} & = & \omega +  (\xi-\frac16)  \frac{3 a''}{a\omega}
   -   \frac{m^2 (a'' a+ a'\,  ^2)}{4 \omega^3}
   + \frac{5 m^4 a'\, ^2 a^2 }{8 \omega^5}
             \nn \\
&&    +  \frac{m^2 (a'''' a   + 4 a''' a'  + 3 a'' a'')}{16  \omega^5}
      - \frac{m^4 \left(19 a^2 a''\, ^2+19 a' \, ^4
         +28 a^2 a'''  a' +122 a a' \, ^2 a'' \right) }{32 \omega^7}
             \nn \\
&&    +   \frac{221  m^6 (a'' a' \,^2 a^3 + a'\, ^4 a^2)}{32 \omega^9}
      -  \frac{ 1105 m^8 a^4 a'\, ^4}{128  \omega^{11}}
         \nn  \\
&& + (\xi -\frac{1}{6})
   \Big[ - \frac{3}{4\omega^3}( \frac{ a''''}{a}-\frac{ a''\, ^2}{a ^2}
     -\frac{2 a'''  a' }{a ^2}+\frac{2 a'\, ^2 a''}{ a^3})
                  \nn \\
&&  + \frac{m^2}{4\omega^5}(15  a' a''' -6  \frac{a'\, ^2 a''}{a}
           + 9  a'' a'')
    - \frac{ 75 m^4 a a'\, ^2  a''}{8\omega^7}     \Big]
    -\frac92  \Big(\xi-\frac16 \Big)^2 \frac{a''\, ^2}{a^2 \omega^3}
\ea
}
which is the same as Bunch's (2.20) \cite{Bunch1980}, and
{\allowdisplaybreaks
\ba\label{W4gen}
(W_k^{(4)})^{-1} & =  &
\frac{1}{\omega }+\frac{m^2  (a a''+ a'\, ^2 )}{4 \omega ^5}
  -\frac{5 m^4 a^2  a'\, ^2}{8 \omega^7}
 -3 (\xi -\frac{1}{6} )  \frac{1}{\omega ^3} \frac{a''}{a}   \nn \\
&&  -\frac{m^2  (3 a''\, ^2 + a''''  a + 4 a''' a' )}{16 \omega ^7}
    +\frac{7 m^4 \left(3 a^2 a''\, ^2 +3  a'\, ^4+18 a  a'\, ^2 a''
        +4 a^2 a'''  a'\right)}{32 \omega ^9} \nn \\
&& -\frac{231  m^6 a^2 \left( a'\, ^4+a  a'\, ^2 a''\right)}{32 \omega ^{11}}
   +\frac{1155 m^8 a^4   a'\, ^4}{128 \omega ^{13}} \nn \\
&& +  (\xi -\frac{1}{6} )
  \Big[ \frac{3}{4 \omega ^5} ( -\frac{a''\, ^2}{a^2}  + \frac{a'''' }{a}
  +2 \frac{ a'\, ^2 a''}{a^3}  -2 \frac{a'''  a'}{a^2})
     \nn \\
&& -\frac{15 m^2   \left( a''\, ^2 +a'''  a'\right)}{4 \omega ^7}
    +\frac{105  m^4 a  a'\, ^2 a''}{8 \omega ^9} \Big]
    + (\xi -\frac{1}{6} )^2 \, \frac{27}{2\omega ^5} \frac{a''\,^2}{a^2}  .
\ea
}
Keeping only up to second order time derivative terms,
  $W_k^{(4)}$ and $(W_k^{(4)})^{-1}$ reduce to
 the 2nd-order expressions  $W_k^{(2)}$ and $(W_k^{(2)})^{-1}$
in (\ref{Wk2nd}) and (\ref{W2xi}),  respectively.
Notice that for the special case $\xi=\frac16 $ and $m=0$,
we have
\be
W_k^{(4)} =W_k^{(2)}=W_k^{(0)}=k,
\ee
which holds for a general RW spacetime,
also implies Eq.(\ref{v1024})  in Sect \ref{section4} for de Sitter space.

From the effective frequency $W_k^{(4)}$ in (\ref{W4s})
and  its reciprocal (\ref{W4gen})
we can calculate the following quantities
that  occur in the 4th-order subtraction terms for
the power spectrum and spectral stress tensor.
{\allowdisplaybreaks
\be\label{v2sub}
|v^{(4)}_k|^2   =  (2 W_k^{(4)})^{-1} ,
\ee
\ba
|v^{(4)}_k \, ' |^2  & = & \frac12
      \Big(  \frac{ (W_k^{(4)\, '})^2 }{4 (W_k^{(4)})^{3} } +  W_k^{(4)}  \Big) \nn \\
&  = &\frac{\omega}{2} +\frac{7  m^4 a^2 a'^2}{16 \omega ^5}
   -\frac{ m^2 (a a''+   a'^2 )}{8\omega ^3}
   +(\xi-\frac16)\frac{3    a''}{2\omega a }
   +\frac{m^2 }{64 \omega ^5}\left(2 a a''''+6 a''^2+8 a' a'''\right)\nn \\
&&
   -\frac{m^4 }{256 \omega ^7}\left(76 a'^4+536 a a'^2 a''+76 a^2 a''^2
   +128 a^2 a' a'''\right) \nn \\
&& +\frac{259 m^6 }{512 \omega ^9} \left(8 a^2 a'^4+8 a^3 a'^2 a''\right)
  -\frac{1365  m^8 a^4 a'^4}{256 \omega ^{11}} \nn \\
&& + (\xi -\frac{1}{6} ) \Big[-\frac{1}{16 \omega ^3}
     (\frac{6 a''''}{a}
   +\frac{12 a'^2 a''}{a^3}-\frac{6 a''^2}{a^2}-\frac{12 a' a'''}  {a^2} ) \nn \\
&& + \frac{ m^2 a^2 }{64 \omega ^5}
    (-\frac{96 a'^2 a''}{a^3}+\frac{72 a''^2}{a^2} +\frac{168 a' a'''}{a^2})
    - \frac{105  m^4 a a'^2 a''}{16 \omega ^7}\Big]
    - (\xi -\frac{1}{6} )^2\frac{9}{4\omega ^3} \frac{a''^2}{a^2} , \nn \\
\ea
}
\ba
v_k ^{(4)}\, ' v_k^{(4)}\, ^*  + v_k^{(4)} v_k^{(4)}\, ^*\, '
& = &  -   \frac{ W_k^{(4)\, '} }{2 ( W_k^{(4)})^2 } \nn \\
& = &  -\frac{m^2aa'}{2\omega^3}
  +\frac{ m^2 (3a'a''+aa''')}{8 \omega ^5}
  -\frac{5m^4 (a  a'^3+a^2a'a'')}{4 \omega ^7}
  +\frac{35 m^6 a^3  a'^3}{16 \omega ^9}    \nn \\
&&  + (\xi -\frac{1}{6} )(\frac{9 m^2 a'a''}{2 \omega ^5}
    +\frac{3a'a''}{2 a^2 \omega ^3}-\frac{3 a'''}{2 a \omega ^3}) ,
\ea
putting together yields
{\allowdisplaybreaks
\ba
|(\frac{v^{(4)}_k}{a})' |^2 &= & \frac{1}{a^2}\Big[|v_k^{(4)'}|^2
 -\frac{a'}{a}(v_k^{(4)'}v_k^{(4)*}+v_k^{(4)}v_k^{(4)*'})
 +(\frac{a'}{a})^2|v_k^{(4)}|^2\Big] \nn \\
&= &    \frac{1}{a^2} \Bigg[\frac{\omega }{2}
+\frac{1}{2  \omega } \frac{a'^2}{ a^2}
+\frac{m^2\left(3a'^2-a a''\right)}{8 \omega ^3}
+\frac{7  m^4 a^2  a'^2}{16 \omega ^5}
+(\xi-\frac16)\frac{3  a''}{2 \omega a } \nn \\
&& +\frac{m^2\left(a^3 a'''' +3a''^2a^2+4 a'^4-8aa'^2 a''\right)}{32 \omega ^5 a^2}
  +\frac{m^4\left(41a'^4-54 aa'^2 a''-19 a^2a''^2-32a^2a'a'''\right)}{64 \omega ^7}
     \nn \\
&&  +\frac{m^6 \left(119 a^2a'^4+259 a^3a'^2 a''\right)}{64 \omega ^9}
    -\frac{1365 a^4 m^8 a'^4}{256 \omega ^{11}}
           \nn \\
&&-(\xi-\frac16)\Big(\frac{1}{8\omega^3}(\frac{3 a''''}{a}-\frac{18a' a'''}{ a^2}
    +\frac{30 a'^2 a''}{ a^3}-\frac{3 a''^2}{a^2})
    +\frac{m^2}{8 \omega ^5  a }(48  a'^2a''-9a a''^2-21 aa'a''') \nn \\
&&  +\frac{105  m^4 a a'^2 a''}{16 \omega ^7} \Big)
    -(\xi-\frac16)^2 \frac{9}{4 \omega^3} \frac{a''^2}{a^2}\Bigg] .
\ea
}
So far the  formulae  given above  are valid for a general flat RW spacetime.

For de Sitter space,
the scale factor satisfies  the following relation
\[
 (-\frac{1}{6})
  \Big[ \frac{3}{4 \omega ^5} ( -\frac{a''\, ^2}{a^2}  + \frac{a'''' }{a}
  +2 \frac{ a'\, ^2 a''}{a^3}  -2 \frac{a'''  a'}{a^2})
             \Big]
    + ( -\frac{1}{6} )^2 \, \frac{27}{2\omega ^5} \frac{a''\,^2}{a^2}
    =0.
\]
By this relation, in the special case $\xi=0$ and $m=0$,
we find that
\be
W_k^{(4)} =W_k^{(2)}, ~~~~  (W_k^{(4)})^{-1} =(W_k^{(2)})^{-1}
\ee
which imply  Eq.(\ref{rhopxi0m0}) in Sect \ref{section3}.

\section{  The subtraction terms for the stress tensor}\label{sectionC}

Substitution of the  0th-order adiabatic mode $v^{(0)}_k$ of (\ref{vkWKB0})
into (\ref{rhok}) (\ref{sprectpressure}) in place of $v_k$
yields the 0th-order subtraction term for stress tensor
{\allowdisplaybreaks
\ba \label{rhoA0}
\rho_{k\, A0}  &= & \frac{ k^3}{4\pi^2 a^4}
 \Big[ |v^{(0)} _k\, '|^2 + k^2  |v^{(0)}_k|^2 +m^2 a^2 |v^{(0)}_k|^2 \nn \\
&&   + (6\xi-1) \Big(
   \frac{a'}{a} (v^{(0)}_k\, '  v^{(0)}\, ^*_k + v_k^{(0)} v^{(0)}\, ^*\, '_k  )
    -  \frac{a'\, ^2}{a^2}  |v^{(0)}_k|^2  \Big) \Big] \nn \\
&&   \simeq   \frac{ k^3}{4\pi^2 a^4}  \omega ,
\ea
\ba \label{pkA0}
p_{k\, A 0}
& = & \frac{k^3}{4 \pi^2 a^4}
  \Bigg[   \frac13 |v^{(0)} _k\, '|^2 + \frac13 k^2  |v^{(0)}_k|^2
      - \frac13 m^2 a^2 |v^{(0)}_k|^2 \nn \\
&&  + 2(\xi-\frac16)\Big( -2 |v^{(0)} _k\, '|^2
    + 3 \frac{a'}{a} (v^{(0)}_k\, '  v^{(0)}\, ^*_k + v_k^{(0)} v^{(0)}\, ^*\, '_k) \nn \\
&&  - 3(\frac{a'}{a})^2 |v_k^{(0)}|^2
     +  2(k^2 + m^2 a^2) |v_k^{(0)}|^2
    + 12 \xi  \frac{a''}{a} |v_k^{(0)}|^2  \Big) \Bigg]  \nn \\
& \simeq &  \frac{k^3}{12 \pi^2 a^4}  \Big[ \omega - \frac{m^2 a^2}{\omega} \Big] ,
\ea
\be\label{trcsb0}
\langle T^\mu\, _\mu \rangle_{k\, A0}
= \rho_{k\, A 0} -3 p_{k\, A 0}
=\frac{m^2 k^3}{2\pi^2 a^2} \frac{1}{2\omega }.
\ee
}
Note that these   contain no time derivatives of $a(\tau)$,
and do not depend on the coupling $\xi$.

Substitution of the 2nd-order adiabatic mode $v^{(2)}_k$ of (\ref{vkWKB2})
into (\ref{rhok}) (\ref{sprectpressure}) in place of $v_k$
yields the  2nd-order subtraction term for
the spectral energy density and pressure
\ba   \label{rhoA22count}
\rho_{k\,A 2}
& = & \frac{k^3}{4\pi^2 a^4}
\Big[  \omega + \frac{m^4 a^4}{8\omega^5} \frac{a'\,^2}{a^2}
  + ( \xi-\frac16) \big( -\frac{3}{\omega} \frac{a'\,^2}{a^2}
                        - \frac{3m^2 a'\, ^2}{\omega^3}  \big) \Big],
\ea
\ba\label{pA2count}
 p_{k\, A 2} & = &   \frac{k^3}{12 \pi^2 a^4}  \Bigg[
\omega - \frac{m^2 a^2}{\omega }
- \frac{m^4 a^4}{8\omega^5} (  \frac{2 a''}{a}+\frac{a'\,^2}{a^2})
  +\frac{5 m^6 a^6}{8 \omega^7} \frac{a'\,^2}{a^2} \nn \\
&&    + (\xi-\frac16) \Big( \frac{1}{\omega} (6\frac{a''}{a}- 9\frac{a'\,^2}{a^2})
       +\frac{6m^2 a^2}{\omega^3}(\frac{a''}{a}-\frac{a'\,^2}{a^2})
      - \frac{9m^4 a^4}{\omega^5}\frac{a'\,^2}{a^2} \Big)
        \Bigg]  .
\ea
They reduce to (\ref{rhoA0}) (\ref{pkA0})
when the  time derivatives of $a(\tau)$
are dropped.

Substitution of the  4th-order adiabatic mode $v^{(4)}_k$
into (\ref{rhok}) (\ref{sprectpressure}) in place of $v_k$,
yields  the 4th-order subtraction  term for spectral energy density
and pressure  \cite{Bunch1980,AndersonParker1987}
{\allowdisplaybreaks
\ba\label{countrh1}
\rho_{k\,A 4}
& = & \frac{k^3}{4\pi^2 a^4}
\Big[
\omega + \frac{m^4 a^4}{8\omega^5} \frac{a'\,^2}{a^2}
-\frac{m^4 a^4}{32\omega^7} (2\frac{a''' a'}{a^2} -\frac{a''\,^2}{a^2}
+\frac{4a''  a'\,^2}{a^3}-\frac{a'\,^4}{a^4}) \nn \\
&& + \frac{7 m^6 a^6 }{16\omega^9} (\frac{a''  a'\,^2}{a^3}+\frac{a'\,^4}{a^4})
  - \frac{105 m^8 a^8 }{128\omega^{11}}  \frac{a'\,^4}{a^4} \nn \\
&&+ (\xi-\frac16) \big[ -\frac{3}{\omega} \frac{a'\,^2}{a^2} - \frac{3m^2 a'\, ^2}{\omega^3}
   + \frac{3m^2 a^2}{4\omega^5}
      (\frac{2a''' a' }{a^2} -\frac{a''\,^2 }{a^2}-\frac{a'\,^4}{a^4}) \nn \\
&&  -\frac{15 m^4 a^4}{8\omega^7}
        (\frac{4a''  a'\,^2}{a^3}+\frac{3a'\,^4}{a^4})
    + \frac{105 m^6 a^6}{8\omega^9}\frac{a'\,^4}{a^4} \big] \nn \\
&& + (\xi-\frac16)^2  \big[ -\frac{9}{2\omega^3}
    ( \frac{2a'''  a'}{a^2} -\frac{a'' \, ^2}{a^2}-\frac{4a''  a'\,^2}{a^3})
    + \frac{27 m^2 a^2}{\omega^5} \frac{a''  a'\, ^2}{a^3}
\big]
\Big] ,
\ea
}
{\allowdisplaybreaks
\ba\label{countpress}
 p_{k\, A 4} &  =  & \frac13 (\frac{k^3}{4\pi^2 a^4})  \Bigg[
   \omega - \frac{m^2 a^2}{\omega }
    - \frac{m^4 a^4}{8\omega^5} (  \frac{2 a''}{a}+\frac{a'\,^2}{a^2})
    \nn \\
&&  +\frac{5 m^6 a^6}{8 \omega^7} \frac{a'\,^2}{a^2}
    +\frac{m^4 a^4}{32\omega^7} (\frac{2 a''''}{a} + 6 \frac{a'''a'}{a^2}
        +7 \frac{a''\,^2}{a^2} -\frac{4a''  a'\,^2}{a^3} +  \frac{a'\, ^4}{a^4} )
         \nn \\
&& -\frac{ m^6 a^6 }{32\omega^9}
       (\frac{28a'''  a'}{a^2}+21 \frac{a''\,^2}{a^2} +\frac{112 a'' a'\,^2}{a^3}
       + 7 \frac{a'\,^4}{a^4})            \nn \\
&& + \frac{21 m^8 a^8 }{128\omega^{11}}
      ( 44\frac{a'' a'\,^2}{a^3} +39\frac{a'\,^4}{a^4} )
   - \frac{1155 m^{10} a^{  10 } }{128 \omega^{13}}  \frac{a'\,^4}{a^4}
         \nn \\
&& + (\xi-\frac16) \Big( \frac{1}{\omega} (6\frac{a''}{a}- 9\frac{a'\,^2}{a^2})
       +\frac{6m^2 a^2}{\omega^3}(\frac{a''}{a}-\frac{a'\,^2}{a^2})
       \nn  \\
&& + \frac{3m^2 a^2}{4\omega^5}
      (-\frac{2 a'''' }{a}    +\frac{2a''' a' }{a^2}
        +\frac{4a'' a'\,^2 }{a^3} -\frac{a''\,^2 }{a^2}-\frac{3 a'\,^4}{a^4})
   - \frac{9m^4 a^4}{\omega^5}\frac{a'\,^2}{a^2}
           \nn \\
&&  +\frac{15 m^4 a^4}{8\omega^7}
     (\frac{8a'''  a'}{a^2}+\frac{6a''\,^2}{a^2} +\frac{12 a''a'\,^2}{a^3}
      -\frac{5a'\,^4}{a^4})
            \nn \\
&&  - \frac{105 m^6 a^6}{2\omega^9}(\frac{2a''a'\,^2}{a^3}+\frac{a'\,^4}{a^4})
         + \frac{945 m^8 a^8}{8\omega^{11}} \frac{a'\,^4}{a^4}
         \Big)
             \nn \\
&& + (\xi-\frac16)^2  \Big(
    \frac{9}{2\omega^3}
    ( \frac{2a''''  }{a}-\frac{10  a''' a'}{a^2}
     -\frac{5 a'' \, ^2}{a^2}   +\frac{16a''  a'\,^2}{a^3} )
      \nn \\
&&  -\frac{27 m^2 a^2}{2\omega^5}
       (\frac{4a''' a'}{a^2} + \frac{3a'' \, ^2}{a^2}-\frac{8a''  a'\,^2}{a^3})
       +\frac{135  m^4 a^4}{\omega^7}\frac{a''  a'\,^2}{a^3}   \Big)
  \Bigg]  .
\ea
}
They reduce to (\ref{rhoA22count}) (\ref{pA2count})
when the fourth order  time derivatives of $a(\tau)$ are dropped.
The subtraction term of the spectral trace is given by
{\allowdisplaybreaks
\ba\label{trcsub}
\langle T^\mu\, _\mu  \rangle_{k\, A 4} & = &
\rho_{k\, A 4 } -3p_{k \, A 4}  \nn \\
& = & \frac{k^3}{4\pi^2 a^4}
\Big[
 \frac{m^2 a^2}{\omega } + \frac{m^4 a^4}{4\omega^5} (\frac{a''}{a}+ \frac{a'\,^2}{a^2})
   -\frac{5 m^6 a^6}{8 \omega^7} \frac{a'\,^2}{a^2}
   -\frac{m^4 a^4}{16\omega^7}
     (\frac{a''''}{a} +4 \frac{a'''a'}{a^2}+ \frac{3a''\,^2}{a^2})
         \nn \\
   && + \frac{ m^6 a^6 }{32\omega^9}
       (\frac{28a'''  a'}{a^2}+\frac{126 a'' a'\,^2}{a^3}
       +21 \frac{a''\,^2}{a^2}+21 \frac{a'\,^4}{a^4}) \nn \\
   &&- \frac{231  m^8 a^8 }{32 \omega^{11}}
      (\frac{a'' a'\,^2}{a^3}+ \frac{a'\,^4}{a^4})
      + \frac{1155 m^{10} a^{10} }{128 \omega^{13}}  \frac{a'\,^4}{a^4}
          \nn \\
&& +(\xi-\frac16)  \big[ -\frac{6}{\omega} (\frac{a''}{a}-\frac{a'\,^2}{a^2})
     - \frac{3m^2 a^2}{\omega^3}(\frac{2a''}{a}-\frac{a'\,^2}{a^2}) \nn \\
&&     + \frac{9m^4 a^4}{\omega^5}\frac{a'\,^2}{a^2}
       + \frac{3m^2 a^2}{2\omega^5}
      (\frac{a'''' }{a} -\frac{2a'' a'\,^2 }{a^3} +\frac{a'\,^4}{a^4}) \nn \\
&&  -\frac{15 m^4 a^4}{4\omega^7}
        (\frac{4a'''  a'}{a^2}+\frac{3a''\,^2}{a^2}
         +\frac{8a''a'\,^2}{a^3}-\frac{a'\,^4}{a^4}) \nn \\
&& + \frac{105 m^6 a^6}{8\omega^9}(\frac{8a''a'\,^2}{a^3}+\frac{5a'\,^4}{a^4})
         - \frac{945 m^8 a^8}{8\omega^{11}} \frac{a'\,^4}{a^4} \big] \nn \\
&& + (\xi-\frac16)^2  \big[ -\frac{9}{\omega^3}
    ( \frac{a''''  }{a}-\frac{4a''' a'}{a^2}
    -\frac{3a''\, ^2}{a^2}+\frac{6a'' a'\,^2}{a^3})
       \nn \\
&&     + \frac{27 m^2 a^2}{2\omega^5}
       (\frac{4a''' a'}{a^2} + \frac{3a'' \, ^2}{a^2}-\frac{6a''  a'\,^2}{a^3})
       -\frac{135  m^4 a^4}{\omega^7}\frac{a''  a'\,^2}{a^3}  \big]   \Big] .
\ea
}
The  formulae (\ref{rhoA0})--(\ref{trcsub})
  are valid for a general flat RW spacetime.

In de Sitter space,
\be\label{desi}
( \frac{2a'''  a'}{a^2} -\frac{a'' \, ^2}{a^2}-\frac{4a''  a'\,^2}{a^3})=0,
\ee
\be\label{desi2}
( \frac{2a''''  }{a}-\frac{10  a''' a'}{a^2}
     -\frac{5 a'' \, ^2}{a^2}   +\frac{16a''  a'\,^2}{a^3} )=0,
\ee
\be
( \frac{a''''  }{a}-\frac{4a''' a'}{a^2}
    -\frac{3a''\, ^2}{a^2}+\frac{6a'' a'\,^2}{a^3})=0,
\ee
so that
the $(\xi-\frac16)^2/\omega^3$ terms in
$\rho_{k\, A 4}$, $p_{k\, A 4}$
and $\langle T^\mu\, _\mu  \rangle_{k\, A 4}$ vanish.
This will simplify calculations
in Sections \ref{section3}, \ref{section4}, \ref{section5}  to some extent.

 In the following we check that
the four-divergence of the subtraction terms is zero, at each adiabatic order.
This is actually implied  by the construction
of WKB mode $v_k^{(n)}$ in (\ref{vn}) which satisfies the equation  (\ref{equWk}),
ie, $v_k^{(n)}$ satisfies the field equation (\ref{equvk}) to the $n$th order.
We still give  explicit calculations as a checking in the following.
By (\ref{rhoA0}) (\ref{pkA0}),
the four-divergence of the 0th-order subtraction terms is zero,
\ba\label{rhoA0cons}
\langle  T^{0\nu  }\, _{; \nu} \rangle_{k\, A 0}
& = & \rho_{k\, A 0}' +  3 \frac{a'}{a} (\rho_{k\, A 0} +  p_{k\, A 0 }) \nn \\
& =&  \frac{k^3}{4\pi^2} \Big( -\frac{ a' \left(3  m^2 a^2+4 k^2\right) }{a^5 \omega}
 +\frac{a'\left(3  m^2 a^2 +4 k^2\right)  }{a^5 \omega} \Big)
= 0.
\ea
Thus, the 0th-order regularized spectral stress tensor
respects the covariant conservation
\be \label{rho0thconsv}
\langle  T^{\mu\nu (0)}\, _{; \nu} \rangle_{k \, reg} =
\langle  T^{\mu\nu }\, _{; \nu} \rangle_{k} -
\langle  T^{\mu\nu  }\, _{; \nu} \rangle_{k\, A 0} =0
\ee
which holds
for a general coupling $\xi$ and a mass $m$.
For the scalar field considered in this paper,
the $k$-modes of stress tensor are independent of each other,
so,  after  $k$-integration of the spectral stress tensor,
the regularized stress tensor also respects the covariant conservation.

Similarly, from  (\ref{rhoA22count}) (\ref{pA2count}) for the 2nd-order,
by explicit calculation,
{\allowdisplaybreaks
\ba \label{consv2nd}
\rho_{k\, A 2}' & = & \frac{k^3}{4\pi^2 a^4}
\Big[
\Big( \frac{ m^2 a a'}{\omega}+\frac{ m^4 a^2 a' a''
   + m^4 a \left(a'\right)^3 }{4\omega ^{5/2}}
   -\frac{5 m^6 a^3 \left(a'\right)^3}{8\omega ^{7/2} } \Big) \nn \\
&& + (\xi -\frac{1}{6} ) \Big(\frac{6  a'\, ^3 -6 a a' a'' }{a^3 \omega }
+\frac{3 m^2 \left(a'\right)^3 -6 m^2 a a' a'' }{a \omega ^{3/2}}
+\frac{9 m^4\text{  }a \left(a'\right)^3}{\omega ^{5/2}}\Big) \nn \\
&& -\frac{4a'}{a} \Big(\omega - (\xi -\frac{1}{6} )
   \Big(\frac{3 a'\, ^2}{\omega a^2} +\frac{3 m^2  a'\, ^2}{\omega ^{3/2}} \Big)
  +\frac{ m^4 a^2  a'\, ^2}{8\omega ^{5/2}} \Big)  \Big]   ,
\ea
}
\be
3 \frac{a'}{a} (\rho_{k\, A 2} +  p_{k, A 2 })
= -  {\text{ r.h.s. of}} ~ (\ref{consv2nd})  ,
\ee
so the four divergence of the 2nd order subtraction term is zero,
\be \label{consvrhoA2}
 \rho_{k\, A 2}' +  3 \frac{a'}{a} (\rho_{k\, A 2} +  p_{k, A 2 }) = 0,
\ee
 and the covariant conservation is respected by the 2nd-order regularized stress tensor,
\be \label{consvrhoreg2}
\langle  T^{\mu\nu (2)}\, _{; \nu} \rangle_{reg} =
\langle  T^{\mu\nu  }\, _{; \nu} \rangle -
\langle  T^{\mu\nu  }\, _{; \nu} \rangle_{A 2} =0 .
\ee

Finally,  from (\ref{countrh1}) and (\ref{countpress})
it is checked  that
the four-divergence of the 4th order subtraction term   is zero,
\be\label{cons4t}
 \rho_{k\, A 4}' +  3 \frac{a'}{a} (\rho_{k\, A 4} +  p_{k, A 4 }) = 0,
\ee
(the explicit expressions of the above two terms
are  lengthy and we do not  list here,)
so  the covariant conservation is respected by
the 4th-order regularized stress tensor,
\be
\langle  T^{\mu\nu (4)}\, _{; \nu} \rangle_{reg} =
\langle  T^{\mu\nu  }\, _{; \nu} \rangle -
\langle  T^{\mu\nu  }\, _{; \nu} \rangle_{A 4} =0 .
\ee
The   covariant conservations
(\ref{rho0thconsv}) (\ref{consvrhoA2}) (\ref{cons4t})
  are valid for a general flat RW spacetime.


\begin{thebibliography}{99}


\bibitem {UtiyamaDeWitt1962} R. Utiyama, and B.S. DeWitt,
          J. Math. Phys.  3, 608 (1962).

\bibitem{FeynmanHibbs1965} R.P. Feynman, and A.R. Hibbs,
     \textit {Quantum mechanics and path integration}  (McGraw-Hill,New York,  1965).

\bibitem{Peebles1993} P.J.E. Peebles, \textit {Principles of Physical Cosmology}
              (Princeton University Press, Princeton, 1993).

\bibitem{WangZhang2017} Y. Zhang,  A\&A. 464, 811  (2007);
          Y. Zhang, and H.X. Miao,  RAA. Vol.9 No.5,  501 (2009);
          Y. Zhang, and Q. Chen,   A\&A.  581, A53 (2015);
          Y. Zhang, Q. Chen, S.G Wu,  RAA. Vol.19 No.4,  53 (2019);
          B. Wang, and Y. Zhang,    Phys. Rev. D  98, 103522 (2017);
                                   Phys. Rev. D 98, 123019  (2018);
                                   Phys. Rev. D 99, 123008  (2019);
          Y. Zhang,  F. Qin, and B. Wang,  Phys. Rev. D 98, 103523 (2017).


\bibitem{MaBertschinger1995} C.P. Ma, and E. Bertschinger,
           Astrophys. J. 455,  7  (1995).


\bibitem{ZhaoZhang2006} W. Zhao, and Y. Zhang,
              Phys. Rev. D  74,  083006  (2006);
             T.Y. Xia,  and  Y. Zhang,
             Phys. Rev. D 78,   123005 (2008);
             Phys. Rev. D 79, 083002 (2009);
             Z. Cai,  and Y. Zhang,  Class. Quant. Grav. 29, 105009  (2012).


\bibitem{CandelasRaine1975}  P. Candelas,  and D.J. Raine,
          Phys. Rev. D 12, 965  (1975).


\bibitem{DowkerCritchley1976}  J.S. Dowker, and R. Critchley,
          Phys. Rev. D 13, 3224  (1976);
          Phys. Rev. D 16, 3390 (1977).

\bibitem{Brown1977} L.S. Brown,   Phys.  Rev.  D  15,  1469 (1977);
         L.S. Brown and J. P. Cassidy,   Phys.  Rev.  D 15,  2810  (1977).


\bibitem{Bunch1979} T.S. Bunch,
        J. Phys. A: Math  Gen.  12,  517 (1979).



\bibitem{Christensen1976} S.M.  Christensen,   Phys.   Rev.  D  14, 2490  (1976).

\bibitem{Christensen1978} S.M.  Christensen,   Phys.   Rev.  D  17, 946 (1978).



\bibitem{BunchChristensenFulling1978} T.S. Bunch, S.M. Christensen, and S.A. Fulling,
    Phys. Rev. D 18, 4435  (1978).

\bibitem{BunchDavies1978} T.S. Bunch, and P.C. W. Davies,
          Proc. R. Soc. Lond.   A360, 117 (1978);
          Proc. R. Soc. Lond.   A357, 381  (1977).


\bibitem{Wald1978} R. Wald,   Phys.  Rev.  D  17,  1477  (1978).

\bibitem{AdlerLiebermanNg1977} S.L. Adler,   J. Lieberman,     and  Y. J.Ng, Ann. Phys. (N.Y.) 106,  279   (1977).

\bibitem{Hawking1977} S. Hawking, Comm. math. Phys. 55, 133 (1977).


\bibitem{Fujikawa1980} K. Fujikawa, Phys. Rev. Lett. 44, 1733 (1980).



\bibitem{Schwinger1951} J.S. Schwinger, Phys. Rev. 82,  664 (1951).

\bibitem{DeWitt1975} S. DeWitt, Phys. Rep. 19C, 295  (1975).

\bibitem{ParkerFulling1974}  L. Parker, and S.A. Fulling,
   Phys. Rev. D 9, 341 (1974).


\bibitem{FullingParkerHu1974} S.A. Fulling,  L. Parker, and B.L. Hu,
   Phys. Rev. D 10, 3905 (1974).

\bibitem{HuParker1978} B.L. Hu, and L. Parker,
   Phys. Rev. D 17, 933 (1978).

\bibitem{BLHu1978}   B.L. Hu, Phys. Rev. D 18, 4460 (1978).


\bibitem{Birrell1978} N.  D.  Birrell,
             Proc. R. Soc. Lond. B. 361,  513  (1978).

\bibitem{Bunch1978} T.S. Bunch, J. Phys. A: Math  Gen. 11,  603 (1978).


\bibitem{BunchParker1979} T.S. Bunch, and L. Parker,
              Phys. Rev. D 20, 2499 (1979).

\bibitem{Bunch1980}  T.S. Bunch, J. Phys. A: Math  Gen. 13, 1297 (1980).



\bibitem{AndersonParker1987}  P.R. Anderson, and L. Parker,
            Phys. Rev. D 36, 2963 (1987).

\bibitem{BirrellDavies1982}N. D.  Birrell, and   P. C. W. Davies,
  \textit{ Quantum Fields in Curved Space}
         (Cambridge  University  Press,  Cambridge, 1982).


\bibitem{ParkerToms}L.  Parker,  and  D. J. Toms,
        \textit { Quantum Field  Theory in Curved Spacetime:
                 Quantized Fields and Gravity}
            (Cambridge  University   Press,   Cambridge, 2009).


\bibitem{Parker2007} L. Parker, arXiv:hep-th/0702216.


\bibitem{MarkkanenTranberg2013} T.  Markkanen, and A. Tranberg,
          JCAP 08,   045  (2013).

\bibitem{Markkanen2018}  T. Markkanen, JCAP 05,   001  (2018).


\bibitem{WangZhangChen2016}  D.G. Wang, Y. Zhang, and  J.W. Chen,
          Phys. Rev. D 94,   044033 (2016).


\bibitem{ZhangWangJCAP2018} Y. Zhang, and B. Wang,
                JCAP 11, 006 (2018).


\bibitem{Zhang1994} Y. Zhang,  Phys. Lett. B 340, 18  (1994).


\bibitem{Watson1958}  G.N.  Watson,
  \textit{ Treatise  on  the  theory  of  Bessel  functions}
         (Cambridge  University   Press,   Cambridge, 1958).


\bibitem{GradshteynRyzhik1980} I.S. Gradshteyn, and I.M. Ryzhik,
   \textit{ Tables of Integrals, Series, and Products} (Academic Press,  San Diego, 1980).


\bibitem{NISTHandbook2010} F.W.J. Olver, D.W. Lozier,  R.F. Boisvert, and C.W. Clark,
         \textit{NIST Handbook of Mathematical Functions}
         (Cambridge  University  Press, Cambridge, 2010).


\bibitem{Chakraborty1973} B. Chakraborty,
           J. Math.   Phys.  14,  188 (1973).


\end{thebibliography}
\end{document}